\documentclass[a4paper,11pt]{book}

\usepackage{ifpdf}

\ifpdf                          
  \usepackage[pdftex]{graphicx} 
  \DeclareGraphicsExtensions{.pdf,.jpg,.png}
  \usepackage[pdftex,colorlinks,plainpages=false,hyperindex,bookmarksopen,
              linkcolor=black,citecolor=black,urlcolor=black,pageanchor=false]{hyperref}
\else
  \usepackage[dvips]{graphicx}  
\fi

\usepackage[total={13cm,19cm},centering,includehead]{geometry} 
\usepackage[english]{babel}     
\usepackage[T1]{fontenc}        
\usepackage{textcomp}           
\usepackage{upgreek}            
\usepackage{fancyhdr}           
\usepackage{mathptmx}
\usepackage[scaled=.90]{helvet}
\usepackage{courier}
\usepackage{url}

\usepackage{bm}
\usepackage{amsmath}
\usepackage{amsthm}
\usepackage{amssymb}
\usepackage{multirow}
\newtheorem{thm}{Theorem}
\usepackage{lettrine}

\providecommand{\autore}{Andrea Modenini}                        
\providecommand{\tutor}{Giulio Colavolpe}                           
\providecommand{\coordinatore}{Marco Locatelli}      
\providecommand{\ciclodottorato}{XXVI}                         
\providecommand{\titolo}{Advanced transceivers for spectrally-efficient communications} 
\providecommand{\meseanno}{Gennaio 2014}                                     

\renewcommand{\chaptermark}[1]{\markboth{Chapter \thechapter{}. #1}
                                        {\thesection{}. #1}}

\makeatletter                 
\def\cleardoublepage{         
\clearpage                    
\if@twoside                   
  \ifodd\c@page               
  \else
    \mbox{}
    \thispagestyle{empty}
    \newpage
    \if@twocolumn
      \mbox{}
      \newpage
    \fi
  \fi
\fi
}
\makeatother

\DeclareMathAlphabet\mathbfcal{OMS}{cmsy}{b}{n}

\begin{document}
\title{Advanced transceivers for spectrally-efficient communications}
\author{Andrea Modenini}
\hyphenation{brush-less}      
\frenchspacing                
\linespread{1.2}\selectfont   
\graphicspath{{images/}}      

     %
%
{
  \thispagestyle{plain}       
  \linespread{1.2}\selectfont 
  \mbox{}
  \vspace{-2cm}
  \begin{center}
  \begin{figure}[h]
  \begin{center}
  \includegraphics[width=2.5cm]{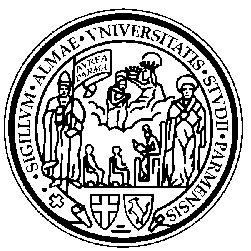}
  \end{center}
  \end{figure}
  \vspace{-5pt}
  \Large\textbf{UNIVERSIT\`A DEGLI STUDI DI PARMA}\\
  \small\textsc{DIPARTIMENTO DI INGEGNERIA DELL'INFORMAZIONE}\\
  \vspace{1.7cm}
  \normalsize\textit{Dottorato di Ricerca in Tecnologie
                     dell'Informazione}\\
  \normalsize\textit{\ciclodottorato{} Ciclo}\\
  \vspace{1.7cm}
  \large\textup{\autore}\\
  \vspace{1.4cm}
  \Large\textbf{\titolo}\\
  \vspace{1.7cm}
  \small\textsc{Dissertazione presentata per il conseguimento}\\
  \small\textsc{del titolo di Dottore di Ricerca}\\
  \vspace{1.2cm}
  \small\textsc{\meseanno}\\
  \end{center}
}
\cleardoublepage

{
  \thispagestyle{plain}       
  \linespread{1.5}\selectfont 
  \mbox{}
  \vspace{-0.8cm}
  \begin{center}
  \vspace{-10pt}
  \Large\textbf{UNIVERSIT\`A DEGLI STUDI DI PARMA}\\
  \vspace{0.3cm}
  \normalsize\textit{Dottorato di Ricerca in Tecnologie dell'Informazione }\\
  \vspace{0.3cm}
  \small\textit{\ciclodottorato{} Ciclo}\\
  \vspace{2cm}
  \Large\textbf{\titolo}\\
  \vspace{2cm}
  \end{center}
  \flushleft
  \normalsize{Coordinatore:}\\
  \textit{Chiar.mo Prof. \coordinatore}\\
  \vspace{0.5cm}
  {Tutor:}\\
  \textit{Chiar.mo Prof. \tutor}\\
  \vspace{0.8cm}
  \flushright \normalsize{{Dottorando: \textit{\autore}}}\\
  \vspace{1cm}
  \begin{center}
  \small\textup{\meseanno}\\
  \end{center}
}

\cleardoublepage
{
  \thispagestyle{plain}       
  \pagenumbering{roman}       
  \mbox{}
  \vspace{5cm}
  \Large\itshape\flushright to my adorable and wonderful wife,\\
  \vspace{-0.5cm}
  \Large\itshape\flushright to my beloved family,\\
  \vspace{-0.5cm}
  \Large\itshape\flushright to my dear friends, \\
  \vspace{-0.5cm}
  \Large\itshape\flushright and all those that supported me through time and believed in me\\
  \vspace{-0.5cm}
  \small\itshape\flushright (e.g. Giulio)\\
  \vspace{-0.5cm}
}

\frontmatter
\cleardoublepage
\renewcommand\contentsname{Contents}
\tableofcontents
\markboth{Contents}{Contents}
\listoffigures
\markboth{List of Figures}{List of Figures}
\listoftables
\markboth{List of Tables}{List of Tables}

\cleardoublepage
\pagenumbering{arabic}
\mainmatter
      \chapter*{Introduction}
\markboth{Introduction}{Introduction}
\addcontentsline{toc}{chapter}{\numberline{}Introduction}
\thispagestyle{empty}

{                                               
\flushright\footnotesize
\textit{Vanity, definitely my favorite sin. }\\
\vspace{0.2cm}
\textrm{-- The Devil's Advocate}\\
\vspace{1.0cm}
}                                               

\lettrine[lines=2]{T}{elecommunications} are a growing field in the global industry,
and their request is increasing more and more every year.
Due to this growing demand, the available bandwidths are getting insufficient to
fulfill the global request.
Independently from the kind of communication system (satellite, wireless, optical),
the desire of every telecommunications operator is to transmit at the highest possible rate
 in the available bandwidth for a given power. 
In more technical words, the aim is to maximize the {\it spectral efficiency}
of the communication systems.

In this thesis, we will consider techniques to improve the spectral efficiency of digital
communication systems, operating on the whole transceiver scheme.
First, we will focus on receiver schemes having detection algorithms with a complexity constraint.
We will optimize the parameters of the reduced detector 
with the aim of maximizing the {\it achievable information rate}. 
Namely, we will adopt the {\it channel shortening} technique (see \cite{FaMa73,RuPr12} and references therein).

Then, we will focus on a technique that is getting very popular in the last years (although presented
for the first time in 1975):
 {\it faster-than-Nyquist} signaling, and its extension which is {\it time packing} (see \cite{Ma75c,MaLa88,BaFeCo09b}
 and references therein).
Time packing is a very simple technique that consists in introducing intersymbol interference
on purpose with the aim of increasing the spectral efficiency of finite order constellations.

Finally, in the last chapters we will combine all the presented techniques, and we will consider
their application to satellite channels.

Although we will not consider here optical communications, we point out that
many of these techniques, can be applied (with suitable {\it tweaks})
also to these scenarios. It is worth to cite \cite{CoFoMoPi11,CoFo13,CoFo13b}.

The remainder of this thesis is organized as follows:
Chapter \ref{ch:basics} will introduce the basics of the work in this
thesis. Chapter \ref{ch:CS}
will focus entirely on the channel shortening technique,
illustrating also many practical detection schemes.
In Chapter \ref{chap:time_pack} we will turn our attention
to the time packing technique, showing its potential.
Finally in Chapters \ref{ch:cs_sat} and \ref{ch:spec_tpack}
we will {\it connect the dots} and apply the proposed techniques
to the satellite channel.

\section*{Publications}
\addcontentsline{toc}{section}{\numberline{}Publications}

The work in this thesis is part of the results of the research activities conducted during
the PhD studies, with the following publications:
\subsection*{Journals}
\begin{itemize}
 
\item A. Modenini, F. Rusek, and G. Colavolpe "Optimal transmit filters for ISI channels under channel shortening detection,"  IEEE Transactions on Communications, vol. 61, pp. 4997-5005, December 2013.

\item A. Piemontese, A. Modenini, G. Colavolpe, and N. Alagha "Improving the spectral efficiency of nonlinear satellite systems through time-frequency packing and advanced processing," IEEE Transactions on Communications, vol. 61, pp. 3404-3412, August 2013.

\item G. Colavolpe, A. Modenini, and F. Rusek, "Channel Shortening for Nonlinear Satellite Channels," Communications Letters, IEEE , vol.16, no.12, pp.1929-1932, December 2012.

\item G. Colavolpe, Tommaso Foggi, A. Modenini, and A. Piemontese, "Faster-than-Nyquist and beyond: how to improve spectral efficiency by accepting interference," Opt. Express 19, 26600-26609 (2011).
\end{itemize}
\subsection*{Conferences}
\begin{itemize}
 \item A. Piemontese, A. Modenini, G. Colavolpe, and N. Alagha, "Spectral Efficiency of Time-Frequency-Packed Nonlinear Satellite Systems," in 31th AAIA International communications satellite systems conference, Florence, Italy, October 2013.

 \item G. Colavolpe, and A. Modenini, "Iterative carrier syncrhonization in the absence of distributed pilots for low SNR applications," in Proc. Intern. Workshop of Tracking Telemetetry and Command System for Space Communications. (TTC'13), European Space Agency, Darmstadt, Germany, September 2013.

 \item A. Modenini, F. Rusek, and G. Colavolpe, "Optimal transmit filters for constrained complexity channel shortening detectors," in Proc. IEEE Intern. Conf. Commun. (ICC'13), Budapest, Hungary, June 2013, pp. 1688-1693.

 \item A. Modenini, G. Colavolpe, and N. Alagha, "How to significantly improve the spectral efficiency of linear modulations through time-frequency packing and advanced processing," in Proc. IEEE Intern. Conf. Commun. (ICC'12), Ottawa, Canada, June 2012, pp. 3430-3434.
\end{itemize}
\subsection*{Patents}
\begin{itemize}
 \item G. Colavolpe, A. Modenini, A. Piemontese, and N. Alagha, "Data detection method and data detector for signals transmitted over a communication channel with inter-symbol interference," assigned to ESA-ESTEC, The Neederlands. International patent application n. F027800186/WO/PCT, December 2012.

\end{itemize}

\clearpage
\section*{Common abbreviations}
\addcontentsline{toc}{section}{\numberline{}Common abbreviations}
\begin{tabbing}
\hspace*{2.3cm} \=  \=  \kill
AWGN \> additive white Gaussian noise \\
AIR \> achievable information rate \\
ASE \> achievable spectral efficiency \\
BCJR \> Bahl Cocke Jelinek Raviv (algorithm) \\
CS \> channel shortening \\
DTFT \> discrete time Fourier transform \\
DFT \> discrete Fourier transform \\
HPA \> high power amplifier \\
ICI \> interchannel interference \\
IMUX \> input multiplexer (filter) \\
IR \> information rate \\
ISI \> intersymbol interference \\
FTN \> faster-than-Nyquist \\
MAP \> maximum a posteriori \\
MF \> matched filter \\
MIMO \> multiple-input multiple-output \\
MMSE \> minimum mean square error \\
OMUX \> output multiplexer (filter) \\
PSK \> phase shift keying (modulation) \\
QAM \> quadrature amplitude (modulation) \\
SE \> spectral efficiency \\
WF \> whitening filter \\
WMF \> whitening matched filter \\

\end{tabbing}

\section*{Mathematical notation}
\addcontentsline{toc}{section}{\numberline{}Mathematical notation}
\begin{tabbing}
\hspace*{2.3cm} \=  \=  \kill
$h$ \> scalar (possibly complex)\\ 
$h^*$ \> complex-conjugated of the complex scalar\\ 
$\Re(h)$ \> real part of of the complex scalar \\
$\Im(h)$ \> imaginary part of the complex scalar\\
$\bm{h}$ \> vector \\
$\bm{h}^T$ \> transposed vector \\
$\bm{h}^\dagger$ \> transposed and conjugated vector (Hermitian)\\
$\bm{H}$ \> matrix \\
$\bm{I}$ \> identity matrix \\
$(\bm{H})_{ij}$ \> scalar entry $(i,j)$ of the matrix \\
$\bm{H}^T$ \> transposed matrix \\
$\bm{H}^\dagger$ \> transposed and conjugated matrix (Hermitian)\\
$\mathrm{Tr}(\bm{H})$ \> trace of the matrix \\

$\mathbf{h}$ \> block vector \\
$\mathbf{h}^T$ \> transposed block vector \\
$\mathbf{h}^\dagger$ \> transposed and conjugated block vector (Hermitian)\\
$\mathbf{H}$ \> block matrix \\
$(\mathbf{H})_{ij}$ \> matrix entry $(i,j)$ of the block matrix \\
$\mathbf{H}^T$ \> transposed block matrix \\
$\mathbf{H}^\dagger$ \> transposed and conjugated block matrix (Hermitian)\\

$\delta_i$ \> Kronecker delta \\
$\otimes$ \> convolution\\
$F[y(\omega)]$ \> functional of $y(\omega)$ \\
$\frac{ \delta F[y(\omega)]}{\delta y}$ \> functional derivative w.r.t. $y(\omega)$\\

$P(c)$ \> probability mass function of a discrete random variable $c$ \\
$H(c)$ \> entropy of a discrete random variable $c$  \\
$p(r)$ \> probability density function of a continuous random variable $r$ \\
$h(r)$ \> entropy of a continous random variable $r$  \\

\end{tabbing}

      \chapter{Basics}\label{ch:basics}
\thispagestyle{empty}

{
}


\lettrine{T}{his} chapter will introduce the basic arguments on which we will mainly focus in this thesis.
The chapter is organized as follows: in \S\ref{sec:lin_modulation} we introduce the notation
for linear modulations. In \S\ref{sec:obs_model} we describe the main observation models.
In \S\ref{sec:opt_det} we describe optimal detection algorithms for the presented observation models.
Finally, in \S\ref{sec:fig_merit} we present the figures of merit that will be used for the performance analysis.

\section{Linear modulations over the AWGN channel}\label{sec:lin_modulation}
In this thesis, we will mainly consider linearly modulated signals whose
complex envelope can be expressed as
\begin{equation}
	s(t)= \sum_{k=0}^{N-1} c_k p(t-kT) \label{eq:lin_signal}
\end{equation}
being $\{c_k\}_{k=0}^{N-1}$ the transmitted symbols, $p(t)$ the shaping pulse, and $T$ the symbol time.
Symbols $\{c_k\}$ will be considered belonging to a $M$-ary constellation in the complex domain.
The transmitted symbols can be either coded or uncoded.
If the signal is transmitted over a channel with additive white Gaussian noise (AWGN),
the complex envelope of the received signal will read
\begin{eqnarray}
	r(t) & = & s(t) + w(t)  \\ 
	& = & \sum_{k=0}^{N-1} c_k p(t-kT) + w(t) \label{eq:rec_signal}
\end{eqnarray}
where $w(t)$ is white Gaussian noise having power spectral density $N_0$.
Without loss of generality, in (\ref{eq:rec_signal}) we considered a frequency flat
channel. Clearly, the extension to frequency selective channels is obtained 
straightforwardly: the received shaping pulse in (\ref{eq:rec_signal}) will be equal to
\begin{equation}
	p(t) \otimes h(t)
\end{equation}
being $h(t)$ the channel impulse response.

\section{Observation models}\label{sec:obs_model}
For detection, we need a discrete-time observation model $\bm{r}$
of the received signal.
The observation model $\bm{r}$ shall be a sufficient statistics,
i.e., a function
of the received signal that does not involve any information loss.
For the received signal (\ref{eq:rec_signal}), different sufficient statistics can be found.

\begin{figure}
	\begin{center}
		\includegraphics{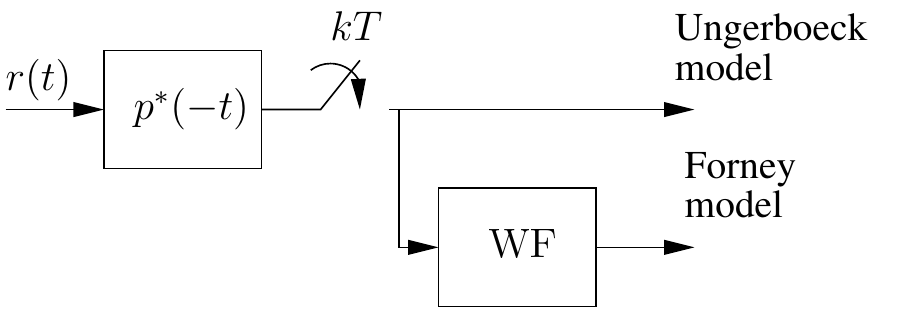}
		\caption{Block diagram of the system which carry out the Ungerboeck observation model and
		the Forney observation model.}\label{fig:bd_ung_fo}
	\end{center}
\end{figure}

The first model that we consider is the {\it Ungerboeck observation model}~\cite{Un74}.
The Ungerboeck observation model is obtained as shown in Figure \ref{fig:bd_ung_fo}.
The received signal passes through a filter matched to the shaping pulse $p(t)$ (matched filter, MF). 
The signal at the output is then sampled with time interval $T$. The
sequence of samples $\bm{r}=[r_0,\dots,r_{N-1}]^T$ results to be
\begin{equation}
	r_k= \sum_{i=-\nu}^\nu c_{k-i}g_i + n_k \label{eq:ung_model}
\end{equation}
where 
\begin{equation}
	g_i=\int_{-\infty}^{\infty} p(t)p^*(t-iT)\mathrm{d}t \,
\end{equation}
are the intersymbol interference (ISI) taps,
which are null for $|i|>\nu$ (being $\nu$ the channel memory), and
$\{n_k\}$ are Gaussian random variables with autocorrelation function
\begin{equation}
	\mathrm{E}\left\{ n_{k+i}n_k^* \right\}=N_0g_i \,.
\end{equation}
The samples (\ref{eq:ung_model}) can be gathered in a useful matrix notation
\begin{equation}
	\bm{r}=\bm{G}\bm{c}+\bm{n} \label{eq:matrix_ungerboeck}
\end{equation}
where $\bm{c}$ and $\bm{n}$ are defined as $\bm{c}=[c_0,\dots,c_{N-1}]^T$, 
$\bm{n}=[n_0,\dots,n_{N-1}]^T$,
and $\bm{G}$ is a Toeplitz matrix (see Appendix \ref{ch:toeplitz_matrix}) with entries \mbox{$(\bm{G})_{\ell m}=g_{\ell-m}$}.

Since white noise is often preferred, the
MF output can be filtered by a whitening filter (WF)\footnote{The cascade of the MF and the WF is called whitened matched filter (WMF).} 
as shown in Figure \ref{fig:bd_ung_fo}.
This yields another sufficient statistics known as {\it Forney observation model} \cite{Fo72b}.
The Forney model reads
\begin{equation}
	r_k= \sum_{i=0}^\nu c_{k-i}h_i + w_k \label{eq:fo_model}
\end{equation}
where $\{w_k\}$ are Gaussian random variables with $\mathrm{E}\left\{ w_{k+i}w_k \right\}=N_0\delta_i$
being $\delta_i$ the Kronecker delta, and $\{h_i\}_{i=0}^{\nu}$ are the ISI taps
such that $g_i= h_i \otimes h^*_{-i}$. 
The value of memory $\nu$ of the Forney model is always equal to the one
of the Ungerboeck model.
The Forney observation model (\ref{eq:fo_model})
can be expressed by means of the matrix notation
\begin{equation}
	\bm{r}=\bm{H}\bm{c}+\bm{w} \label{eq:forney_matrix}
\end{equation}
where $\bm{H}$ is a Toeplitz and lower triangular matrix with entries $(\bm{H})_{\ell m}=h_{\ell-m}$.
Moreover it holds $\bm{H}^{\dagger}\bm{H}=\bm{G}$.

Both the Ungerboeck and Forney models show that a discrete-time channel equivalent to the continuous-time one can be found.
In addition to the Ungerboeck and Forney models, there exist other sets of sufficient statistics (although not consider in this thesis).
Another example of sufficient statistics is the one described in \cite{MeOePo94}.


\section{Optimal MAP detection: the BCJR algorithm}\label{sec:opt_det}
Optimal maximum a posteriori (MAP) symbol detection is based on the strategy 
\begin{equation}
	\hat{c}_k = \arg \max_{c_k} P(c_k | \bm{r}) \qquad k=0,\dots,N-1 \,. \label{eq:MAP}
\end{equation}
The {\it a posteriori probabilities} $P(c_k|\bm{r})$ can be effectively computed by means of
the Bahl-Cocke-Jelinek-Raviv (BCJR) algorithm \cite{BaCoJeRa74}.
Let us define $\sigma_k$ as the {\it state} of the channel at the discrete time $k$,
and gather all the states in the vector $\bm{\sigma}=[\sigma_0,\dots,\sigma_{N-1}]^T$. 
The basic hypothesis of the BCJR algorithm is that it exists a definition of the state
$\sigma_k$ which allows to factorize the probability $P(\bm{c},\bm{\sigma}|\bm{r})$ as
\begin{equation}
	P(\bm{c},\bm{\sigma}|\bm{r}) \propto \prod_{k=0}^{N-1} \Lambda_k(r_k,c_k,\sigma_k,\sigma_{k+1})P(c_k) \label{eq:fact_bcjr}
\end{equation}
where $P(c_k)$ is the {\it a priori probability} on the symbol $c_k$ at time $k$,
and $\Lambda_k(r_k,c_k,\sigma_k,\sigma_{k+1})$ is the {\it metric} at time $k$.
The metrics are not necessarily probability mass functions.

\begin{figure}
	\begin{center}
		\includegraphics{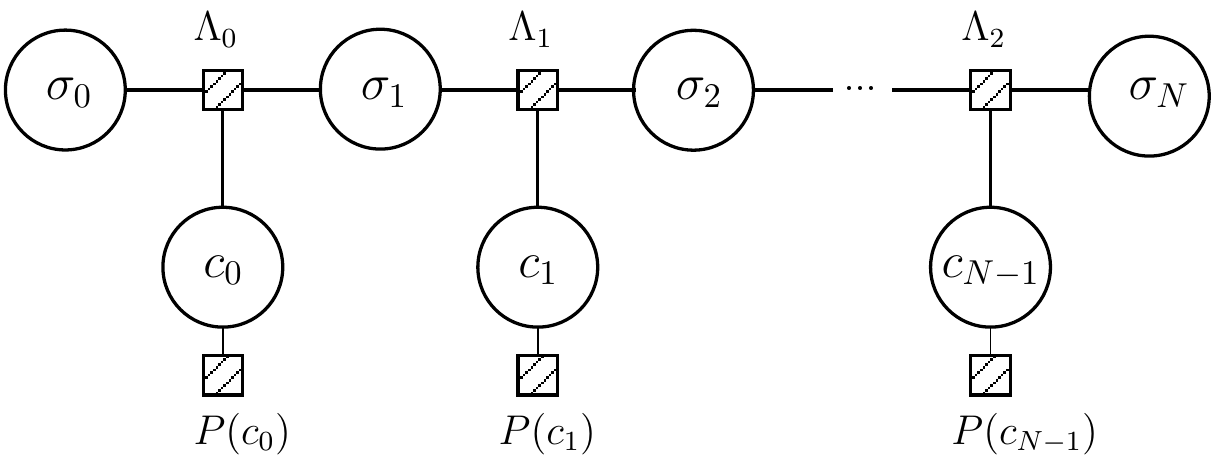}
		\caption{Factor graph of the BCJR algorithm.}\label{fig:fg_bcjr}	
	\end{center}
\end{figure}

\begin{figure}
	\begin{center}
		\includegraphics[width=0.35\paperwidth]{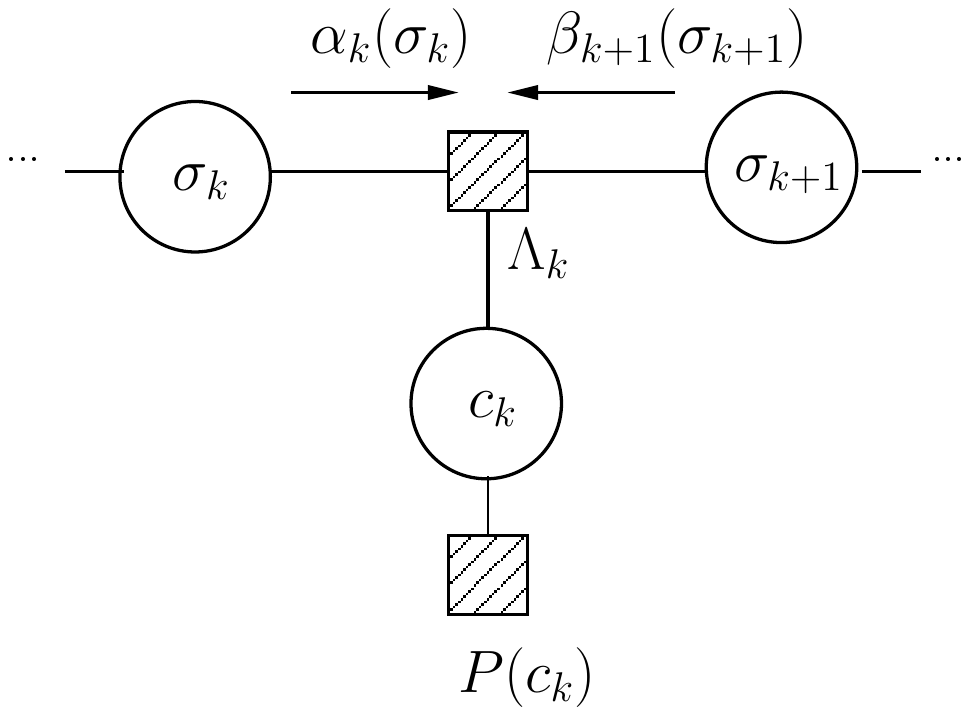}
		\caption{Message passing algorithm on a section of the BCJR factor graph.}\label{fig:fg_bcjr_zoom}	
	\end{center}
\end{figure}
The BCJR algorithm can be derived as follow. 
Equation (\ref{eq:fact_bcjr}) can be represented by the factor graph (FG, see \cite{KsFrLo01}) in Figure \ref{fig:fg_bcjr}.
The application of the sum product algorithm (SPA) to the FG in Figure \ref{fig:fg_bcjr} 
gives a unique message passing algorithm
to compute any marginal probability
of (\ref{eq:fact_bcjr}).
In particular, we are interested in the a posteriori probabilities in (\ref{eq:MAP})
that are obtained with the following marginalization:
\begin{equation}
	P(c_k|\bm{r})= \sum_{\sim\{c_k\}}P(\bm{c},\bm{\sigma}|\bm{r}) \,,
\end{equation}
where $\sum_{\sim\{c_k\}}$ denotes the sum with respect to all the variables,
except $c_k$.
Denoting the messages on the graph by $\alpha_k(\sigma_k)$ and $\beta_k(\sigma_k)$ as shown in in Figure \ref{fig:fg_bcjr_zoom},
it can be shown that the a posteriori probabilities are obtained with the following message passing algorithm:
\begin{itemize}
 \item Initialize the algorithm as
	\begin{eqnarray}
		\alpha_0(\sigma_0) & = & 1  \\
		\beta_N(\sigma_N) & = & 1   \,.
	\end{eqnarray}

 \item {\it Forward recursion}: for each $k=0,\dots,N-1$ compute the messages $\alpha_{k+1}(\sigma_{k+1})$  as
	\begin{equation}
		\alpha_{k+1}(\sigma_{k+1}) =  \sum_{c_k,\sigma_k} \alpha_{k}(\sigma_{k})\Lambda_k(r_k,c_k,\sigma_k,\sigma_{k+1})P(c_k) \,.\label{eq:alphabcjr} 
	\end{equation} 
 \item {\it Backward recursion}: for each $k=N-1,N-2,\dots,0$ compute the messages $\beta_{k}(\sigma_{k})$ as
	\begin{equation}
		\beta_{k}(\sigma_{k})  =  \sum_{c_k,\sigma_{k+1}} \beta_{k+1}(\sigma_{k+1})\Lambda_k(r_k,c_k,\sigma_k,\sigma_{k+1})P(c_k) \,. \label{eq:betabcjr}
	\end{equation}
 \item {\it Completion}: for each $k=0,\dots,N-1$ compute the a posteriori probabilities as
	\begin{equation}
		P(c_k|{\rm r}) \propto P(c_k)\sum_{\sigma_k,\sigma_{k+1}} \alpha(\sigma_k)\Lambda_k(r_k,c_k,\sigma_k,\sigma_{k+1})\beta(\sigma_{k+1}) \,. \label{eq:gammabcjr}
	\end{equation}
\end{itemize}
The complexity of the BCJR algorithm is proportional to $\mathcal{O}(MS)$, being $M$ the cardinality 
of the transmitted symbols $\{c_k\}$ and $S$ the cardinality of the state $\{\sigma_k\}$.
The algorithm is conveniently implemented in the logarithmic domain~\cite{RoViHo97}.
The reader can notice that the BCJR algorithm performs a trellis processing similar to the Viterbi algorithm.
In fact the algorithms are stricly related. 
It can be shown that working in the logarithm domain, and by substituting the 
logarithm of a sum of exponentials with the max of the arguments,
the Viterbi algorithm is obtained \cite{KsFrLo01}. 

For the sake of clarity we finally show the BCJR algorithm for the observation models presented in \S\ref{sec:obs_model}.
For the Forney model, it is easy to notice from (\ref{eq:fo_model}) that the state can be defined 
by the vector of the past symbols $\bm{\sigma}_k=[c_{k-1},\dots,c_{k-\nu}]$. 
Thus, the conditional probability factorizes with terms
\begin{eqnarray}
	\Lambda_k(c_k,\bm{\sigma}_k,\bm{\sigma}_{k+1}) & = & p(r_k|c_k,\bm{\sigma}_k)\mathcal{I}(c_k,\bm{\sigma}_k,\bm{\sigma}_{k+1})  \\
	& = & \mathrm{exp}\left\{ -\frac{\left|y_k - \sum_{i=0}^\nu h_i c_{k-i} \right|^2}{N_0} \right\}\mathcal{I}(c_k,\bm{\sigma}_k,\bm{\sigma}_{k+1})  \,   
\end{eqnarray}
where $\mathcal{I}(c_k,\bm{\sigma}_k,\bm{\sigma}_{k+1})$ is an indicator function: it
is equal to 1 if the transition \mbox{$(c_k,\sigma_k)\rightarrow \sigma_{k+1}$} is valid, and to 0 otherwise.

Since the cardinality of the state is $S=M^\nu$, the complexity of the algorithm increases exponentially with $\nu$.
Moreover it can be shown that for the Forney model the forward and backward recursions have the following probabilistic meaning~\cite{FeBaCo07}
\begin{eqnarray}
	\alpha_k(\sigma_k) & = & P(\sigma_k | \bm{r}_0^{k-1}) \quad \forall k=0,\dots,N  \\
	\beta_k(\sigma_k) & = & p(\bm{r}_k^{N-1}| \sigma_k) \quad \forall k=0,\dots,N 
\end{eqnarray}
where $\bm{r}_a^b$ denotes either the vector $[r_a,\dots,r_b]$ for any $a \leq b$, or the empty set otherwise.

For the Ungerboeck observation model it can be shown (see \cite{CoBa05b}) that the state
can be defined again as $\bm{\sigma}_k=[c_{k-1},\dots,c_{k-\nu}]$
and the conditional probability factorizes with terms
\begin{equation}
	\Lambda_k(c_k,\bm{\sigma}_k,\bm{\sigma}_{k+1})  =  \mathrm{exp}\left\{ \frac{2\Re\left( c_k^*r_k\right)- |c_k|^2g_0 -2c^*_k\sum_{i=1}^\nu g_ic_{k-i}}{N_0} \right\}  \mathcal{I}(c_k,\bm{\sigma}_k,\bm{\sigma}_{k+1})  \,.
\end{equation}
Since the memory $\nu$ of the Ungerboeck model in (\ref{eq:ung_model})
is always equal to the memory of the Forney model, also the complexity of the algorithms is the same.
We point out that for the Ungerboeck model the $\Lambda_k(c_k,\bm{\sigma}_k,\bm{\sigma}_{k+1})$ are not probability density functions
as for the Forney case.

\section{Performance analysis}\label{sec:fig_merit}
We will consider different figures of merit for the performance analysis.
The first figure of merit is the {\it spectral efficiency} (SE) that can be achieved by a given modulation and 
coding format (MODCOD) on a given channel defined as
\begin{equation}
	\mathrm{SE}= \frac{r\log_2(M)}{T W} \quad \mathrm{[bit/s/Hz]} \label{eq:SE}
\end{equation}
where $r$ is the rate of the adopted channel code, and $W$ is the reference bandwidth.
The reference bandwidth can be the available bandwidth of the channel, the transmitted signal bandwidth,
or any other bandwidth definition depending on the considered communication system.
In a frequency division multiplexed (FDM) system, it can be defined as the distance between the carriers
of two adjacent channels \cite{PiMaCo12}.
The achieved SE will be often placed in the Shannon plane as a function of the signal-to-noise ratio (SNR).
For each SE, the corresponding SNR is the value which guarantees 
a reliable communication. 
In many practical applications the communication is considered
reliable if the packet error rate (PER) is below a given threshold,
for example $10^{-5}$.

The second figure of merit that we consider is the {\it achievable information rate} (AIR).
For a channel with channel law $p(\bm{r}|\bm{c})$, and a particular modulation format,
the information rate (IR) is defined as \cite{Sh48}
\begin{eqnarray}
	I(\bm{c};\bm{r}) & =&  h(\bm{r})-h(\bm{r}|\bm{c})  \\ 
	& = & \mathrm{E}\{ -\log_2 p(\bm{r}) \} - \mathrm{E}\{ -\log_2 p(\bm{r}|\bm{c}) \}  \, \label{eq:AIR},
\end{eqnarray}
and measures the highest rate achievable on the channel with the adopted modulation format.
In many applications, however, the receiver could consider
a mismatched channel law $q(\bm{r}|\bm{c})$ (also denoted as {\it auxiliary channel}) different from the actual channel law $p(\bm{r}|\bm{c})$.
We define thus, the {\it achievable information rate} as the highest rate achievable on the channel
with the mismatched receiver \cite{MeKaLaSh94,GaLaTe00}. It reads 
\begin{eqnarray}
	I_{\mathrm R} & =&  \mathfrak{h}(\bm{r})-\mathfrak{h}(\bm{r}|\bm{c})  \\ 
	& = & \mathrm{E}\{ -\log_2 q(\bm{r}) \} - \mathrm{E}\{ -\log_2 q(\bm{r}|\bm{c}) \} \label{eq:mis_AIR}
\end{eqnarray}
where $q(\bm{r})= \sum_{\bm{c}} q(\bm{r}|\bm{c})P(\bm{c})$.
We point out that in (\ref{eq:mis_AIR}) the average is computed with respect to the actual statistics $p(\bm{r}|\bm{c})$,
and the mismatched entropies are explicitly denoted by $\mathfrak{h}$ to distinguish them from the standard entropies $h$.
The AIR is always upper bounded as 
\begin{equation}
	I_{\mathrm R} \leq I(\bm{c};\bm{r}) 
\end{equation}
with equality if and only if $q(\bm{r}|\bm{c})=p(\bm{r}|\bm{c})$, or in other words,
if and only if the receiver performs optimal detection and decoding \cite{ArLoVoKaZe06}.

Since the bandwidth in many applications is getting more and more a
limited resource, it is of interest 
to evaluate the
{\it achievable spectral efficiency} (ASE) which is defined as
\begin{equation}
	\eta = \frac{I_{\mathrm R}}{T W} \, \quad \mathrm{[bit/s/Hz]}  \, . \label{eq:ASE}
\end{equation}
The ASE, in simple words,
is the maximum SE that can be achieved with a joint detection and decoding scheme.
It can be computed independently of the adopted coding scheme~\cite{ArLoVoKaZe06}, avoiding long PER simulations. 


      \chapter{Channel shortening}\label{ch:CS}
\thispagestyle{empty}


\lettrine{T}{he} complexity of the optimal detection
increases exponentially with the memory taken into account by the detector (see \S\ref{sec:opt_det}). 
Thus, for many practical communication schemes, optimal detection can
be prohibitive, since the complexity is unmanageable.
In this chapter, we consider detectors with reduced complexity.
The complexity reduction techniques can be classified mainly in two
families:
\begin{itemize}
 \item techniques that perform detection on the original trellis
but only a fraction of the available paths is explored (e.g. the $\mathcal{M}$-BCJR in \cite{PrAn12} and sphere decoding \cite{BoGrBrFo03}).
 \item techniques
that work on a reduced trellis which is then fully processed.
\end{itemize}

We consider the {\it channel shortening} (CS), a complexity reduction technique which belongs to the second family.
Channel shortening is a technique originally proposed in 1972 by Falconer and
Magee \cite{FaMa73}, and recently improved by Rusek and Prlja \cite{RuPr12}.
In this chapter we will first review the CS technique
proposed by Rusek and Prlja and then, the previous works on CS in \S\ref{sec:previous}.
In \S\ref{sec:adaptive} we will derive an adaptive version of CS.
The optimization of the transmit filter for CS detectors will be discussed
in \S\ref{sec:opt_pulse}. Finally in \S\ref{sec:MIMO_ISI_CS}
and \S\ref{sec:cont_time_cs} we extend the CS to other channels.

\section{CS algorithm}\label{sec:CS_alg}
Let us consider the discrete-time ISI channel with AWGN
\begin{equation}
	r_k= \sum_{i=0}^\nu h_i c_{k-i} + w_k \label{eq:cs_sismodel}
\end{equation}
where $\{c_k\}$ are the transmitted symbols belonging to a properly normalized $M$-ary constellation,
$\{h_i\}_{i=0}^\nu$ are the ISI taps, $\nu$ is the channel memory,
and $\{w_k\}$ are independent Gaussian random variables with variance $N_0$.

The observable $\{r_k\}$ can be filtered
with a discrete-time filter matched to $\{h_i\}$.
The resulting observable is the Ungerboeck observation model  (\ref{eq:ung_model}), having $g_i=\sum_{k} h^*_{k-i}h_k$.
The optimal detection is performed by means of the BCJR algorithm.
Using the matrix notations (\ref{eq:matrix_ungerboeck}), the channel law can be expressed as
\begin{equation}
	p(\bm{r}|\bm{c}) \propto \mathrm{exp}\left\{ \frac{\Re \left( \bm{c}^\dagger \bm{H}^\dagger \bm{r} \right)- \bm{c}^\dagger\bm{G}\bm{c} }{N_0}   \right\} \,
\end{equation}
where $\bm{G}=\bm{H}^\dagger\bm{H}$ and is semi-positive definite.
We now consider a reduced-complexity detector, 
which considers a mismatched channel law
\begin{equation}
	q(\bm{r}|\bm{c}) \propto \mathrm{exp}\left\{ \Re \left( \bm{c}^\dagger (\bm{H}^r)^\dagger \bm{r}\right) - \bm{c}^\dagger\bm{G}^r\bm{c}    \right\} \,, \label{eq:mis_CS}
\end{equation}
where  $\bm{H}^r$ is the new front end filter, named {\it channel shortener}, and
$\bm{G}^r$ is the ISI to be set at detector, named  {\it target response}.
The superscript $r$ denotes that they are solely considered at {\it receiver},
and are different from the actual $\bm{H}$ and $\bm{G}$.
For simplicity the matrix $\bm{H}^r$ and $\bm{G}^r$ in~(\ref{eq:mis_CS}) 
include also the noise variance $N_0$.
Let $L\leq \nu$ the memory taken into account by the detector.
Due to this constraint on the complexity, the target response must be such that
\begin{equation}
	(\bm{G}^r)_{ij}=0 \quad \forall |i-j|>L \label{eq:CS_con} \,.
\end{equation}
The matrix $\bm{G}^r$ does not need to be semi-positive definite \cite{RuFe12}.

The achievable information rate (AIR) of the mismatched detector (see \S\ref{sec:fig_merit}) is
\begin{equation}
	I_{\mathrm R} = \mathfrak{h}(\bm{r}) - \mathfrak{h}(\bm{r}|\bm{c}) \,.
\end{equation}
The aim of CS is, for a given $L$, find the $\bm{H}^r$
and $\bm{G}^r$ which maximize the AIR. Namely we want to solve
the following maximization problem
\begin{equation}
	I_{\mathrm{OPT}}=\max_{\bm{H}^r,\bm{G}^r} I_{\mathrm{R}} 
\end{equation}
under the constraint (\ref{eq:CS_con}).

The optimization for achievable information rate was completely solved
in \cite{RuPr12} under the assumption that $\bm{c}$ are independent Gaussian symbols.
Closed-form expressions for $\bm{G}^r$, $\bm{H}^r$ for the ISI channel
can be found with the following algorithm:
\begin{itemize}
 \item Compute the sequence $\{b_i\}_{i=-L}^L$ as
	\begin{eqnarray}
		b_i&=&\frac{1}{2\pi}\int_{-\pi}^{\pi}\frac{N_0}{|H(\omega)|^2+N_0} e^{j\omega i} \mathrm{d}\omega \label{eq:bi}
	\end{eqnarray}
	where $H(\omega)$ is the discrete time Fourier transform (DTFT) of $\{h_i\}$.
 \item Compute the real-valued scalar
	\begin{equation}
		\label{cc} \mathcal{C} = b_0 -{\bm b}{\bm B}^{-1}{\bm b}^{\dagger}, 
	\end{equation}
 	where ${\bm b}=[b_1,b_2,\ldots, b_{L}],$
	and ${\bm B}$ is $L\times L$ Toeplitz with entries $(\bm{B})_{ij}=b_{j-i}$.
 \item Define the vector $\bm{u}=  { \frac{1}{ \sqrt{\mathcal{C}} }} [1,\, -{\bm b}{\bm B}^{-1}]$ and find the optimal
       target response as
   \begin{equation}
      G^r(\omega)= |U(\omega)|^2 - 1 \, ,
   \end{equation}
   where $U(\omega)$ is the DFT of $\{u_i\}$.
 \item Finally, the optimal channel shortener is found as
    \begin{equation}
      H^r(\omega)=\frac{ H(\omega)}{|H(\omega)|^2 + N_0 }(G^r(\omega)+1) ~~. \label{eq:Hr}
    \end{equation}
\end{itemize}
By using the optimal channel shortener and the target response $I_{\mathrm{OPT}}$ results to be
\begin{equation} 
I_{\mathrm{OPT}}=-\log_2(\mathcal{C}) \, .  \label{eq:I_opt}
\end{equation}
The proof is shown in \cite{RuPr12}.

Clearly, when $L=\nu$, the trivial solution $G^r(\omega)=|H(\omega)|^2$,
$H^r(\omega)=H(\omega)$ is found and the achievable rate simplifies to the famous
formula
\begin{equation}
	-\log_2(\mathcal{C})=\int_{-\infty}^{\infty}\log_2\left(1+ \frac{|H(\omega)|^2}{N_0}\right) \mathrm{d}\omega \,.
\end{equation}

Although the algorithm here is limited to the ISI channel,
we point out that CS can be applied also to multiple-input multiple-output
(MIMO) channel.
In fact \cite{RuPr12} worked on a slightly general model which 
represents either the ISI channel (\ref{eq:cs_sismodel}) or a the MIMO channel.

Now, the first question that the reader could ask is, why should we be interested
in the optimal solution for Gaussian inputs, when practical communication schemes
use finite cardinality alphabets?
Although we cannot give a proof, in our experience we saw that employing the solution
for Gaussian inputs for finite low-order cardinality alphabets, the resulting AIR
is still excellent (see \cite{RuPr12,CoMoRu12,MoRuCo13}).

\begin{figure}
	\begin{center}
		\includegraphics[width=0.75\columnwidth]{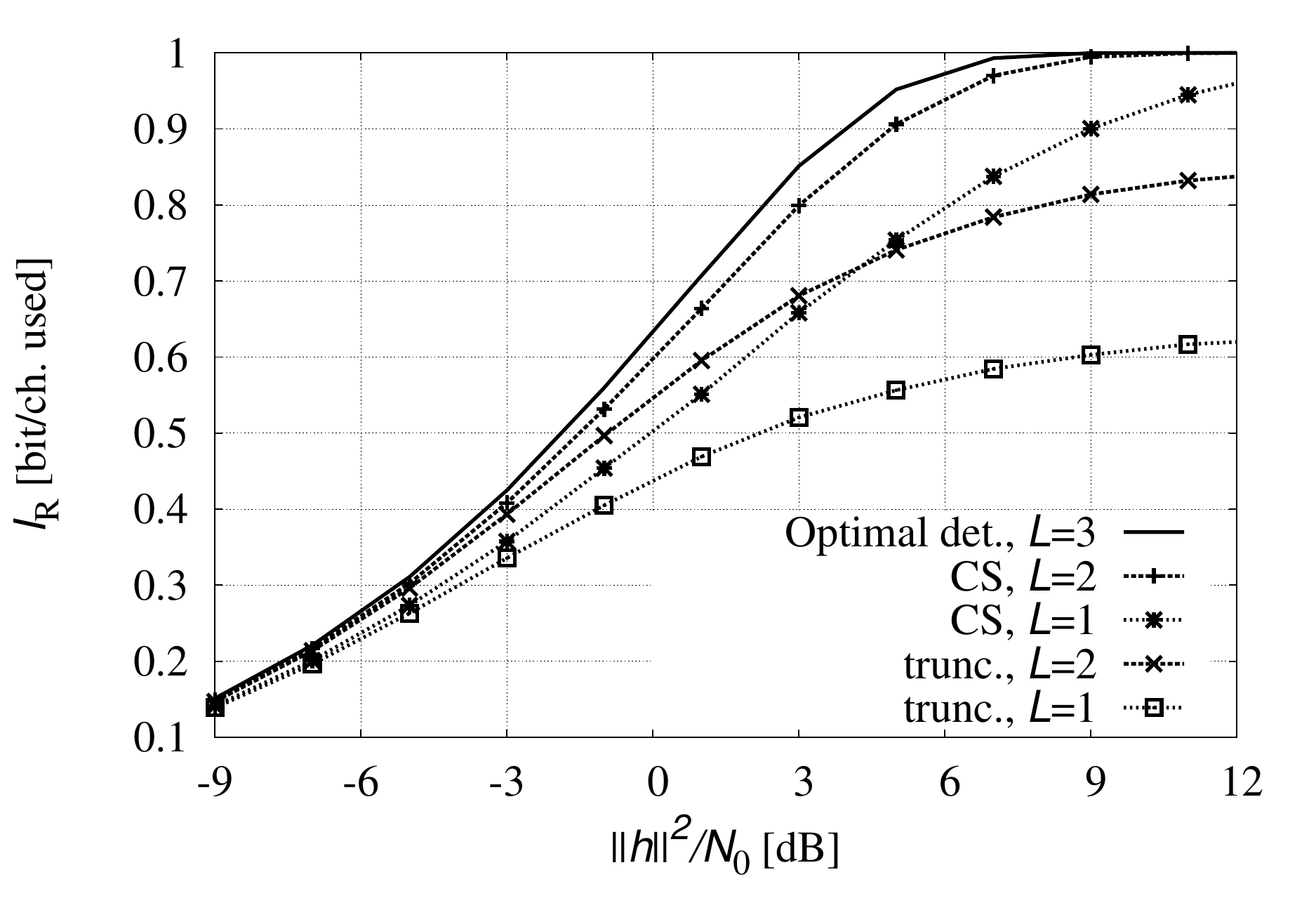}
		\caption{AIRs of the CS detector on the EPR4 channel.}\label{eq:epr4_cs_example}
	\end{center}
\end{figure}

Before going ahead let us show an example for the sake of clarity. 
We considered an EPR4 channel having channel response $\bm{h}=[0.5\,,0.5\,,-0.5\,,-0.5]$.
Figure \ref{eq:epr4_cs_example} shows the AIR by employing a BPSK modulation
and the CS detector.
For comparison the figure shows also the AIR by employing
the na\"{\i}ve technique of truncating the considered ISI at detector to $L$ values,
and the AIR for optimal detection ($L=\nu=3$), for which the CS technique
and truncation are the same algorithm.
The AIR were computed by means of the Monte Carlo method described in~\cite{ArLoVoKaZe06}.
From the figure it can be seen that CS outperforms
the truncation method, with SNR gains beyond 3~dB.

\section{Previous works on CS}\label{sec:previous}
The original paper in 1973 by Magee and Falconer \cite{FaMa73} proposed channel shortening detectors
optimized from a minimum mean-square-error (MMSE) perspective, and many papers in the literature followed 
the same approach (e.g. \cite{AlCi96,Al01}).
In \cite{FaMa73}, the detector considered a mismatched channel law 
\begin{equation}
	q(\bm{r}|\bm{c}) \propto \exp \left\{ -\frac{\left|\bm{W}\bm{r}-\bm{Q}\bm{c} \right|^2}{N_0} \right\} \label{eq:q_original_CS}
\end{equation}
where $\bm{W}$ and $\bm{Q}$ are Toeplitz matrix, representing
the channel shortener and the target response respectively.
However, two main flaws in this approach can be found: first of all, 
minimizing the mean-square-error does not directly correspond to achieving the highest
information rate (in the Shannon sense) that can be supported
by a shortening detector. Second, it can observed that (\ref{eq:q_original_CS}) can
be equivalently expressed as
\begin{equation}
	q(\bm{r}|\bm{c}) \propto \exp \left\{ \frac{2\Re(\bm{c}^\dagger\bm{Q}^\dagger\bm{W}\bm{r}) - \bm{c}^\dagger\bm{W}^\dagger\bm{W}{c}}{N_0} \right\}
\end{equation}
which is equal to (\ref{eq:mis_CS}) by setting $\bm{H}^r=\bm{W}^\dagger \bm{Q}$
and $\bm{G}^r=\bm{W}^\dagger\bm{W}$.
Thus, it can be noticed that the traditional CS detector requires $\bm{G}^r$ to be semi-positive definite
and $\bm{H}^r$ to have a specific structure.
In \cite{RuPr12} and \cite{Pr13}, it is shown that the new CS presented in the previous section, 
outperforms the traditional CS.

Other papers on CS instead adopted other figures of merit: for example~\cite{MeYoRo96} 
proposed {\it maximum shortening signal-noise-ratio} (MSSNR) 
which minimizes the energy outside a window of interest and holds the
energy inside fixed. The work in \cite{BaMaJo03} instead considered
the {\it sum-squared auto-correlation} (SA) of the combined channel-equalizer response,
and tries to minimize the SA outside a window of interest.
However, all these techniques do not imply a maximization of the AIR.

Before Rusek and Prlja publication \cite{RuPr12} in 2012, 
in almost forty years of CS, the only work
who adopted an information theoretic approach was \cite{AbLa00} by Abou-Faycal and Lapidoth,
presented in a conference in 2000.
 
Unfortunately, although \cite{AbLa00} is an excellent work and has very similar
results to~\cite{RuPr12}, the full version of the paper was not available on the web
and it had very few citations. In fact, Rusek and Prlja were not aware of \cite{AbLa00} and worked independently.
The common points and differences between the two works can be summarized as follow:
\begin{itemize}
 \item 	Both works choose the channel shortener and target ISI by maximizing the achievable information rate. 
	The target ISI $\bm{G}^r$ is found with the same algorithm (although it looks different in the two works).
 \item  The work \cite{AbLa00} considered the Forney detection instead of Ungerboeck detection. Thus, if $\bm{G}^r$
	is positive definite, the two techniques are the same. In the case $\bm{G}^r$ is not positive definite,
	 \cite{AbLa00} does not give a clear answer on how the channel shortener can be computed.
 \item  The framework of \cite{RuPr12} was also for MIMO channel.
 \item  The proof of \cite{RuPr12} gave a closed formula for the channel shortener. Instead \cite{AbLa00} left a parameter
	to be optimized in the formula. 
\end{itemize}

\section{Adaptive CS detector for unknown channels}\label{sec:adaptive}
The optimal CS detector in \S\ref{sec:CS_alg} is carried out by assuming a perfect knowledge
of the ISI channel. In this section, we consider the case of unknown  ISI taps $\{h_i \}$
of the channel model (\ref{eq:cs_sismodel}), and we will derive an adaptive CS detector.

The derivation of the adaptive CS detector relies on the following observations.
First, the optimal channel shortener (\ref{eq:Hr}) is the combination of two filters: 
a MMSE filter and a filter with frequency response $G^r(\omega)+1$. 
Second, the sequence $\{b_i\}_{i={-L}}^L$ is
the autocorrelation of the error 
\begin{equation}
	b_i=\mathrm{E} \left\{ e_{i+k}e^*_k \right\}
\end{equation}
being $e_k=c_k-\hat{c}_k$ and $\hat{c}_k$ the output of the MMSE filter.
This is found by observing that $N_0/(|H(\omega)|^2+N_0)$ is the power spectral density of the error at
the output of a MMSE filter \cite{CiOBCh3}.

The adaptive CS detector can be summarized in the following steps:
\begin{itemize}
	\item the transmitter sends a training sequence, known at receiver side.
	\item the receiver computes the MMSE filter by means of the training sequence.
	\item the receiver estimates the error correlation $b_k$.
\end{itemize}

The first two steps can be easily done by means of the least mean square (LMS) algorithm, or the recursive 
least square (RLS) algorithm \cite{Ha96}.
The last step, the error-correlation estimation, can be easily done by computing at receiver
the error sequence $\{e_k\}$, and using standard estimators (e.g., the {\it xcorr} in Matlab).

\section{Optimized transmit filter for CS detector}\label{sec:opt_pulse}

In \S\ref{sec:CS_alg} we showed the algorithm to derive the optimal CS detector when
the considered memory $L$ is lower then the actual memory $\nu$.

In this section, we extend the CS algorithm by designing a proper transmit filter 
to be employed jointly with a channel-shortening detector with the aim of further improving
the achievable information rate. In other words,
we consider to adopt, at the receiver side, a channel-shortening detector and then solve for the optimal transmit filter to be used jointly with it.
When the use of the optimal full-complexity receiver is allowed, the answer to this question is the classical waterfilling filter. We are generalizing the waterfilling concept to the case of reduced-complexity channel-shortening detectors, i.e., we
essentially redo Hirt's derivations \cite{Hi88}, but this time with the practical
constraint of a given receiver complexity.

Our results are not as conclusive as in the
unconstrained receiver complexity case. With functional analysis, we
can prove that, for real channels, the optimal transmit filter has a frequency response
described by $L+1$ real-scalar values. In general, for complex channels,
the optimal transmit filter is described by $L+1$ complex scalar values.
The transmit filter optimization thereby becomes a problem of finite dimensionality, and a
numerical optimization provides the optimal spectrum. Note that, in
practice, $L$ is limited to rather small values and $L=1$ is an
appealing choice from a complexity perspective. This essentially leads to very effective numerical optimizations.

\subsection*{Problem formulation}
We consider the channel model (\ref{eq:cs_sismodel}).
The transmitted symbols $\{c_k\}$ are a precoded version of the information symbols $\{a_k\}$
as
\begin{eqnarray}
	c_k & = & a_k \otimes p_k \label{eq:prec} \\
	 & = & \sum_i a_{k-i}p_i
\end{eqnarray}
where $\{p_i\}$ is a transmit filter subject to the power constraint $\sum_i |p_i|^2=1$.
Using a matrix notation we can express (\ref{eq:prec}) as
\begin{equation}
	\bm{c}= \bm{P}\bm{a}
\end{equation}
where $\bm{P}$ is a Toeplitz matrix with entries $(\bm{P})_{ij}=p_{i-j}$.
The combined channel-precoder thus reads
\begin{eqnarray}
	\bm{r} & = & \bm{H}\bm{c}+\bm{w} \\
	 & = & \bm{H}\bm{P}\bm{a}+\bm{w} \\
	 & = & \bm{V}\bm{a}+\bm{w} \,, \label{eq:ccp}
\end{eqnarray}
where $\bm{V}=\bm{H}\bm{P}$. Equivalently, (\ref{eq:ccp}) can be expressed by means of the scalar notation
\begin{equation}
	r_k = \sum_{i=0}^{\nu_C} v_i a_{k-i} + w_k  \label{eq:scalar_ccp}
\end{equation}
where $v_i = h_i \otimes p_i$, and $\nu_C$ is the combined memory.
If the CS detector with memory $L$ is used for detection on the combined channel precoder $\bm{H}\bm{P}$, the AIR (for Gaussian symbols) 
reads
\begin{equation}
	I_{\mathrm{OPT}}= -\log_2(\mathcal{C}) \label{eq:I_opt_2}
\end{equation}
where $\mathcal{C}$ is the real-valued scalar (\ref{cc}), function of
the coefficients $\{b_i\}_{i=-L}^L$ which read
\begin{eqnarray}
	b_i & = & \frac{1}{2\pi}\int_{-\pi}^{\pi}\frac{N_0}{|V(\omega)|^2+N_0} e^{j\omega i} \mathrm{d}\omega \label{eq:b_V} \\
	& = & \frac{1}{2\pi}\int_{-\pi}^{\pi}\frac{N_0}{|P(\omega)|^2|H(\omega)|^2+N_0} e^{j\omega i} \mathrm{d}\omega \,.
\end{eqnarray}

The problem we aim at solving is to maximize the AIR $I_{\mathrm{OPT}}$ of (\ref{eq:I_opt_2}) over
the transmit filter $P(\omega)$, i.e., the DTFT of
$\bm{p}$. Thus, we have the following optimization problem at hand
\begin{eqnarray}\label{eq:opt_problem}
 &\min_{P(\omega)} \mathcal{C}[P(\omega)]& \nonumber \\
&\mathrm{such\; that}&  \label{eq:const_pw} \\
& \int_{-\pi}^{\pi} |P(\omega)|^2\mathrm{d}\omega =2\pi&  \nonumber \,.
\end{eqnarray}

In (\ref{eq:opt_problem}), we have made explicit the
dependency of $\mathcal{C}$ on $P(\omega)$, but not on $N_0$ and $H(\omega)$,
since these are not subject to optimization. Since the starting point is the expression of the AIR when the optimal channel-shortening detector is employed, we are thus jointly optimizing the channel shortening filter, 
the target response, and the transmit filter, although for Gaussian inputs only. 
However, as shown in the numerical results, when a low-cardinality discrete alphabet is employed, a significant performance 
improvement is still observed (see also~\cite{RuPr12}).

The optimization problem (\ref{eq:opt_problem}) is an instance of
calculus of  variations. We have not been able to solve it in
closed form, but we can reduce the optimization problem into an $L+1$
dimensional problem, which can then  efficiently be solved by standard
numerical methods. The main result of this optimization is the following theorem.

\begin{thm}\label{thm:opt_pulse}
The optimal transmit filter with continuous spectrum for the channel $H(\omega)$ with a memory
$L$ channel-shortening detector satisfies

\begin{equation}
 |P(\omega)|^2 =\max\left(0,\frac{N_0}{\sqrt{|H(\omega)|^2}}\sqrt{\sum_{\ell=-L}^L A_{\ell}
      e^{j\ell \omega}}-\frac{N_0}{|H(\omega)|^2}\right)\,,\label{eq:Pw_opt}
\end{equation}
where $\{A_{\ell}\}$ are complex-valued scalar constants with Hermitian symmetry, i.e.
\begin{equation}
A_{\ell}=A^*_{-\ell} \,.
\end{equation}
\end{thm}

For the proof see the Appendix \ref{ch:app_th_opt_p}.

Theorem \ref{thm:opt_pulse} gives a general form of the optimal transmit filter to be used
for a memory $L$ channel shortening detector. By definition, it 
becomes the classical waterfilling filter when $L=\nu_C$.
Hence, it also provides
an insight to the behavior of the transmit filter for the classical
waterfilling algorithm. We remind the reader that $\nu+1$ denotes the duration of the channel impulse
response and $\nu_C+1$ denotes the duration of the combined transmit
filter and channel response. We summarize our finding in the following
\begin{thm}\label{th:memory}
Let $P(\omega)$ be the transmit filter found through the
waterfilling algorithm. Then, 
$$ \nu_C \geq \nu.$$
\end{thm}
For a proof, see the Appendix \ref{ch:app_cs_memory}.

Whereas the statement is trivial when the transmit filter and the channel
have a finite impulse response (FIR), the theorem proves that this fact holds
also when they have infinite impulse responses (IIR). 
Thus, for a FIR channel response, the waterfilling solution cannot contain any pole that cancels a zero of the channel, while, for IIR channels, 
the waterfilling solution cannot contain any zero that cancels a pole. Thus,
the overall channel cannot be with memory shorter than the original one.

Theorem \ref{th:memory} reveals the interesting fact that the waterfilling
algorithm trades  a rate gain for detection complexity. By using the
optimal transmit filter, a capacity gain is achieved, but the
associated decoding complexity (of a full complexity detector) must
inherently increase. Thus, with waterfilling, it is not possible to
achieve both a rate gain and a decoding complexity reduction at the
same time.

\subsection*{Numerical results for the optimized filter}

Theorem \ref{thm:opt_pulse} provides a general form of the optimal transmit filter for
channel shortening detection of ISI channels. What remain to be
optimized are the $L+1$ complex-valued constants $\{A_{\ell}\}$. A closed
form optimization seems out of reach since the constraint
in (\ref{eq:const_pw})
has no simple analytical form in $\{A_{\ell}\}$. 
In fact, the integral
$$\int\sqrt{1+A\cos(x)}dx$$ is an instance of the incomplete elliptic
integral of the second kind, for which no closed form is known to
date.

We have applied a straightforward numerical optimization of the
variables $\{A_{\ell}\}$ under the constraints
in (\ref{eq:const_pw})
and
\begin{equation}
	\sum_{\ell=-L}^{L} A_{\ell}e^{j\ell\omega }\geq 0.
\end{equation}
With a standard workstation and any randomly generated channel
impulse response, the optimization is stable, converges to the
same solution no matter the starting position as long as the signal-to-noise-ratio (SNR) is not very high or very low, and is altogether a matter of fractions of a second.

\begin{figure}
	\begin{center}
		\includegraphics[width=0.75\columnwidth]{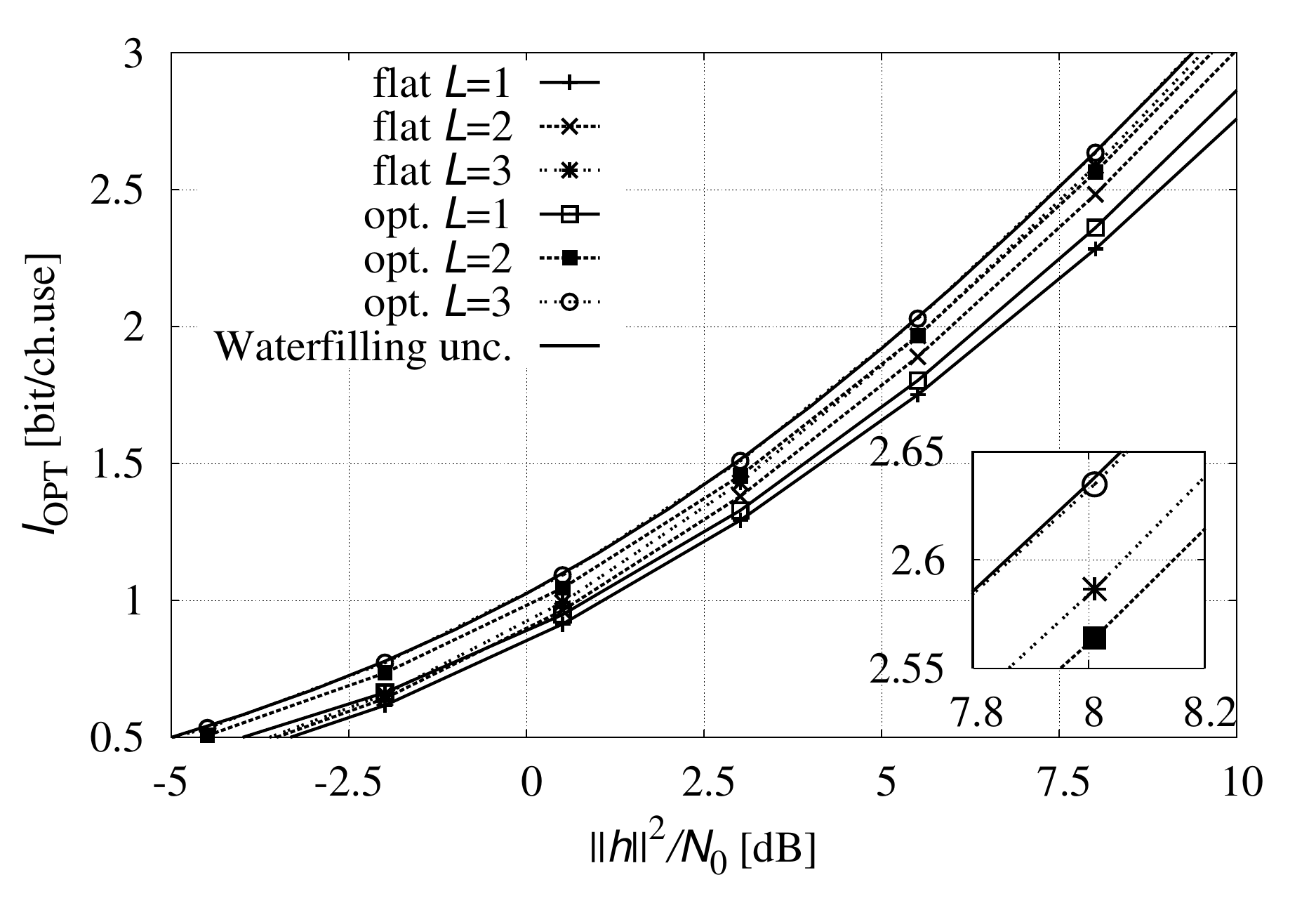}
		\caption{AIRs for Gaussian inputs when different values of the memory $L$ are considered at receiver.}\label{fig:AIR_Gauss_epr4}
	\end{center}
 \end{figure}

We now describe some illuminating examples. In all cases, the transmit power is the same both in the absence and presence of the optimal transmit filter. 
We first consider the complex channel $\bm{h}=[0.5,0.5,-0.5,-0.5j]$
with memory $\nu=3$.\footnote{Other examples can be found in~\cite{MoRuCo13}.}
Fig.~\ref{fig:AIR_Gauss_epr4} shows the AIR $I_{\mathrm{OPT}}$ for Gaussian
inputs when the transmit filter is optimized for different values of the memory $L$
considered by the receiver. 
For comparison, the figure also gives $I_{\mathrm{OPT}}$ for a flat transmit power spectrum  (i.e., no transmit filter at all) and
the channel capacity (i.e., when using the spectrum obtained by means of the waterfilling algorithm and assuming a receiver with unconstrained complexity). It can be seen that
using an optimized transmit filter for each $L$, 
significant gains are achieved w.r.t. the flat power spectrum
at all SNRs. The flat spectrum
reaches its maximum information rate when $L=\nu$ but suffers a loss 
from the channel capacity.
On the other hand, we can see that the optimized transmit filter 
when $L=\nu$ achieves an achievable rate which is close to the channel capacity.
However, there is not an exact match. This loss is due to the fact that 
$\nu$ must be lower than the combined channel-precoder memory $\nu_C$ as stated by Theorem~\ref{th:memory}.

\begin{figure}
	\begin{center}
		\includegraphics[width=0.75\columnwidth]{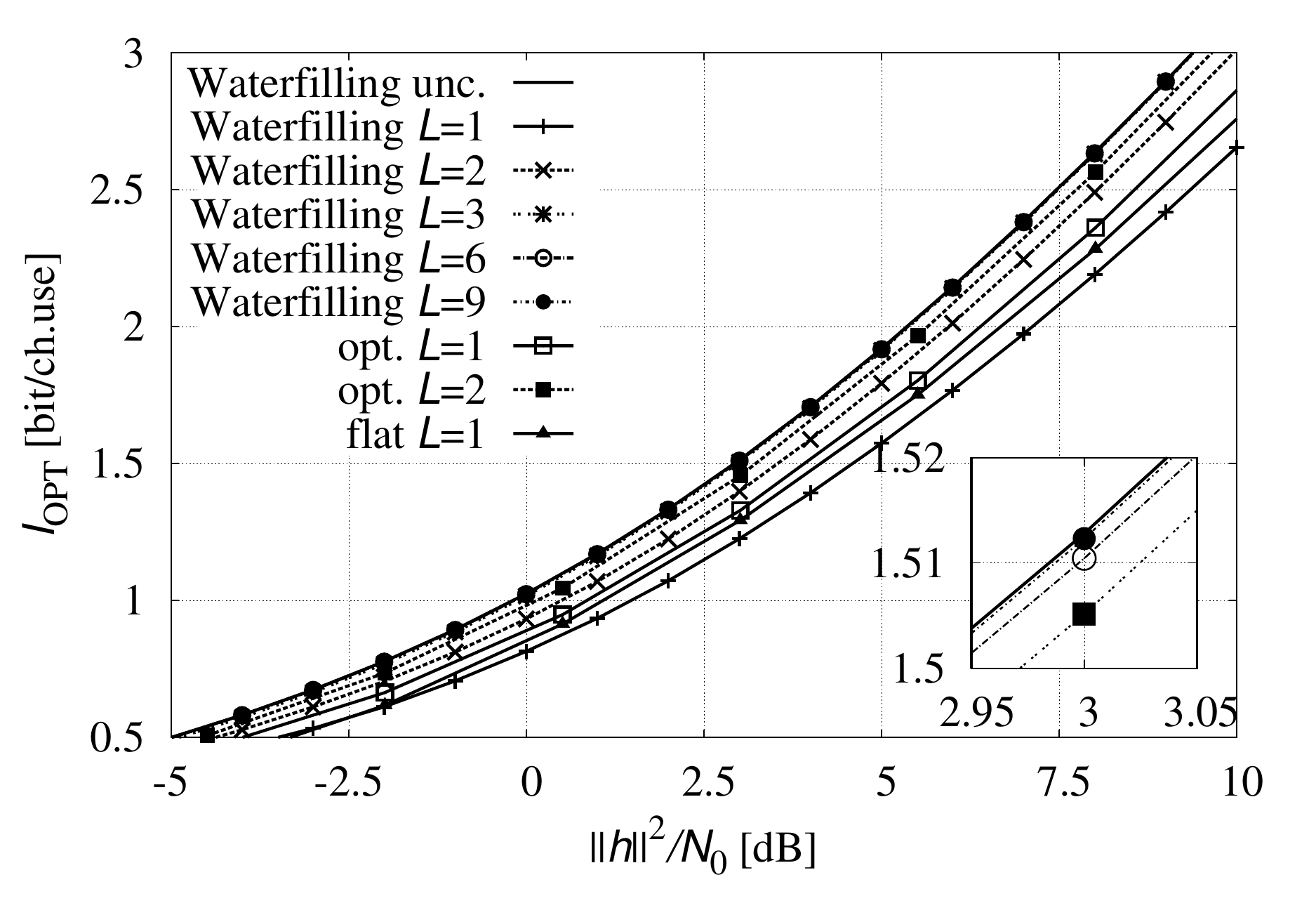}
		\caption{AIRs for Gaussian inputs with the waterfilling-solution power spectrum, when different values of the memory $L$ are considered at receiver.}\label{fig:theorem2}
	\end{center}
\end{figure}

This behavior is clearly illustrated by Fig.~\ref{fig:theorem2}, which plots the information rate when the transmit filter is found through the waterfilling algorithm and the receiver complexity is constrained with values of the memory $L$. It can be seen that when the memory $L$ is increased more and more, even above $\nu$, the information rate becomes closer and closer to the channel capacity.
Moreover, it is important to notice 
that if, na\"{\i}vely, a transmit filter found through the waterfilling algorithm is used when the receiver complexity is constrained, 
a loss w.r.t. the optimized case occurs and it may even be better to not have any transmit filter at all for high SNR values. 

\begin{figure}
	\begin{center}
		\includegraphics[width=0.75\columnwidth]{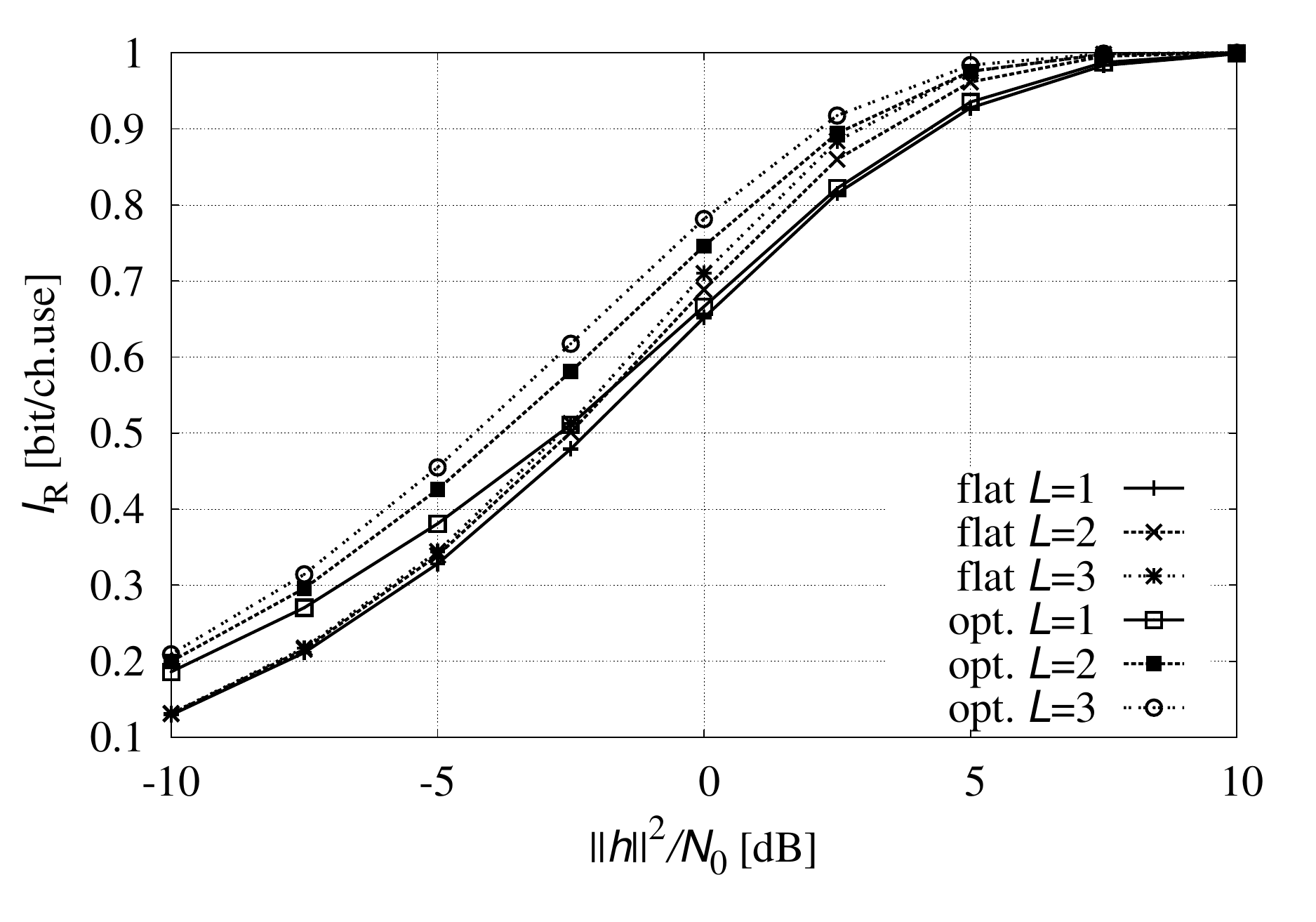}
		\caption{AIRs for BPSK modulation when different values of the memory $L$ are considered at receiver.}\label{fig:AIR_bpsk_epr4}
	\end{center}
\end{figure}

\begin{figure}
	\begin{center}
		\includegraphics[width=0.75\columnwidth]{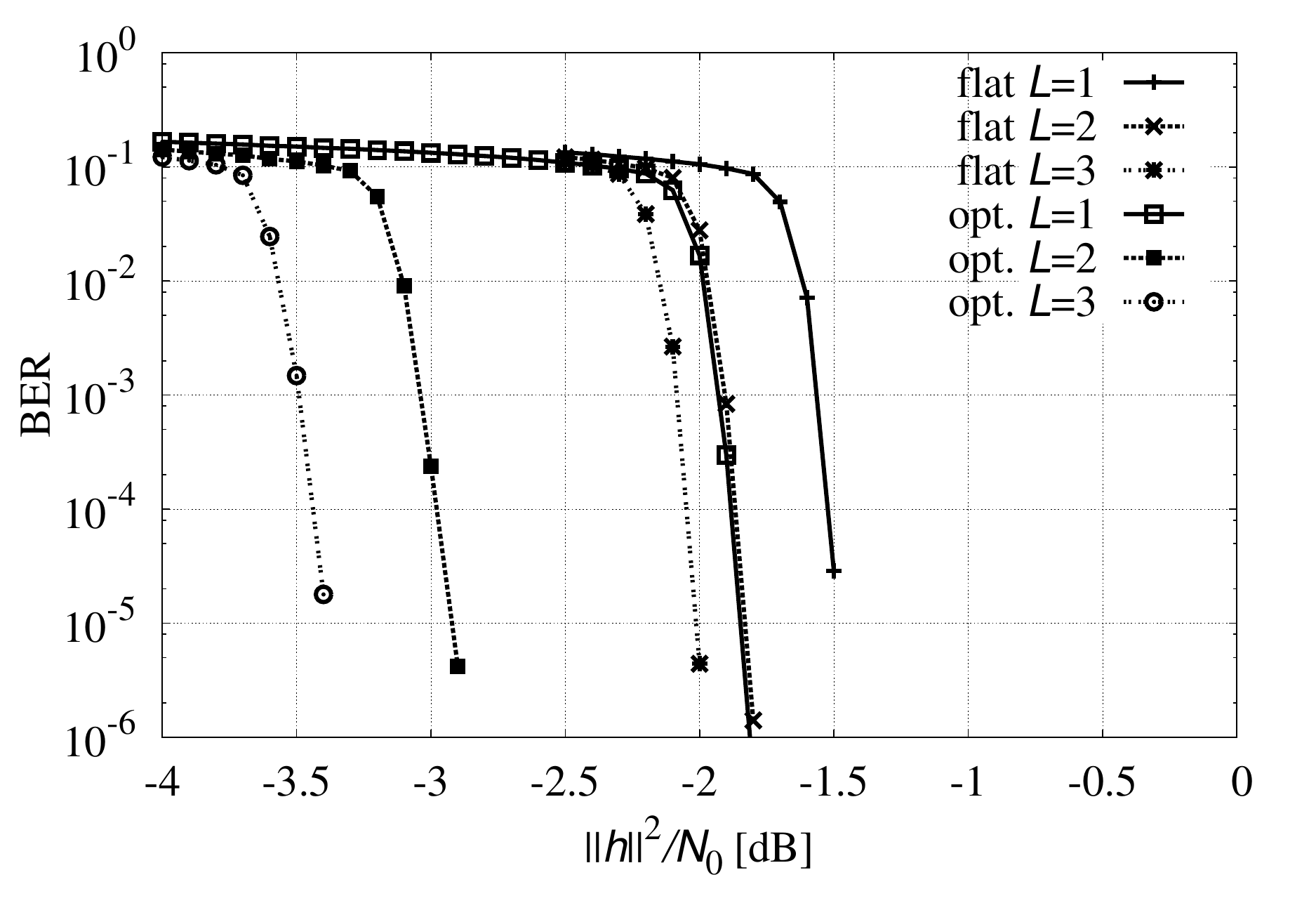}
		\caption{Bit error rate for BPSK modulation for different values of the memory $L$ considered at receiver.}\label{fig:BER_bpsk_epr4}
	\end{center}
\end{figure}

Although the results were so far presented only for Gaussian symbols, 
we now show that when the optimized transmit filter and detector for Gaussian inputs 
are used for low-cardinality discrete alphabets, the ensuing $I_{\mathrm{R}}$
is still excellent.\footnote{We remind the reader that
  $I_{\mathrm{OPT}}$ refers to an optimized detector while $I_{\mathrm{R}}$ refers to the achievable rate for a non optimized detector. Since
  the filters have been optimized for Gaussian inputs, but we are
  using here low-cardinality constellations, the filters could be
  further optimized and for these reason we use the notation $I_{\mathrm{R}}$.}
Fig.~\ref{fig:AIR_bpsk_epr4} shows the AIR for a binary phase shift keying (BPSK)
modulation.
It can be noticed that the behavior among the curves for BPSK  reflects the behavior for Gaussian symbols.
The AIR can be approached in practice with proper modulation and coding  formats. Fig.~\ref{fig:BER_bpsk_epr4} shows
the bit error rate (BER) of a BPSK-based system using the
DVB-S2 low-density parity-check code with rate 1/2. 
In all cases, 10 internal iterations  within the LDPC decoder and 10
global iterations were carried out. It can be noticed 
that the performance is in accordance with the AIR results.
All simulations that we have presented were also carried out
for other channels (e.g., EPR4, Proakis B and C)
and our findings for those channels are in principle identical to those for the channel here presented.

\begin{figure}
	\begin{center}
	\includegraphics[width=0.55\columnwidth]{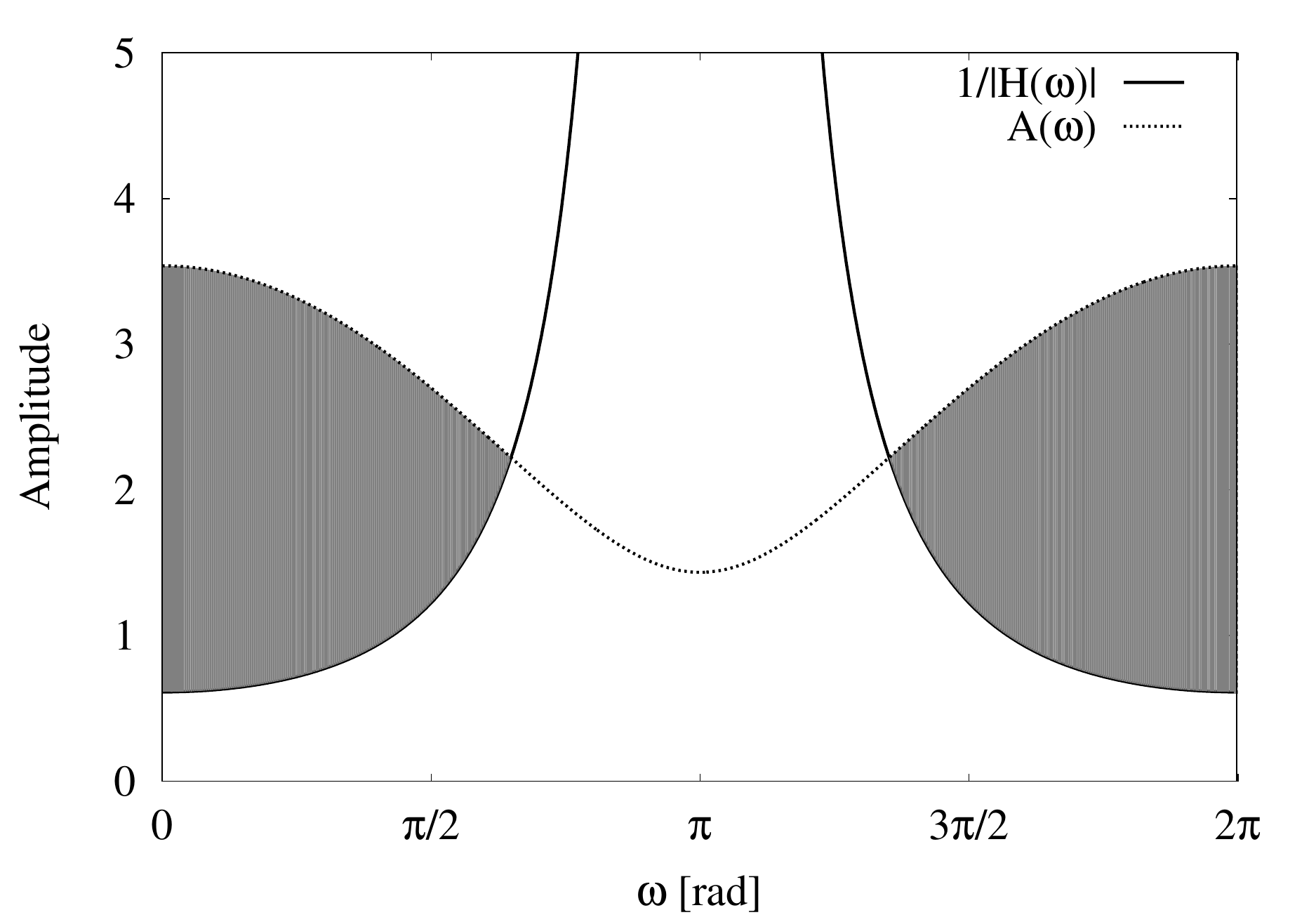} 
	\includegraphics[width=0.55\columnwidth]{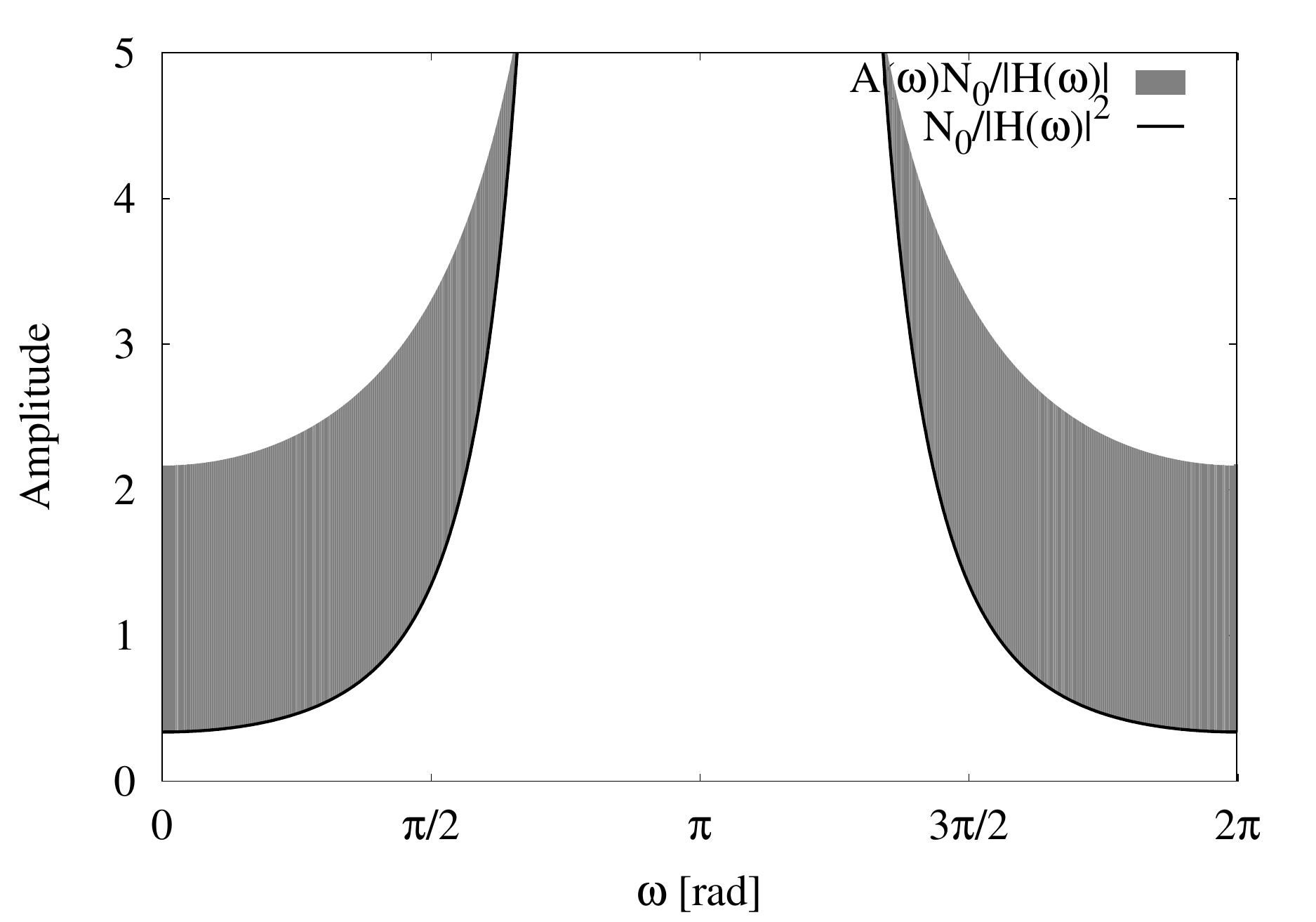}
	\includegraphics[width=0.55\columnwidth]{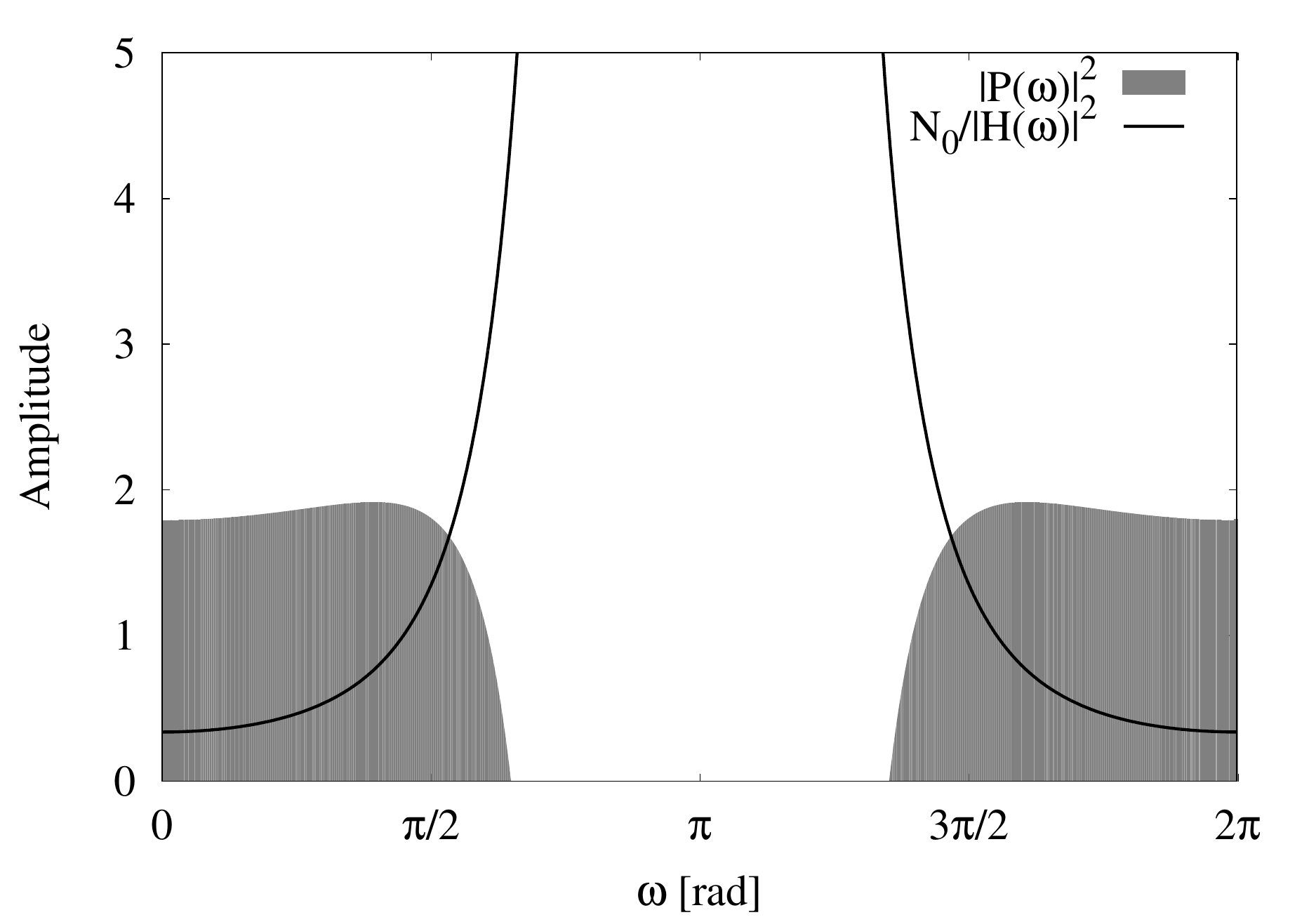}

	\caption{Graphical interpretation for the derivation of the optimal transmit
	filter for the case of the Proakis B channel, with $N_{0}=0.9$, and
	receiver constrained to memory $L=1$.}\label{fig:graph_pw}
	\end{center}
\end{figure}

Finally we show that Theorem \ref{thm:opt_pulse}, similarly to waterfilling, has a graphical interpretation,
although not effective as well.
Let us define $A(\omega)=\sqrt{\sum_{\ell=-L}^{L}A_{\ell}e^{j\ell\omega}}$.
It can be seen from (\ref{eq:Pw_opt}) that $|P(\omega)|^{2}\neq0$ when $A(\omega)\geq\frac{1}{|H(\omega)|}$.
Let us now consider as an example the Proakis B channel, for which
$\bm{h}=[0.407,0.815,0.407]$ and suppose that we are constrained
at the receiver side to $L=1$. Since the channel is real, $A(\omega)$
can be expressed as a function of the two real parameters $A_{0}$
and $A_{1}$ as $\mbox{\ensuremath{A(\omega)=\sqrt{A_{0}+2A_{1}cos(\omega)}}}$.
The magnitude of $\frac{1}{|H(\omega)|}$, is depicted in Figure \ref{fig:graph_pw} (top). 
If we also report the optimal expression of $A(\omega)$
(the optimal coefficients are $A_{0}\simeq7.3$ and $A_{1}\simeq5.2$
when $N_0=0.9$)
in the same figure, we obtain that the non-zero part of the frequency
response of the transmit filter is given by the gray filled difference
in the figure. The final amount of power spent on each frequency will
be obtained by weighing the gray filled curve by $N_{0}/|H(\omega)|$
(center part of the figure), and by reporting it over the abscissa
(bottom part of the figure). 

\clearpage
\section{Extension to MIMO-ISI channels}\label{sec:MIMO_ISI_CS}
We extend now the CS framework to MIMO-ISI channels of the form
\begin{equation}
	\bm{r}_k = \sum_{i=0}^\nu \bm{H}_i \bm{c}_{k-i} + \bm{w}_k \label{eq:MIMO_ISI}
\end{equation}
where the ISI taps $\{\bm{H}_i\}_{i=0}^\nu$ are $K\times K$ matrix.  
Vectors $\bm{r}_k$, $\bm{c}_{k}$, and $\bm{w}_k$  are, respectively, the observable, the transmitted symbols (independent identically distributed, IID)
and the white noise having autocorrelation function
\begin{equation}
	\mathrm{E}\left\{ \bm{w}_{k+i}\bm{w}^\dagger_k\right\}=N_0\bm{I}\delta_i
\end{equation}
where $\bm{I}$ is the identity matrix, and $\delta_i$ the Kronecker delta.
All vectors are column vectors with size $K$.
Without loss of generality in (\ref{eq:MIMO_ISI}) we assumed that the number of transmitting antennas
is equal to the number of receiving antennas, since any non-square channel, can be decomposed
into a square equivalent channel by means of the QR factorization \cite{RuPr12,Pr13,HoJo85}.

All vectors in (\ref{eq:MIMO_ISI}) can be gathered in a block-matrix notation as
\begin{equation}
	\mathbf{r} = \mathbf{H}\mathbf{c} + \mathbf{w} \label{eq:forney_block_matrix}
\end{equation}
where $\mathbf{H}$ is block Toeplitz, with submatrix $(\mathbf{H})_{\ell m}=\bm{H}_{\ell-m}$.
Notice that for $K=1$, (\ref{eq:forney_block_matrix}) becomes 
the matrix notation (\ref{eq:forney_matrix}) of the scalar case.
For the matrix response $\{\bm{H}_i\}$, we define the discrete time Fourier transform (DTFT) as
\begin{equation}
	\bm{H}(\omega)= \sum_i \bm{H}_i \mathrm{e}^{-j\omega i}
\end{equation}
which is equivalent to take scalar DTFT of each entry in the matrix $\{\bm{H}_i\}$.
The anti-trasform is thus define as
\begin{equation}
	\bm{H}_i = \frac{1}{2\pi}\int_{-\pi}^{\pi} \bm{H}(\omega) \mathrm{e}^{j\omega i} \mathrm{d}\omega \,. \label{eq:dtft_Hi}
\end{equation}

The optimal detection for the channel (\ref{eq:MIMO_ISI}) can be done with a generalized 
version of the Ungerboeck BCJR algorithm described in \S\ref{sec:opt_det}.
The Ungerboeck observation model is derived by filtering the samples $\{\bm{r}_k \}$
with a  filter matched to the channel, having DTFT $\bm{H}^\dagger(\omega)$.
Then, BCJR detection is performed based on the metric
\begin{equation}
	\Lambda_k(\bm{c}_k,\bm{\sigma}_k,\bm{\sigma}_{k+1})  =  \mathrm{exp}\left\{ \frac{\Re\left( 2\bm{c}_k^\dagger \bm{r}_k\right)- \bm{c}_k^\dagger \bm{G}_0\bm{c}_k -2 \bm{c}^\dagger_k\sum_{i=1}^\nu \bm{G}_i\bm{c}_{k-i}}{N_0} \right\}  \mathcal{I}(\bm{c}_k,\bm{\sigma}_k,\bm{\sigma}_{k+1})  \,,
\end{equation}
where $\bm{\sigma_k}=[\bm{c}_{k-1},\dots,\bm{c}_{k-\nu}]$ is a block vector,
and $\{\bm{G}_i\}_{i=-\nu}^\nu$ reads
\begin{equation}
	\bm{G}_i= \sum_{k=\max(0,i)}^{\min(\nu,\nu+i)} \bm{H}^\dagger_{k-i} \bm{H}_k \,.
\end{equation}

\begin{figure}
	\begin{center}
		\includegraphics[width=0.55\paperwidth]{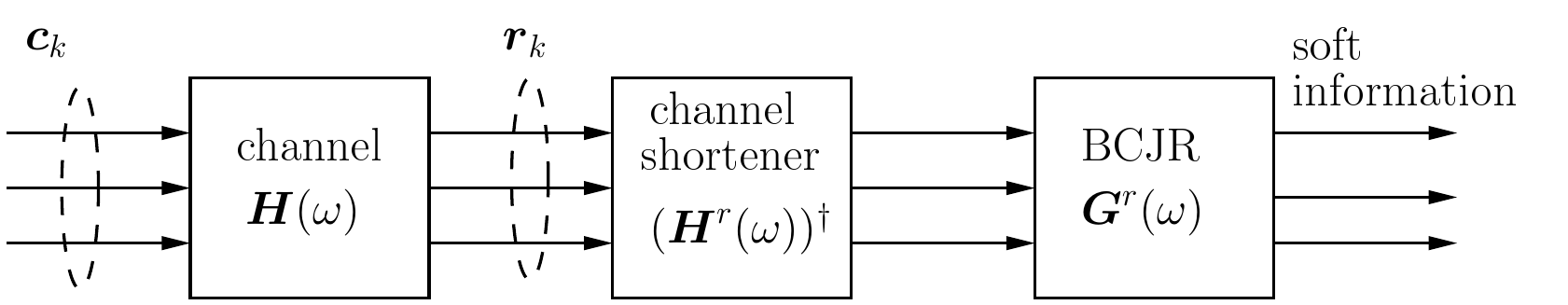}
		\caption{Block diagram of the CS transceiver scheme over a MIMO-ISI channel for $K=3$.}\label{fig:cs_det_mimoisi}
	\end{center}
\end{figure}

Optimal detection has complexity $\mathcal{O}(M^{(\nu+1) K})$.
We consider instead a channel shortening detector as in Figure~\ref{fig:cs_det_mimoisi}
where the front-end $\bm{H}^r(\omega)$ is a matrix filter with size $K\times K$.
The detection is performed on a target ISI $\{\bm{G}^r_i\}_{i=-L}^L$, being $L\leq \nu$ the memory
taken into account at detector.
The proposed CS detector performs detection on a shorter ISI, but 
by fully processing each matrix ISI tap $\bm{G}^r_i$ with size $K\times K$. The ensuing complexity 
of the CS detector is $\mathcal{O}(M^{(L+1)K})$.

The optimal front-end filter $\{\bm{H}_i^r\}$
and target response
$\{\bm{G}_i^r\}$ are obtained in closed form through the following steps:
\begin{itemize}
  \item Compute
        \begin{equation}
         \bm{B}(\omega) = N_0\bm{H}^\dagger(\omega)  
             \left[\bm{H}(\omega)\bm{H}^\dagger(\omega)+ N_0\bm{I} \right]^{-1}(\bm{H}^{\dagger}(\omega))^{-1} \,.
        \end{equation}
        Applying the anti trasform to $\bm{B}(\omega)$ yields the matrix sequence \{$\bm{B}_i$\}.
  \item Find 
   \begin{equation}
      \mathbfcal{C}= \bm{B}_0 - \mathbf{\underline{B}} \mathbf{B}^{-1}\mathbf{\underline{B}}^\dagger 
   \end{equation}
  where we defined the block matrix $\mathbf{\underline{B}}=[\bm{B}_1,...,\bm{B}_L]$ and
  the block Toeplitz $\mathbf{B}$ constructed on \{$\bm{B}_i$\} as
  \begin{equation}
	\mathbf{B}=\left( 
		\begin{array}{cccc}
		 \bm{B}_0 & \bm{B}_1 & \dots & \bm{B}_{L-1} \\ 
		 \bm{B}^\dagger_1 & \bm{B}_0 & \dots & \bm{B}_{L-2} \\ 
		 \vdots &  & \ddots & \vdots \\
		 \bm{B}^\dagger_{L-1} & \bm{B}^\dagger_{L-2} & \hdots  & \bm{B}_0 
		\end{array}
	\right) \,.
  \end{equation}

  \item Define the sequence \{$\bm{U}_i$\} where $\bm{U}_0$ is the Cholesky decomposition of $\mathbfcal{C}$, namely 
        $\mathbfcal{C}=\bm{U}_0^\dagger\bm{U}_0$, and $\bm{U}_i$ for $1 \leq i \leq L$ is
        the $(1,i)$ matrix entry of
        \begin{equation}
		\mathbf{\underline{U}}  =  -\bm{U}_0\mathbf{\underline{B}}\mathbf{B}^{-1} \,.
        \end{equation}

  \item Set 
    \begin{equation}
      \bm{G}^r_i= \sum_{k=\max(0,i)}^{\min(L,L+i)} \bm{U}_{k-i}^\dagger \bm{U}_{k} -  \delta_i\bm{I}\,.
    \end{equation}
  \item The optimal front-end filter is given by
    \begin{equation}
        \bm{H}^r(\omega)=  \left[\bm{H}(\omega)\bm{H}^\dagger(\omega)+ N_0\bm{I} \right]^{-1}\bm{H}(\omega)\left( \bm{G}^r(\omega)+\bm{I}\right)  \,.  
    \end{equation}
\end{itemize}
The proof is given in Appendix \ref{ch:app_cs}.

\subsection*{Optimization of the transmit filter for MIMO-ISI channels}

On MIMO-ISI channels, a transmit filter can be adopted with the aim of further improving the performance 
(as did for the scalar case in \S\ref{sec:opt_pulse}). Namely,
we consider the transmitted symbols $\{\bm{c}_k\}$, a precoded version of the information symbols $\{\bm{a}_k\}$.
We will show that the advantages of a transmit filter are twofold: we
can further improve the achievable information rate, and detection can be performed
as for $K$ independent parallel channels with complexity $\mathcal{O}(M^{L+1})$.

\begin{figure}
	\begin{center}
		\includegraphics[width=1.0\columnwidth]{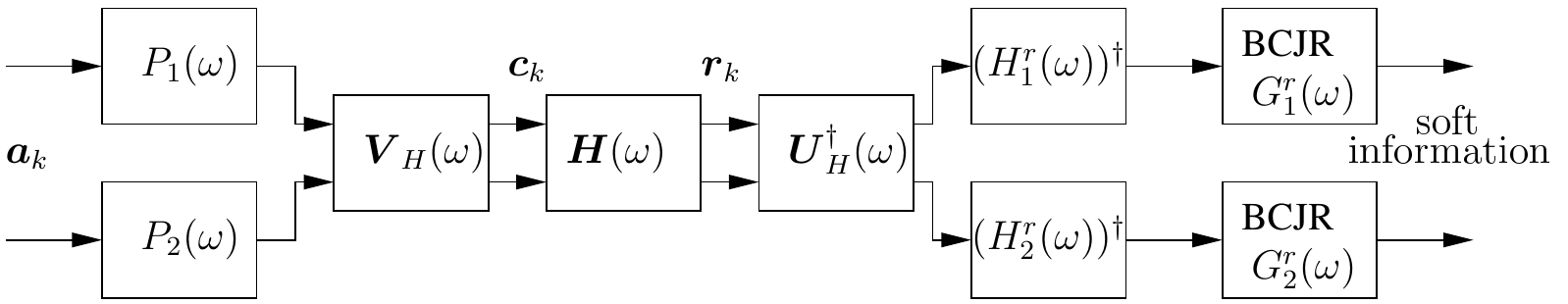}
		\caption{Block diagram of the transceiver for $2\times 2$ MIMO-ISI channels.}\label{fig:bd_awgn}
	\end{center}
\end{figure}

It is well known that a $K\times K$ MIMO channel can
be decomposed into $K$ independent parallel channels by means of singular value decomposition (SVD) \cite{Te99}. 
With a similar approach, the DTFT of $\{\bm{H}_i\}$ in (\ref{eq:dtft_Hi}),
can be factorized by means of SVD as
$$\bm{H}(\omega) = \bm{U}_H(\omega) \bm{\Sigma}(\omega) \bm{V}^{\dagger}_H(\omega)\,,$$
where $\bm{U}_H(\omega)$ and $\bm{V}_H(\omega)$ are unitary matrices and $\bm{\Sigma}(\omega)$ is a diagonal matrix
with elements $\{\Sigma_{i} (\omega)\}_{i=1}^K$.
By adopting the MIMO filter $\bm{V}_H(\omega)$ at the transmitter and the filter
$\bm{U}^{\dagger}_H(\omega)$ at the receiver, without any information loss we obtain $K$ independent parallel channels with channel
responses $\{\Sigma_{i} (\omega)\}_{i=1}^K$.
The transceiver block diagram is as shown in Fig.~\ref{fig:bd_awgn} for the case $K=2$.
The objective function to be maximized is
\begin{equation} 
	I_{\rm OPT} = \sum_{i=1}^K -\log_2(\mathcal{C}_i) 
\end{equation}
under the constraint
\begin{equation}
	\sum_{i=1}^K \frac{1}{2\pi}\int_{-\pi}^{\pi} |P_i(\omega)|^2 \mathrm{d}\omega = K
\end{equation}
where $\mathcal{C}_i$ is given in (\ref{cc}) and $P_i(\omega)$ is the precoder  for the channel $\Sigma_{i} (\omega)$.
By solving the Euler-Lagrange equation, the optimal precoders have spectra of the form~\eqref{eq:Pw_opt}.

\clearpage
\subsection*{Numerical results for MIMO-ISI channels}
We now consider a $2\times2$ MIMO-ISI channel, with \mbox{$\nu=3$} and taps
\begin{eqnarray}
	\bm{H}_0 & = & 
		\left(\begin{array}{rr}
		  -0.080302 & \phantom{-}0.256280 \\
		  0.385964 & 0.353422
		\end{array}\right) \\
	\bm{H}_1 & = & 
		\left(\begin{array}{rr}
		  \phantom{-}0.440662 & -0.168631 \\
		  0.159813 & -0.338684
		\end{array}\right) \\
	\bm{H}_2 & = & 
		\left(\begin{array}{rr}
		  -0.358555 & -0.303972 \\
		  -0.084969 & 0.668917
		\end{array}\right) \\
	\bm{H}_3 & = & 
		\left(\begin{array}{rr}
		  \phantom{-}0.669006 & 0.066229 \\
		  0.347376 & -0.207065
		\end{array}\right) \,.
\end{eqnarray}

\begin{figure}
	\begin{center}
	\includegraphics[width=0.75\columnwidth]{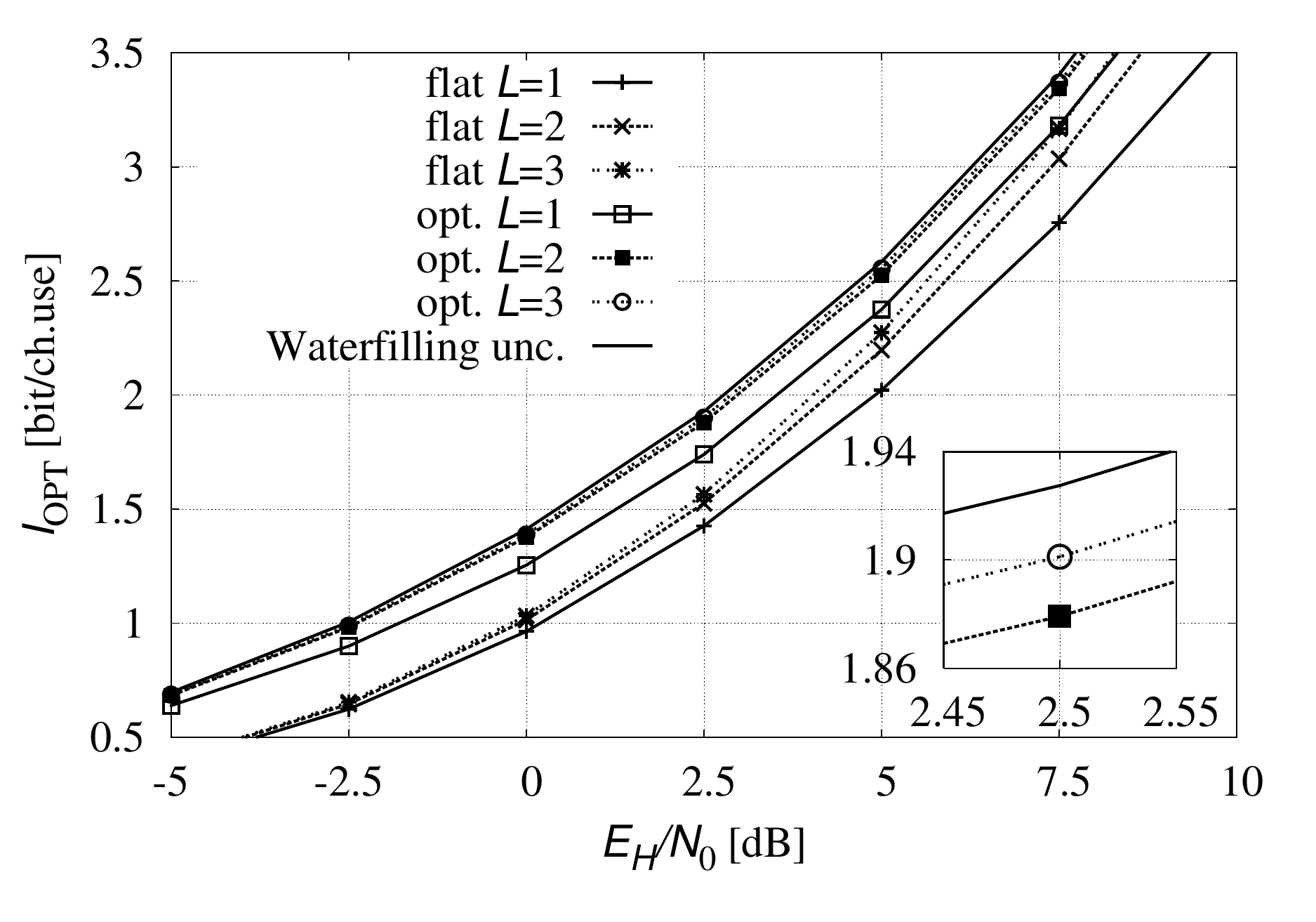}
	\caption{AIRs for Gaussian inputs over a \mbox{MIMO-ISI} channel with $K=2$ and $\nu=3$, when different values of the memory $L$ are considered at the receiver.}\label{fig:AIR_MIMO_ISI} 
	\end{center}
\end{figure}

Fig.~\ref{fig:AIR_MIMO_ISI} shows the AIR $I_{\mathrm{OPT}}$ for Gaussian
inputs as a function of $E_H/N_0$, being \mbox{$E_H=\sum_{\ell} \mathrm{Tr}(\bm{H}_\ell \bm{H}_\ell^\dagger)$}.
The transmit filters are optimized for the equivalent channels $\Sigma_1(\omega)$ and $\Sigma_2(\omega)$
for different values of the memory $L$ considered by the receiver. 
For comparison, the figure also gives $I_{\mathrm{OPT}}$ for flat transmit power spectra (i.e., $\bm{c}_k=\bm{a}_k$ and $\mathrm{E}\{\bm{a}_{k+i} \bm{a}_{k}^\dagger \}=\bm{I}\delta_i$ (where
$\bm{I}$ is the identity matrix and $\delta_i$ is the Kronecker delta) and
the channel capacity (i.e., when using the spectra obtained by means of the waterfilling algorithm and assuming a receiver with unconstrained complexity). It can be seen that
conclusions for scalar ISI channels also hold for MIMO-ISI. However, we found that, for MIMO-ISI channels, the objective function seems to have some local maxima, and thus the optimization
can depend on the starting position. This problem can be easily solved by running the optimization more times 
(three times were always enough in all our tests) and keeping the maximum value. 

\section{Channel shortening for continuous-time channels}\label{sec:cont_time_cs}
This section shows that the CS framework, although derived for discrete-time channel models, can be easily extended
to continuous-time AWGN channels.
Namely we extend the channel shortening technique to the single carrier scenario,
and a multicarrier scenario, with frequency division multiplexing (FDM).

\subsection*{Linear modulation on the continuous-time AWGN channel}
We consider a linear modulation over a continuous-time AWGN channel.
The received signal reads
\begin{equation}
	r(t)= \sum_{k=0}^{N-1} c_k \tilde{p}(t-kT) +w(t) \label{eq:cont_time_cs}
\end{equation}
where $\{c_k\}_{k=0}^{N-1}$ are the $N$ transmitted symbols, which are independent and identically distributed (IID).
The $\tilde{p}(t)$ is the shaping pulse, 
$T$ the symbol time, and $w(t)$ is white Gaussian noise with power
spectral density $N_0$.
The shaping pulse $\tilde{p}(t)$ is constrained  to have bandwidth $W$ and energy
\begin{equation}
	\int_{-W/2}^{W/2} |\tilde{P}(f)|^2 \mathrm{d}f=1
\end{equation}
being $\tilde{P}(f)$ the Fourier transform of $\tilde{p}(t)$.
The channel is assumed perfectly known at the
receiver and time-invariant. The channel frequency response is assumed flat over
$W$, although the generalization to the case of a frequency-selective channel is straightforward.

As explained in \S\ref{sec:obs_model}, a sufficient statistics
for detection of (\ref{eq:cont_time_cs}) can be carried out by using a whitening matched filter (WMF).
The ensuing observable, is the Forney observation model and reads as (\ref{eq:scalar_ccp}), 
where the ISI $\{ v_i \}$ of the combined channel precoder is such that
\begin{equation}
	|V(\omega)|^2  =  \frac{1}{T}\sum_i\left|\tilde{P}\left(\frac{\omega}{2\pi T}- \frac{i}{T}\right)\right|^2 \,,
\end{equation}
and optimal detection can be performed
with the BCJR algorithm (see \S\ref{sec:opt_det}).
Clearly, this discrete-time model will depend on the adopted shaping pulse,
its bandwidth, the employed symbol time, and the channel impulse response if the channel is frequency selective.

\begin{figure}
	\begin{center}
		\includegraphics[width=1.0\columnwidth]{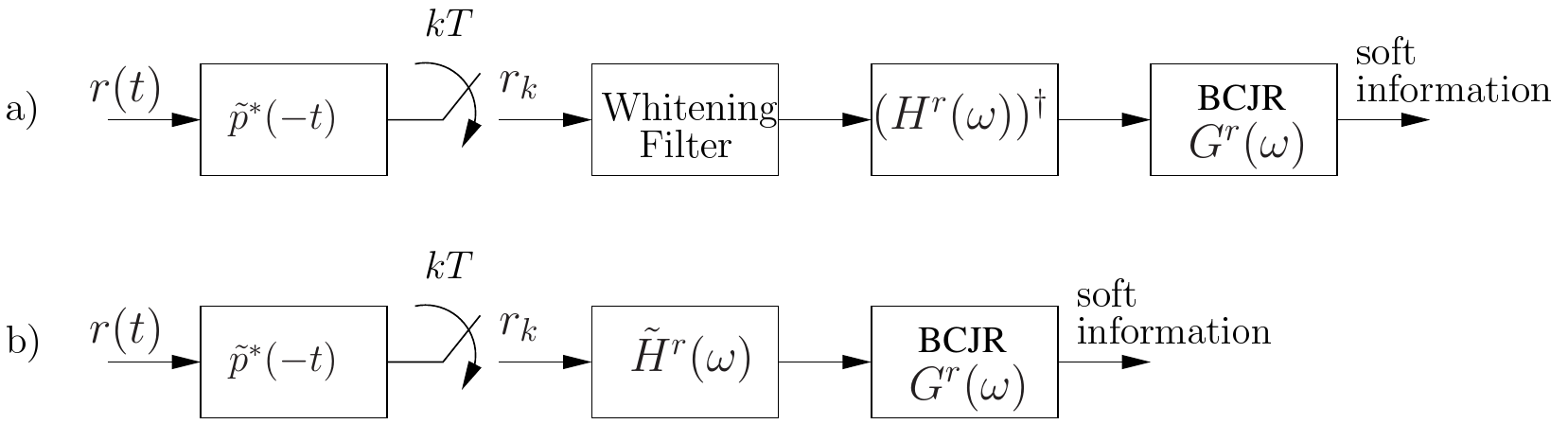}
		\caption{Block diagram of the CS receiver for continuous-time AWGN channels.}\label{fig:rec_cs_ct}
	\end{center}
\end{figure}

Instead of optimal detection, we want to consider a CS detector with memory $L$.
The optimal target ISI and channel shortener
are derived again through (\ref{cc})--(\ref{eq:Hr})
by means of the $\{b_i\}_{i=-L}^L$ in (\ref{eq:b_V}).
The corresponding channel shortening receiver is shown in Figure \ref{fig:rec_cs_ct}a.
Since the WMF can be implemented as a cascade of
a continuous-time matched filter followed by a discrete-time
whitening filter, this latter filter can be combined with the
channel shortening filter obtaining a single discrete-time filter
with frequency response
\begin{equation}
	\tilde{H}^r(\omega) = \frac{G^r(\omega)+1}{|V(\omega)|^2+N_0} \,.
\end{equation}
The corresponding channel shortening receiver is shown in Figure \ref{fig:rec_cs_ct}b.

The shaping pulse $\tilde{p}(t)$ can be also optimized by means of the framework in \S\ref{sec:opt_pulse}.
The DTFT of $\{v_k\}$ can be decomposed as
\begin{equation}
	|V(\omega)|^2=|P(\omega)|^2|H(\omega)|^2
\end{equation}
where
\begin{equation}\label{eq:channel_opt_pulse}
	H(\omega)= \begin{cases}
			1 & |\omega| \leq 2 WT\pi  \\
			0 & \mathrm{otherwise}  
	           \end{cases} \,,\,\, \omega \in [-\pi,\pi] \,.
\end{equation}

Thus the optimization problem is still given by (\ref{eq:opt_problem}) where the optimal shaping pulse is such that
\begin{equation}
	|\tilde{P}(f)|^2= T |P(2\pi Tf)|^2
\end{equation}
with $|P(\omega)|^2$ given in \eqref{eq:Pw_opt}. 

Clearly, when $2 WT \geq 1$, the optimal solution is trivial and $|P(\omega)|^2$
is flat. Thus, for $2 WT=1$ the $\tilde{p}(t)$ is a $\mathrm{sinc}$ function, whereas for
$2 WT > 1$ the $\tilde{p}(t)$ can be a pulse whose spectrum has vestigial symmetry (e.g., pulses
with a root raised cosine (RRC) spectrum).
For $2WT < 1$, the  symbol time is such that the Nyquist condition for the absence of ISI cannot be satisfied.
Thus, we are working in the domain of the {\it faster-than-Nyquist} (FTN) paradigm~\cite{Ma75c,LiGe03,RuAn05} or its extension 
represented by time packing~\cite{BaFeCo09b,MoCoAl12}.
Note that, as said before, the discrete-time channel model, will depend on the values of $W$ and $T$. When changing the values 
of $W$ and/or $T$, the corresponding optimal pulse will change and so the maximum value of the AIR for the given allowed complexity. 
In general, when reducing the value of $WT$, the maximum AIR value will decrease. However, the spectral efficiency, defined as the 
ratio between the AIR and the product $WT$ could, in principle, increase~\cite{Ma75c,LiGe03,RuAn09,RuAn05,MoCoAl12,BaFeCo09b}.
This is the rationale behind FTN/time packing that allows to improve the spectral efficiency by accepting interference. 
The optimal value of $T$ is, in that case, properly optimized to maximize the spectral efficiency. This optimization can be now 
performed by also using, for each value of $T$, the corresponding optimal shaping pulse. In other words, we can find the optimal 
pulse for a constrained complexity detector
when FTN/time packing is adopted.

We point out that, for this scenario, the numerical computation of the optimal shaping pulse
in the time-domain can require the adoption of some windowing technique or the use of Parks-McClellan algorithm \cite{OpSc89} 
to obtain a practical pulse since $H(\omega)$ has a spectrum with an ideal frequency cut.

We finally point out that the optimization of the shaping pulse to frequency selective AWGN channels is done straightforwardly
by properly defining  the channel~(\ref{eq:channel_opt_pulse}).

\subsection*{Numerical results for the optimized shaping pulse}
\begin{figure}
	\begin{center}
		\includegraphics[width=0.75\columnwidth]{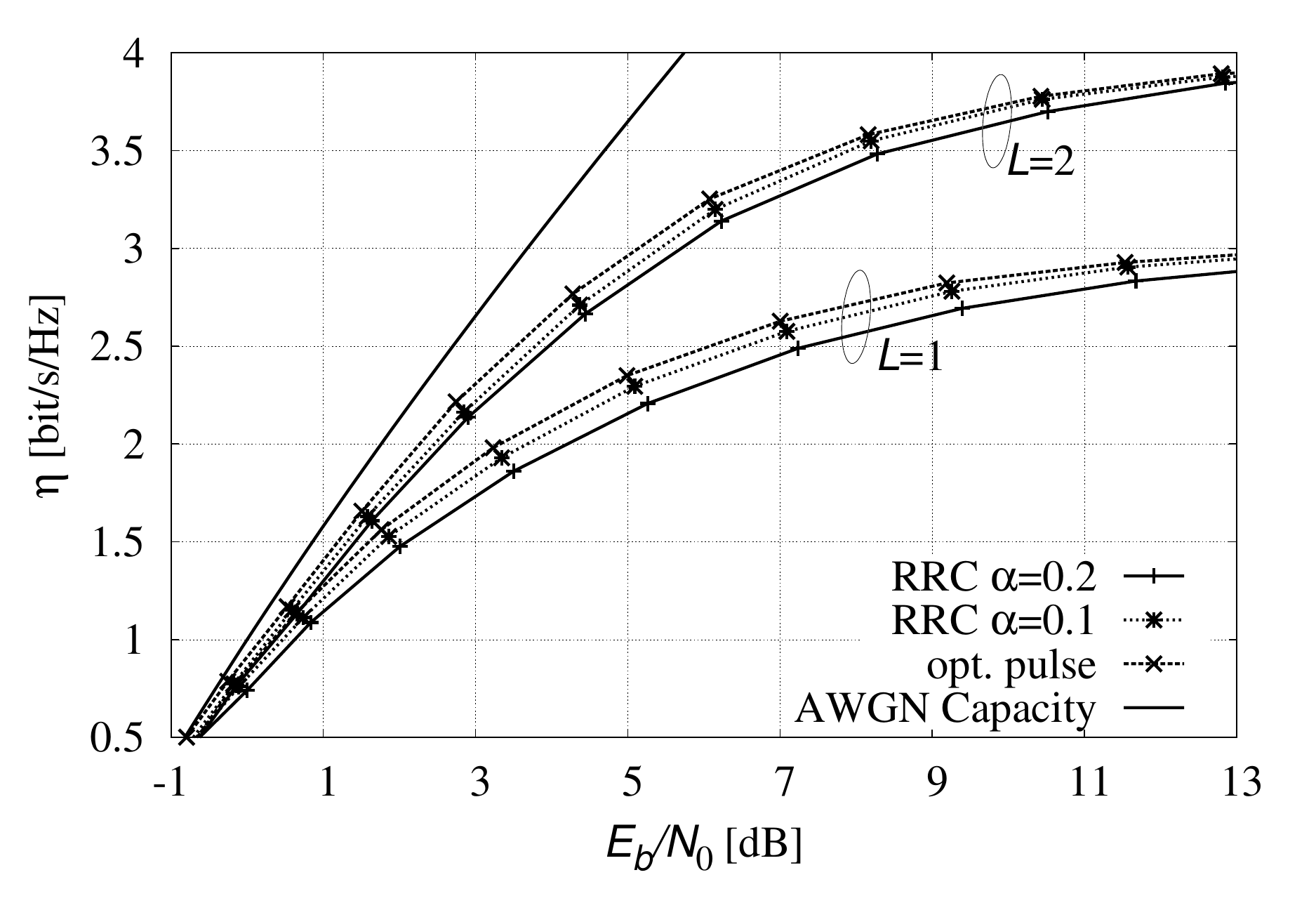}
		\caption{ASE for a BPSK modulation by using the optimized pulse for two values of the memory $L$ considered at receiver.}\label{eq:se_tpack}
	\end{center}
\end{figure}

We computed the optimal shaping pulse on a bandlimited AWGN channel when 
the bandwidth $W$ and the symbol time $T$ are such that  $2WT=0.48$.
Hence, we are in the realm of FTN/time packing and the considered ISI is only due to the adoption of such a technique.
Fig.~\ref{eq:se_tpack} shows the achievable spectral efficiency (ASE) 
\mbox{$\eta=I_{\mathrm{R}}/WT$} for a BPSK modulation on the \mbox{continuous-time} AWGN channel
as a function of the ratio $E_b/N_0$, $E_b$ being the received signal energy per information bit. 
Two values of the memory, namely $L=1$ and $L=2$  are considered at the detector.
For comparison, the figure also gives the ASE for pulses with RRC spectrum
and roll-off $\alpha=0.1$ or $\alpha=0.2$, 
and the unconstrained capacity for the AWGN channel.
It can be seen that the optimized pulse outperforms the other pulses.

\subsection*{FDM on the continous-time AWGN channel}
We consider a scenario with $K$ carriers, each transmitting the symbols
$\{c_{k}^{(\ell)}\}_{k=0}^{N-1}$, being $\ell$ the index of the carrier.
The received signal reads 
\begin{equation}
	r(t)= \sum_{\ell=0}^{K-1}\sum_{k=0}^{N-1} c_k^{(\ell)} p_\ell(t-kT)e^{j2\pi F_\ell t} + w(t) \label{eq:r_fdm}
\end{equation}
where $F_\ell$ is the frequency of the $\ell$-th carrier, $p_\ell(t)$ its shaping pulse, $T$ the symbol time, and $w(t)$
white Gaussian noise with power spectral density $N_0$.

\begin{figure}
	\begin{center}
		\includegraphics[width=1.0\columnwidth]{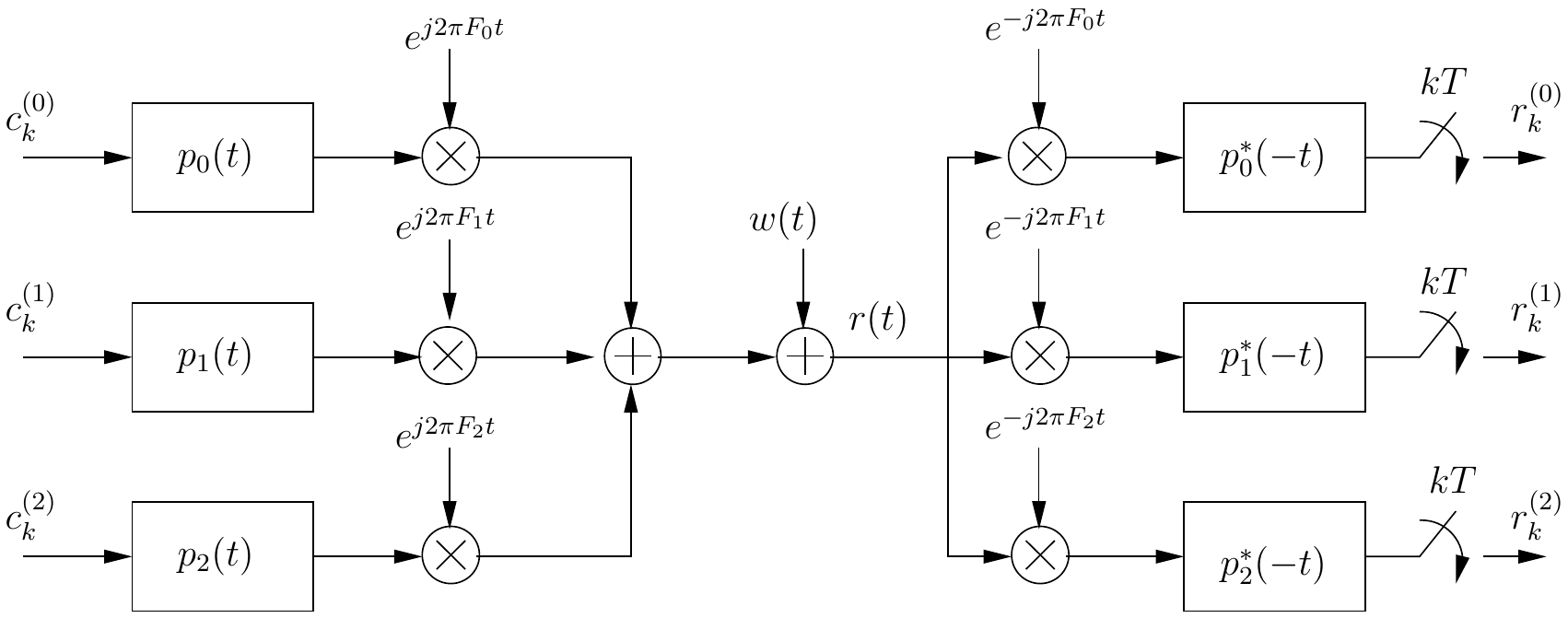}
		\caption{Block diagram of the FDM transceiver scheme over the AWGN channel, when $K=3$.}\label{fig:bd_fdm}
	\end{center}
\end{figure}

\begin{figure}
	\begin{center}
		\includegraphics[width=0.65\columnwidth]{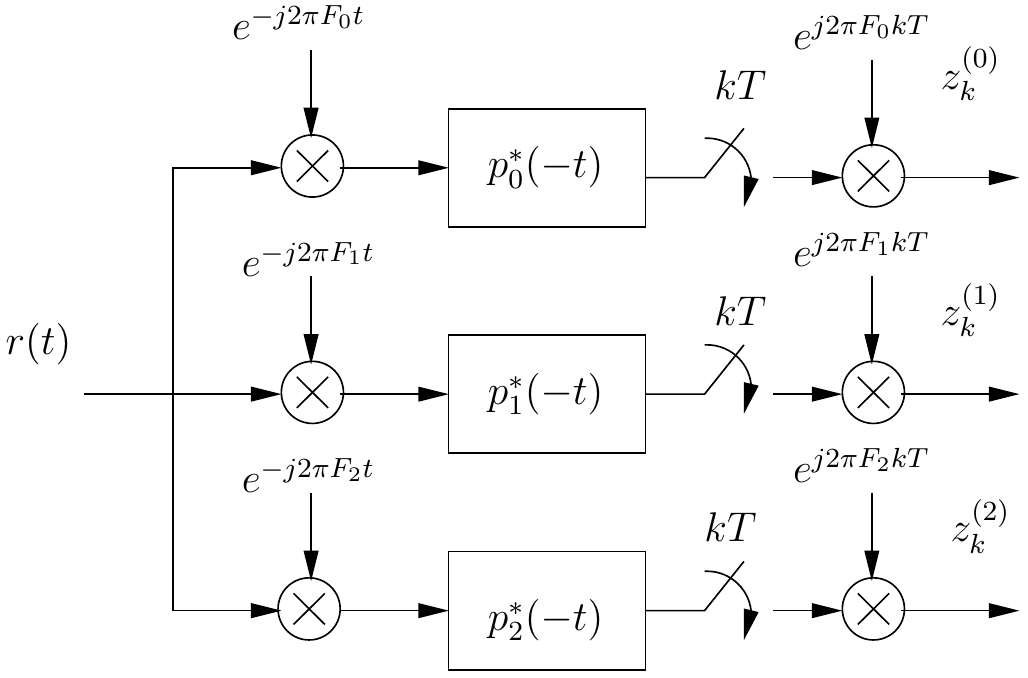}
		\caption{Modified receiver to obtain an equivalent stationary FDM channel.}\label{fig:mod_fdm}
	\end{center}
\end{figure}

A sufficient statistics $\bm{r}_k=[r^{(0)}_k,\dots,r^{(K-1)}_k ]^T$, for the detection of (\ref{eq:r_fdm}) is found by adopting a bank of matched filter to 
the received signal.
The block diagram for $K=3$ is shown in Figure \ref{fig:bd_fdm}.
It can be shown that the observable $\bm{r}_k$ reads
\begin{equation}
	\bm{r}_k = \sum_i \bm{G}_{k,k-i} \bm{c}_{k-i} + \bm{n}_k \label{eq:fdm_sample}
\end{equation}
where $\bm{c}_k=[c_k^{(0)},\dots,c_k^{(K-1)}]^T$, $\bm{n}_k$ is colored Gaussian noise with autocorrelation
\begin{equation}
	\mathrm{E}\left\{ \bm{n}_{k+i}\bm{n}_k \right\} = N_0 \bm{G}_{k+i,k}
\end{equation}
and $\bm{G}_{k,j}$ is a $K\times K$ matrix with entries
\begin{equation}
	\left( \bm{G}_{k,j} \right)_{\ell,u} = e^{-j2\pi(F_\ell-F_u)jT}\int_{-\infty}^{\infty} p_u(t)p_{\ell}^*(t-(k-j)T)e^{-j2\pi(F_\ell-F_u)t}\mathrm{d}t \,.
\end{equation}
Clearly the channel (\ref{eq:fdm_sample}) is not stationary,
since the matrix $\bm{G}_{k,j}$ are time variant.
However if the receiver is modified as in Figure \ref{fig:mod_fdm}
we obtain the equivalent stationary channel
\begin{equation}
	\bm{z}_k = \sum_i \bm{\tilde{G}}_{i} \bm{x}_{k-i} + \bm{\tilde{n}}_k \label{eq:stat_FDM}
\end{equation}
where
\begin{equation}
 \left( \bm{\tilde{G}}_{i} \right)_{\ell,u} = e^{-j2\pi(F_\ell-F_u)iT}\int_{-\infty}^{\infty} p_u(t)p_{\ell}^*(t-iT)e^{-j2\pi(F_\ell-F_u)t}\mathrm{d}t \,,
\end{equation}
\begin{equation}
	\bm{x}_k=\bm{c}_k \circ \left[  
	\begin{array}{c}
		e^{j2\pi F_0 k T } \\
		\vdots \\
		e^{j2\pi F_{K-1} k T }
	\end{array}
	\right]\,,
\end{equation}
and $\circ$ is the Hadamard product.
The equivalent stationary channel does not involve any information loss.
In fact, it is easy to prove that
\begin{equation}
	I(\bm{c};\bm{r})  =  I(\bm{c};\bm{z})  \,.
\end{equation}

Finally, it can be noticed that (\ref{eq:stat_FDM}) is the Ungerboeck observation model of a MIMO-ISI channel.
Thus the CS detector can be designed as described in \S\ref{sec:MIMO_ISI_CS}.
      \chapter{Time packing}\label{chap:time_pack}

\lettrine{I}{n} satellite links for broadcasting and broadband applications, orthogonal
signaling, that ensures absence of intersymbol interference (ISI),
is often adopted. As an example, in the 2nd-generation satellite digital
video broadcasting (DVB-S2) standard~\cite{DVB-S2-TR}, a conventional
square-root raised-cosine (RRC) pulse shaping filter is specified
at the transmitter. In an additive white Gaussian noise channel and
in the absence of other impairments, the use of a matched filter (MF)
at the receiver and proper sampling ensure that optimal detection
can be performed on a symbol-by-symbol basis. On the other hand, it
is known that, when finite-order constellations are considered {[}e.g.,
phase-shift keying (PSK){]}, the efficiency of the communication system
can be improved by giving up the orthogonality condition, thus accepting
interference. For example, {\it faster-than-Nyquist signaling} (FTN, see
\cite{Ma75c,MaLa88}) is a well known technique consisting
of reducing the spacing between two adjacent pulses in the time-domain
well below the Nyquist rate, thus introducing ISI. If the receiver
is able to cope with the interference, the efficiency of the communication
system will be increased. In the original papers on FTN signaling
\cite{Ma75c,MaLa88}, this optimal time spacing is obtained as the
smallest value giving no reduction of the minimum Euclidean distance
with respect to the Nyquist case. This ensures that, asymptotically,
the ISI-free performance is reached, at least when the optimal detector
is adopted. The i.u.d. capacity or information rate, i.e., the average
mutual information when the channel inputs are independent and uniformly
distributed (i.u.d.) random variables, is then computed, still assuming
the adoption of the optimal detector~\cite{RuAn06,RuAn07}. However,
the complexity of this optimal detector easily becomes unmanageable,
and no hints are provided on how to perform the optimization in the
more practical scenario where a reduced-complexity receiver is employed. 

In~\cite{BaFeCo09b}, a different approach for improving the spectral
efficiency, that relies on both {\it time packing} of adjacent symbols and
reducing the spacing of the adjacent channels when applicable (multi-carrier
transmission), has been considered. It is assumed that, at the receiver
side, a symbol-by-symbol detector working on the samples at the MF
output is adopted, and the corresponding information rate is computed,
by also optimizing time and frequency spacings to maximize the \emph{achievable
spectral efficiency} (ASE). Hence, rather than the minimum distance,
the \emph{ASE} \emph{is the performance measure} and, in addition,
a low-complexity detection algorithm, characterized by a given allowable
complexity \emph{irrespectively of the interference set size}, is
considered at the receiver rather than the optimal detector employed
in~\cite{Ma75c,MaLa88,RuAn06,RuAn07}. Although the MF output
represents a set of sufficient statistics for optimal detection, a
suboptimal symbol-by-symbol receiver is considered in~\cite{BaFeCo09b}.
Hence, the ASE can be improved by employing more sophisticated detection
algorithms. In this chapter, we will consider two cases: (i) a proper
filtering of the MF output plus a symbol-by-symbol detector and (ii)
the maximum \emph{a posteriori} (MAP) symbol detector that, in order
to limit the receiver complexity, takes into account only a limited
amount of interference. 

This technique represents a good alternative, for low-order constellations,
to the shaping of the transmitted symbol distribution~\cite{CaOz90},
providing spectral efficiencies that cannot be reached when orthogonal
signaling is employed. Improving the ASE without increasing the constellation
order can be considerably convenient since the larger the constellation
size, the higher the decoding complexity and the lower the robustness
to channel impairments such as time-varying phase noise and non-linearities.
In the case of frequency packing, a further improvement could be achieved
by adopting, at the receiver side, a multi-user detector.
The remainder of this chapter is organized as follows. The system model
is described in~\S\ref{sec:System-Model}. In~\S\ref{sec:Spectral-Efficiency-optimization},
we compute and optimize the spectral efficiency considering detectors
with different complexity. Numerical results are reported in~\S\ref{sec:Simulation-Results},
where we also show the performance of some efficient modulation and
coding formats (MODCODs) designed accordingly.

\section{System model\label{sec:System-Model}}

We consider an additive white Gaussian noise (AWGN) channel and a
frequency-division multiplexed system where perfectly synchronized
(downlink assumption) adjacent channels employ the same linear modulation
format, shaping pulse $p(t)$, and symbol interval (or time spacing)
$T$. The shaping pulse is assumed to have unit energy. The received
signal can be expressed as
\begin{equation}
r(t)=\sqrt{E_{s}}\sum_{k,\ell} c_{k}^{(\ell)}p(t-kT)e^{j2\pi \ell Ft}+w(t)\label{eq:rec_signal2}
\end{equation}
where $E_{s}$ is the symbol energy, $c_{k}^{(\ell)}$ the symbol transmitted
over the $u$-th channel during the $\ell$-th symbol interval, $F$
the frequency spacing between adjacent channels, and $w(t)$ a circularly
symmetric zero-mean white Gaussian noise process with power spectral
density $N_{0}$. The transmitted symbols $\{c_k^{(\ell)}\}$ are independent
and uniformly distributed and belong to a given zero-mean $M$-ary
complex constellation $\chi$ properly normalized such that $E\{|c_{k}^{(\ell)}|^{2}\}=1$.
Note that the summations in~\eqref{eq:rec_signal2} extend from $-\infty$
to $+\infty$, namely an infinite number of time epochs and carriers
are employed. For the spectral efficiency computation, we will consider
the central user only using $F$ as a measure of the signal bandwidth.

The base pulse $p(t)$ has often RRC-shaped spectrum (RRC pulse in
the following) with roll-off factor $\alpha$. In addition to it,
we will consider other transmit pulses, e.g., a pulse whose spectrum
is raised-cosine (RC) shaped (RC pulse in the following) and a Gaussian
pulse. In general, we will consider the case of time-frequency packing
and we will optimize the frequency separation $F$ between two adjacent
users and the symbol interval $T$ in order to maximize the ASE. In
the case of bandlimited pulses (i.e., RRC and RC pulses), we will
also consider time packing only. In this case, adjacent users are
not allowed to overlap in frequency (i.e., $F=(1+\alpha)/T$ for RRC
and RC pulses) and we may assume that only the user with $\ell=0$
is transmitted. In satellite communications, this can correspond to
the use of a single carrier occupying the entire transponder bandwidth.
This is of particular interest since the on-board power amplifier
can operate closer to saturation and hence improve the efficiency.

\section{Spectral efficiency optimization\label{sec:Spectral-Efficiency-optimization}}

In this section, we shown how to compute the ASE for a given receiver
and how to optimize the values of $T$ and $F$.

\subsection*{Symbol-by-Symbol detection\label{sec:Symbol-by-Symbol}}

\begin{figure}
	\begin{center}
		\includegraphics[width=1.0\columnwidth]{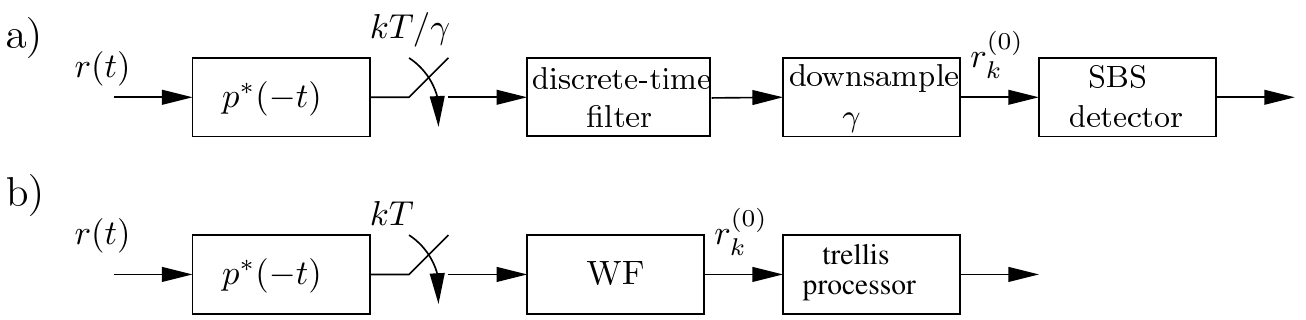}
		\caption{Some considered receivers: (a) symbol-by-symbol detector and (b) single-user
		detector based on trellis processing.\label{fig:BlockDiagramReceiver}}
	\end{center}
\end{figure}

Let us consider the central user (i.e., that for $\ell=0$). We first
consider the case shown in Figure \ref{fig:BlockDiagramReceiver}(a)
of a receiver composed by a filter matched to the shaping pulse $p(t)$,
followed by a proper discrete-time filter, that works on $\gamma\geq1$
samples per symbol interval, and a symbol-by-symbol (SBS) detector.
Although the discrete-time filter could be, in general, fractionally-spaced
(FS, i.e., $\gamma>1$), the detector will operate on one sample per
symbol interval. These samples will be denoted by $\{r_{k}^{(0)}\}$ and
can be expressed as
\begin{equation}
r_{k}^{(0)}=  \sqrt{E_{s}}c_{k}^{(0)}h(0,0,k) +\sqrt{E_{s}}\sum_{(n,\ell)\neq(0,0)}c_{k-n}^{(\ell)}h(n,\ell,k)+z_{k}\label{eq:sample_yk}
\end{equation}
in which $h(n,\ell,k)$ is the residual interference at time $kT$
due to the $\ell$-th user and the $(k-n)$-th transmitted symbol,
and $\{z_{k}\}$ is the additive noise term, in general colored unless
a whitening filter (WF) is employed after the MF. The discrete-time
filter is assumed properly normalized such that the noise variance
is $N_{0}$. The dependence of coefficients $h(n,\ell,k)$ on $k$
is through a complex coefficient of unit amplitude which disappears
for $\ell=0$ (hence $h(n,0,k)$ is independent of $k$) and is due
to the fact that $F$ is not an integer multiple of $1/T$.

Eq.~\eqref{eq:sample_yk} shows the two different impairments experienced
by the receiver, namely the background noise and the interference.
Instead of simply neglecting the interference due to adjacent symbols
and users, we pursue here a more general approach, which consists
of modeling the interference as a zero-mean Gaussian process with
power spectral density equal to $N_{I}$, of course independent of
the additive thermal noise---we point out that this approximation
is exploited only by the receiver, while in the actual channel the
interference is clearly generated as in~\eqref{eq:sample_yk}. Note
that the interference is really Gaussian distributed only if the transmitted
symbols $c_{k}^{(\ell)}$ are Gaussian distributed as well. However, especially
when the interference set is small, e.g., when $T$ and $F$ are large,
the actual interference distribution may substantially differ from
a Gaussian distribution.

We define \emph{auxiliary channel} the channel model assumed by the
receiver. With the above mentioned Gaussian approximation, the auxiliary
channel is 
\begin{equation}
r_{k}^{(0)}=\sqrt{E_{s}}c_{k}^{(0)}h(0,0,k)+\mathfrak{v}_{k}\label{e:auxiliary}
\end{equation}
where $\{\mathfrak{v}_{k}\}$ are independent and identically distributed zero-mean
circularly symmetric Gaussian random variables, with variance $N_{0}+N_{I}$.
It turns out that 
\begin{equation}
N_{I}=E_{s}\sum_{(n,\ell)\neq(0,0)}|h(n,\ell,k)|^{2}\label{eq:N_i}
\end{equation}
which results to be independent of $k$, as can be easily shown. We
are interested in evaluating the ultimate performance limits achievable
by a symbol-by-symbol receiver designed for the auxiliary channel~\eqref{e:auxiliary}
when the actual channel is that in~\eqref{eq:sample_yk}, in terms
of information rate (or spectral efficiency). This issue is an instance
of \emph{mismatched} detection~\cite{MeKaLaSh94} (see also~\cite{ArLoVoKaZe06}).
The achievable information rate (AIR), measured in bit per channel
use, for this mismatched receiver is 
\begin{equation}
	I_{\mathrm{R}} = \mathfrak{h}(r_k^{(0)}) - \mathfrak{h}(r_k^{(0)}|c_k^{(0)}) \label{e:AIR}
\end{equation}
where
\begin{eqnarray}
	\mathfrak{h}(r_k^{(0)}) & = & -\mathrm{E}\left\{ \log_2 \left( \sum_{c\in\chi}q(r_{k}^{(0)}|c)\frac{1}{M} \right) \right\} \\
	\mathfrak{h}(r_k^{(0)}|c_k^{(0)}) & = & -\mathrm{E}\left\{ \log_2  q(r_{k}^{(0)}|c_{k}^{(0)}  ) \right\} 
\end{eqnarray}
where $q(r_{k}^{(0)}|c_{k}^{(0)})$ is a Gaussian probability
density function (PDF) of mean $c_{k}^{(0)}$ and variance $(N_{0}+N_{I})$
(in accordance with the auxiliary channel model),
while the outer statistical average, with respect to $c_{k}^{(0)}$ and
$r_{k}^{(0)}$, is carried out according to the real channel model~\eqref{eq:sample_yk}~\cite{ArLoVoKaZe06}.
Eq.~\eqref{e:AIR} can be evaluated efficiently by means of a Monte
Carlo average~\cite{ArLoVoKaZe06}. From a system viewpoint, the
spectral efficiency, that is the amount of information transmitted
per second and per Hertz, is a more significant quality figure than
the information rate. Under the assumption of infinite transmission,
the ASE is defined as 
\begin{equation}
\eta=\frac{I_\mathrm{R}}{FT}\quad[
\textrm{b/s}/\textrm{Hz} ]\,.\label{e:spectral_efficiency}
\end{equation}

For a given constellation and shaping pulse, it is possible to find
the spacings $T$ and $F$ that provide the largest ASE. In general,
we could expect that the optimal spacings depend on the signal-to-noise
ratio (SNR). In fact, it is possible to show that, as the SNR increases,
not only does the ASE increase, but also the optimal values of the
spacings change. The properties of the function $\eta(T,F,E_{S}/N_{0})$
cannot be easily studied in closed form, but it is clear, by physical
arguments, that it is bounded, continuous in $T$ and $F$, and tends
to zero when $T,F\to0$ or $T,F\to\infty$. Hence, the function $\eta(T,F)$
has a maximum value---according to our findings, in most cases there
are no local maxima other than the global maximum. Formally, for a
given modulation format, shaping pulse, and value of~$E_{S}/N_{0}$,
the optimization problem consists of finding the maximum of $\eta(T,F,E_{S}/N_{0})$
varying $T$ and $F$. This problem can be solved by evaluating $\eta(T,F,E_{S}/N_{0})$
on a grid of values of $T$ and $F$ (coarse search), followed by
an interpolation of the obtained values (fine search).

A measure of the SNR more significant than $E_{s}/N_{0}$ is given
by $E_{b}/N_{0}$, being $E_{b}$ the mean energy per information
bit, for which $E_{s}=I(E_{s})E_{b}$ holds.
The optimization problem becomes
\begin{equation}
\eta_{\textrm{M}}(E_{b}/N_{0})=\max_{T,F>0}\eta(T,F,E_{b}/N_{0})\,.\label{optEb-1}
\end{equation}
In order to solve it for a given value of $E_{b}/N_{0}$, we employed
the following technique. The AIR is first evaluated for some values
of the couple $(T,F)$, and $E_{s}/N_{0}$. The two sets, including
their cardinalities, must be designed so as to ensure an accurate
sampling of the AIR, when the latter is interpreted as a function
of $T$, $F$, and $E_{s}/N_{0}$. For each couple $(T_{i},F_{j})$,
cubic spline interpolation can be used to obtain a continuous function
of $E_{s}/N_{0}$ (fine search), denoted as $I(T_{i},F_{j},E_{s}/N_{0})$.
Then, given a value of $E_{b}/N_{0}$ the following fixed-point problems
are solved in $E_{s}/N_{0}$ for different couples $(T_{i},F_{j})$,

\begin{equation}
\frac{E_{s}}{N_{0}}=I\left(T_{i},F_{j},\frac{E_{s}}{N_{0}}\right)\frac{E_{b}}{N_{0}}
\end{equation}
and the AIRs corresponding to the solutions are denoted by $I(T_{i},F_{j},E_{b}/N_{0})$.
Further improvements could be achieved by adding $N_{I}$ as variable
in eq. \eqref{optEb-1}. However, we have found by numerical results
that choosing $N_{I}$ as in \eqref{eq:N_i} is almost optimal. 

The spectral efficiency depends on the employed discrete-time filter.
Since the optimization of this filter with the aim of maximizing the
spectral efficiency is a hard task, we restricted our analysis to
the cases of a WF, that will be also considered in~\S\ref{sec:BCJR-receiver},
and of a minimum mean square error (MMSE) feedforward equalizer, possibly
fractionally spaced (FS) with at most 22 taps.

\subsection*{Single-User Trellis Processing\label{sec:BCJR-receiver}}

Improved, still achievable, lower bounds can be obtained by relaxing
the constraint on the adopted detection algorithm. In other words,
we can consider a more complex receiver able to cope with (a portion
of) the interference introduced by the adoption of the time-frequency
packing. The receiver considered in this section will not cope with
the interference due to the adjacent users---a single-user receiver
is still adopted.

For a general channel with finite intersymbol interference, an optimal
MAP symbol detector can be designed working on the samples at the
WF output as shown in Figure \ref{fig:BlockDiagramReceiver}(b). These
samples, denoted to as Forney observation model (see \S\ref{sec:obs_model}), can
still be expressed as in~\eqref{eq:sample_yk} with a proper expression
of coefficients $h(n,\ell,k)$. We assume to adopt the optimal receiver
for the following auxiliary channel:
\begin{equation}
r_{k}^{(0)}=\sqrt{E_{s}}\sum_{0\le n\le L}f_{n}c_{k-n}^{(0)}+\mathfrak{v}_{k}\label{e:auxiliary_bcjr}
\end{equation}
where $\{f_{n}\}_{n\geq0}$ are such that $f_{n}=h(n,0,k)$ and, as
mentioned, are independent of $k$, whereas the noise samples $\{\mathfrak{v}_{n}\}$,
that take into account the white noise and the residual interference,
are assumed independent and identically distributed zero-mean circularly
symmetric Gaussian random variables with variance $(N_{0}+N_{I})$,
with
\begin{equation}
N_{I}=\sum_{n>L}E_{s}|f_{n}|^{2}+\sum_{n}\sum_{\ell\neq0}E_{s}|h(n,\ell,k)|^{2}\,.\label{eq:N_i_bcjr}
\end{equation}
which is still independent of $k$. The corresponding MAP symbol detector
takes the form of the classical algorithm by Bahl, Cocke, Jelinek
and Raviv (BCJR)~\cite{BaCoJeRa74} working on a trellis whose state
takes into account $L$ interfering symbols only, according to a given
maximal allowable receiver complexity. The number of trellis states
is equal to $S=M^{L}$. 

Let us define $\bm{c}=[c_{0}^{(0)},c_{1}^{(0)},...,c_{N-1}^{(0)}]$ and
$\bm{r}=[r_{0}^{(0)},r_{1}^{(0)},...,r_{N-1}^{(0)}]$, $N$ being a proper
integer. The simulation-based method described in~\cite{ArLoVoKaZe06}
allows to evaluate the AIR for the mismatched receiver, i.e., 
\begin{eqnarray}
I_{\mathrm{R}} & = & \lim_{N\rightarrow+\infty}\frac{1}{N}I(\bm{c};\bm{r})\nonumber \\
 & = & \lim_{N\rightarrow+\infty}\frac{1}{N}E\left\{ \log_{2}\frac{q(\bm{r}|\bm{c})}{q(\bm{r})}\right\} \;\left[\frac{\mathrm{bit}}{\mathrm{ch.\, use}}\right]\,.\,\,\,\,\label{eq:AIR_2}
\end{eqnarray}
In \eqref{eq:AIR_2}, $q(\bm{r}|\bm{c})$ and $q(\bm{r})$
are PDF according to the auxiliary channel model, while the outer
statistical average is with respect to the input and output sequences
evaluated according to the actual channel model~\cite{ArLoVoKaZe06}.
Eq. \eqref{eq:AIR_2} can be evaluated recursively through the forward
recursion of the BCJR detection algorithm matched to the auxiliary
channel model~\cite{ArLoVoKaZe06}. Once the AIR has been computed,
the spectral efficiency can be derived and the optimal time and frequency
spacings optimized accordingly, as described in the previous section.
For channels with finite ISI, optimal MAP symbol
detection can be equivalently implemented by working directly on the
MF output~\cite{CoBa05b}, i.e., on the so-called Ungerboeck observation
model (see \S\ref{sec:obs_model}). The equivalence does not hold when reduced-complexity
detection is considered and interference from adjacent channels arises.
Since it is difficult to predict which is the most convenient observation
model, it is of interest to evaluate the ASE when both models are
employed and this can be done as described for the Forney model (see
also \cite{RuFe09} for details).

\subsection*{Multi-User Detection}

Although the assumption of a single-user auxiliary channel gives very
useful results, tighter lower bounds can be obtained by using a more
general auxiliary channel model. In fact, we can consider a receiver
for the central user (that with $\ell=0$) that, in addition to the
interference taken into account by the receivers in~\S\ref{sec:BCJR-receiver},
also takes into account the $J$ adjacent signals on each side as
well (multi-user receiver)\textemdash{}we again point out that this
approximation is exploited only by the receiver, while in the actual
channel the interference is generated as in \eqref{eq:rec_signal2}.
The exact MAP receiver for the multi-user auxiliary channel can be
easily derived and employed to find the ASE in the new scenario. The
benefit of employing the multi-user auxiliary channel model when evaluating
the ASE is two-fold: first, it allows to evaluate the performance
degradation due to the use of single-user receivers, despite the presence
of a strong adjacent channel interference, with respect to a more
involved multi-user receiver, which is more {\it matched}
to the real channel. Second, it gives a practical performance upper
bound when low-complexity approximate multi-user receivers, for example
based on linear equalization or interference cancellation, are employed
(as examples, those in \cite{CoFePi11} and references therein). Obviously,
in this case some (limited) degradation must be expected.

\section{Channel shortening detection for time packing}
In the previous section, we adopted mismatched detectors which
consider just a limited amount of interference. The interference
considered at the detector is a truncation of the actual interference (both in time and frequency).
Clearly, the truncation does not maximize the ASE, which is instead what time packing aims to.
A better approach than truncation, is the adoption of channel shortening (CS) detectors, as described in \S\ref{sec:cont_time_cs}.
Namely, for a given memory $L$ considered at detector, we set the ISI at detector and the front-end filter as
the ones which maximize the achievable spectral efficiency.
Moreover, as shown in \S\ref{sec:cont_time_cs}, the shaping pulse can be also optimized to improve the performance.

\clearpage
\section{Numerical results\label{sec:Simulation-Results}}
In this section, we report the optimized spectral efficiency $\eta_{\textrm{M}}$
as a function of $E_{b}/N_{0}$ for different modulation formats and
shaping pulses. The considered modulation formats are the quaternary
and octal PSK (QPSK and 8PSK).

\begin{figure}
	\begin{center}
		\includegraphics[width=0.8\columnwidth]{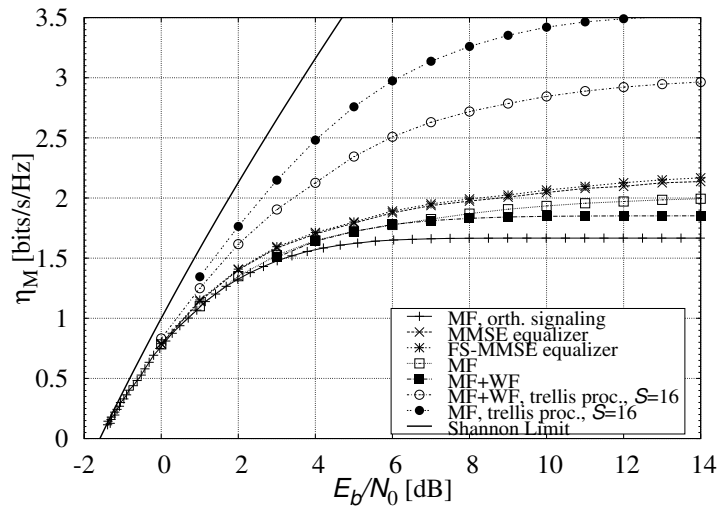}
		\caption{ASE for QPSK with Gray mapping and a RRC pulse having $\alpha=0.2$.\label{fig:ASEs-for--QPSK-RRCgray}}
	\end{center}
\end{figure}

\begin{figure}
	\begin{center}
		\includegraphics[width=0.8\columnwidth]{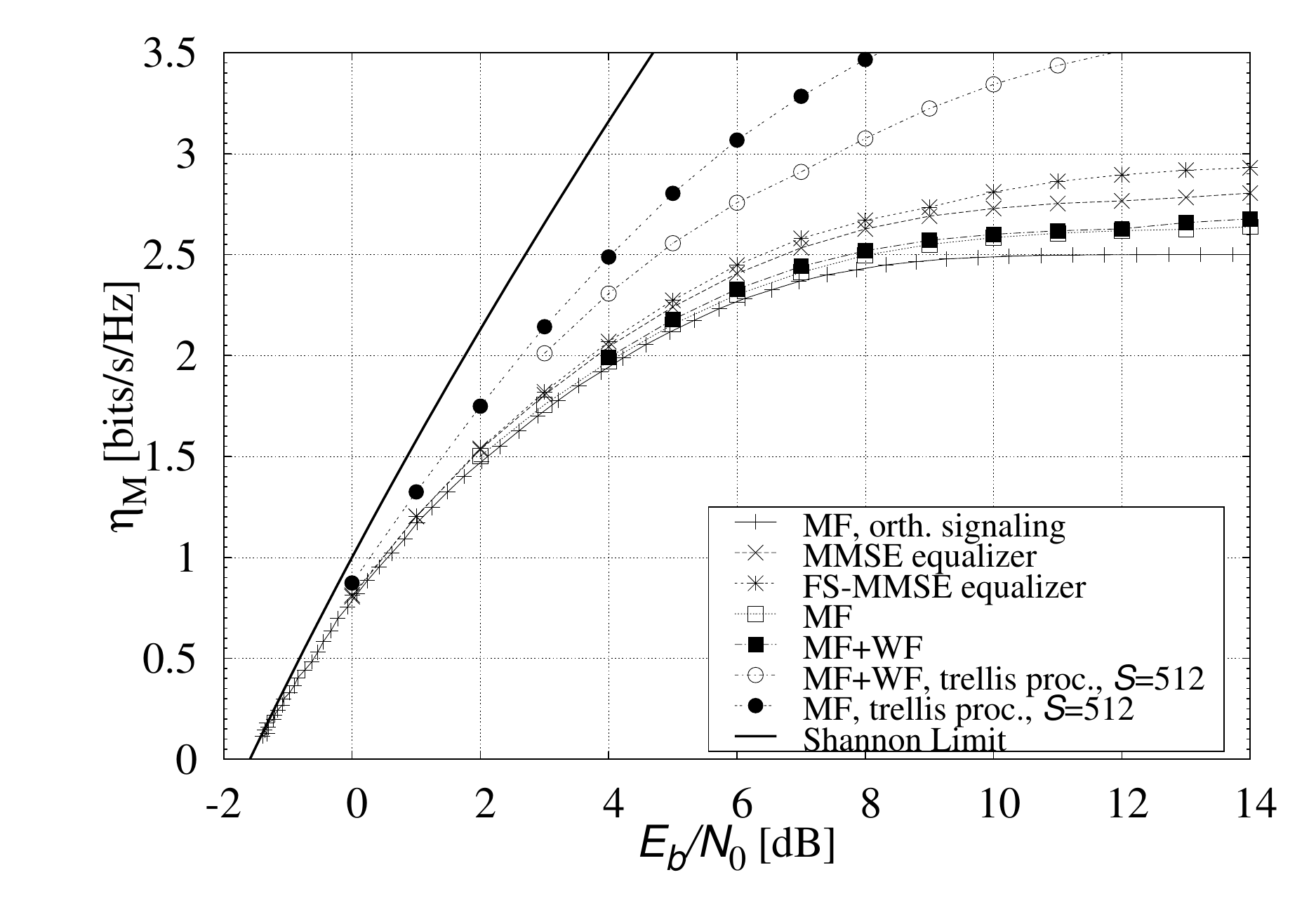}
		\caption{ASE for 8PSK with a RRC pulse having $\alpha=0.2$.\label{fig:ASEs-for-8PSK}}
	\end{center}
\end{figure}

\begin{figure}
	\begin{center}
		\includegraphics[width=0.8\columnwidth]{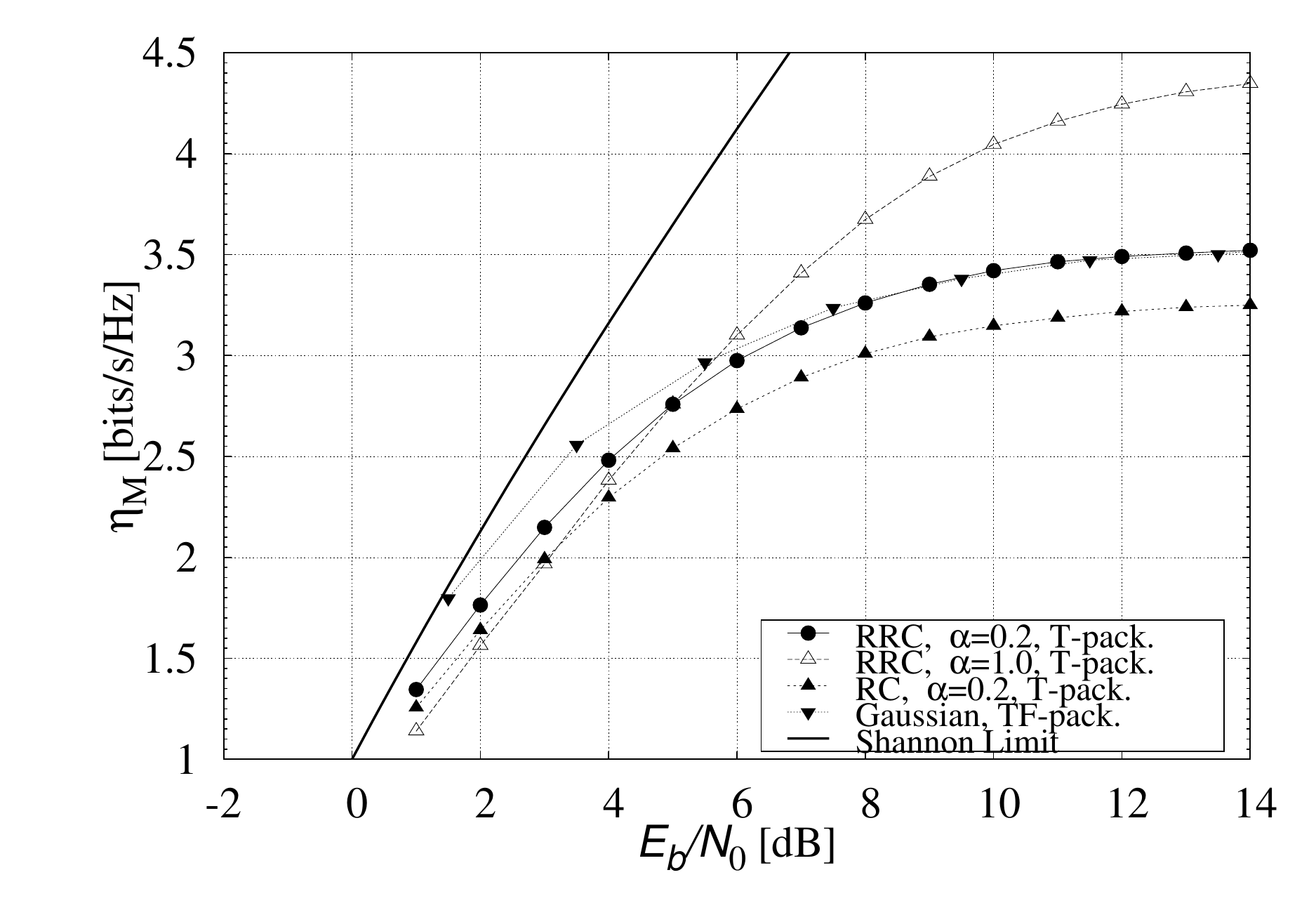}
		\caption{ASE for QPSK with Gray mapping by using different pulses. At the receiver, a MF front end and trellis processing with $S=16$ is considered.\label{fig:ASE-for-QPSKotherpulses}}
	\end{center}
\end{figure}

Fig.~\ref{fig:ASEs-for--QPSK-RRCgray} shows the optimized ASE in
case of time packing only for the QPSK modulation with a RRC pulse
with roll-off $\alpha=0.2$. Both symbol-by-symbol detection and trellis
processing (this latter taking into account $L=4$ interfering symbols)
are considered assuming Gray mapping. In this case, at the receiver
side we may use two identical and independent detectors, one working
on the in-phase and the other one on the quadrature component. This
is beneficial in case of adoption of a MAP symbol detector. In fact,
when $L$ interfering symbols are taken into account, we have two
detectors working on a trellis with $2^{L}$ states instead of a single
detector working on a trellis with $4^{L}$ states. Hence, for a given
complexity, a larger number of interferers can be taken into account.
The curve related to the absence of time packing (i.e., in case of
orthogonal signaling) and the Shannon Limit for AWGN~\cite{Sh48},
are also shown for comparison. It can be observed that the time-packing
technique allows to improve the spectral efficiency for each $E_{b}/N_{0}$
value with respect to the case of orthogonal signaling. Moreover it
can noticed that, in case of use of a symbol-by-symbol detector, the
FS-MMSE equalizer seems the best option whereas the Ungerboeck observation
model is more suited in case of trellis processing. Similar considerations
hold for the 8PSK modulation with a RRC pulse of $\alpha=0.2$. The
relevant results are shown in Fig.~\ref{fig:ASEs-for-8PSK}. 
Still considering QPSK with Gray mapping and trellis processing with
$S=16$, we evaluated the effect of different shaping pulses. In particular,
RRC and RC pulses with different roll-off factors have been considered
along with prolate spheroidal wave functions~\cite{SlPo61} and the Gaussian pulse. In these
two latter cases, frequency packing is also employed. Fig. \ref{fig:ASE-for-QPSKotherpulses}
shows the performance of some of the considered pulses. In particular,
RRC pulses with $\alpha$ equal to $0.2$ and $1.0$ outperform all
other pulses at low and high $E_{b}/N_{0}$ values, respectively.
In particular, an impressive asymptotic spectral efficiency of 4.3
bit/s/Hz is obtained with QPSK and $\alpha=1$.%
\footnote{This is due to the fact that the shaping pulse is smoother and so,
for a given value of $T$, the introduced interference is lower.}

\begin{figure}
	\begin{center}
		\includegraphics[width=0.8\columnwidth]{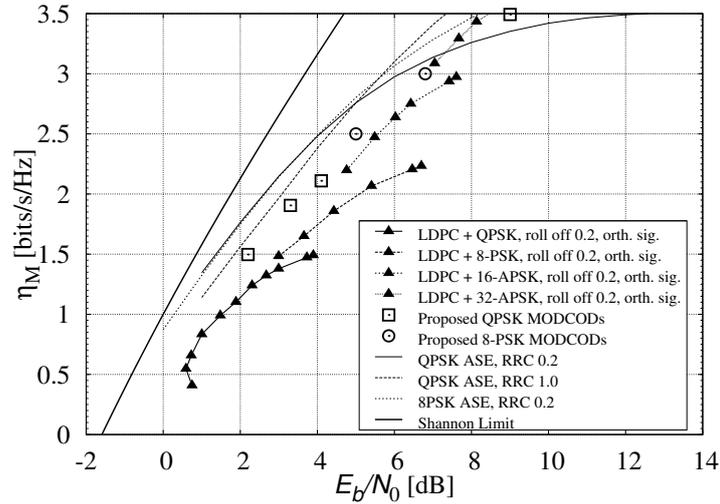}
		\caption{Designed MODCODs for QPSK and 8PSK with RRC pulses.\label{fig:Designed-MODCODs}}
	\end{center}
\end{figure}

What information theory promises can be approached by using proper
coding schemes. We considered MODCODs using the low-density parity-check
(LDPC) codes with length 64,800 bits of the DVB-S2 standard~\cite{DVB-S2-TR},
properly combined with QPSK and 8PSK modulations with time packing.
RRC pulses with $\alpha=0.2$ or $\alpha=1$ are considered. The corresponding
packet error rate (PER) have been computed by means of Monte Carlo
simulations and the results are reported in the spectral efficiency
plane in Fig. \ref{fig:Designed-MODCODs} using, as reference, an
MPEG PER of $10^{-4}$. In the same figure, the performance of the
MODCODs based on the same LDPC codes with orthogonal signaling and
employing QPSK, 8PSK, and the amplitude phase-shift keying (APSK)
modulation with 16 and 32 symbols (16- and 32APSK)~\cite{DVB-S2-TR},
are also shown for comparison. We can observe that we can reach, with
QPSK, values of spectral efficiency that, in case of orthogonal signaling,
cannot be reached even with 16APSK. 

\begin{figure}
	\begin{center}
		\includegraphics[width=0.8\columnwidth]{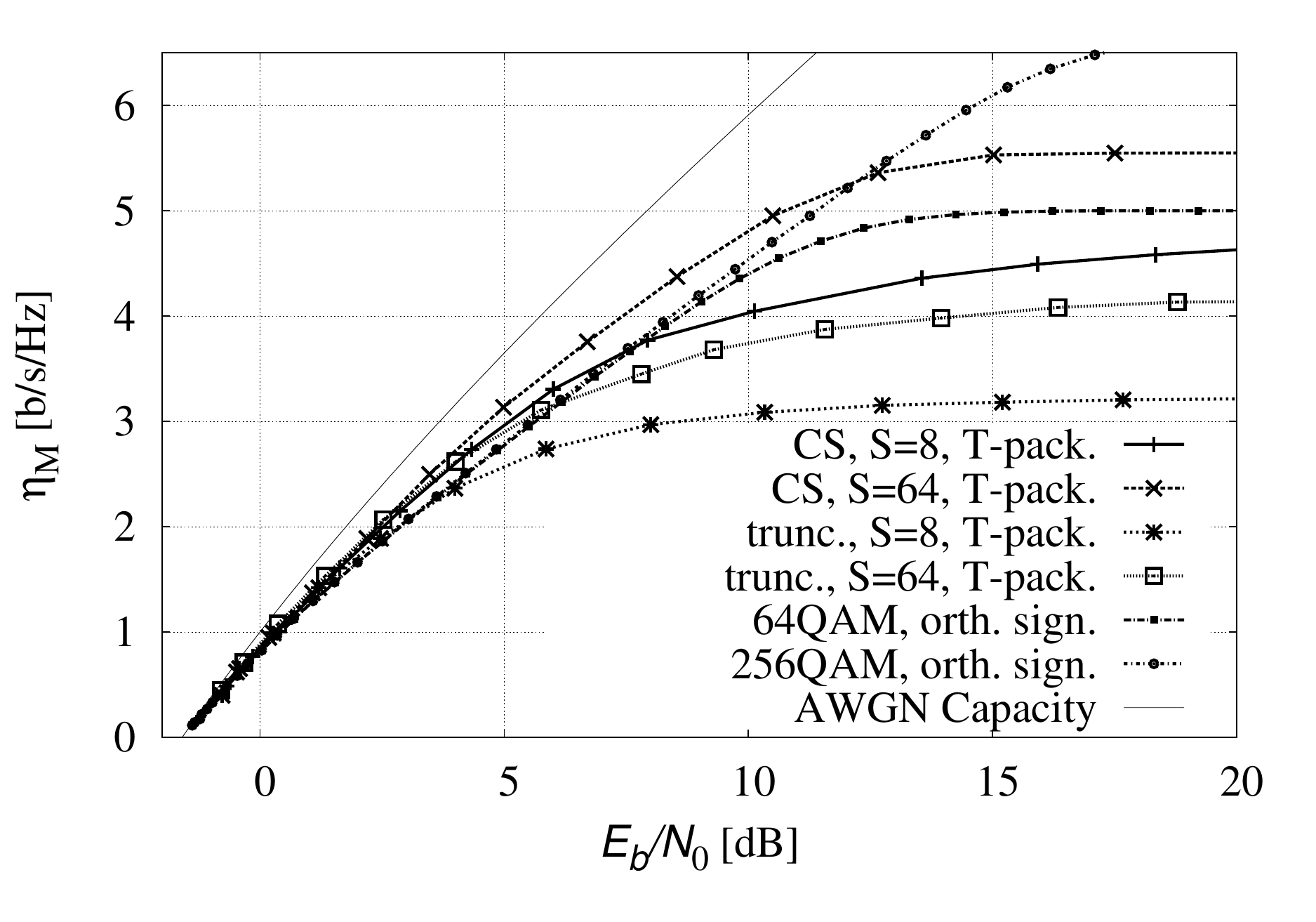}
		\caption{ASE for time packing and CS detection when the modulation is QPSK, with Gray mapping and RRC pulse $\alpha=0.2$.\label{fig:tpack_CS}}
	\end{center}
\end{figure}

The performance can be improved by adopting a CS detector.
Figure \ref{fig:tpack_CS} shows the optimized ASE for RRC and QPSK modulation with Gray mapping and $\alpha=0.2$
by adopting the time packing technique and single-user trellis processing.
Detection is performed by adopting two identical and independent detectors, one working
on the in-phase and the other one on the quadrature component. The figure shows the ASE obtained by single-user trellis processing in either case 
of CS or truncation of the ISI. The considered numbers of states are $S=8$ and $S=64$. In the case of ISI truncation, 
we optimized also the noise variance at detector, in order to achieve the best performance.
For comparison purpose, we also showed the ASE by quadrature amplitude (QAM) with cardinality 64 and 256
with orthogonal signaling.
It can be noticed that CS outperforms the truncation and exhibits an excellent ASE, which for some SNR values is even higher than the
one achieved by QAM modulations.

      \chapter{Detection for satellite channels}\label{ch:cs_sat}

\lettrine{S}{atellite} channels are affected by nonlinear distortions and by intersymbol interference (ISI).
The former originate from the presence
of a high power amplifier (HPA), whereas the latter is introduced
by the input and output multiplexing (IMUX and OMUX) filters placed before and after the HPA.
During the last decades, the nonlinear effects and the  channel memory have been coped with nonlinear compensation and data predistortion at the transmitter side~(see \cite{AnLoRe96} and references therein) or with advanced detection techniques~(see \cite{CoPi12} and references therein).


When the channel memory is too large to be taken into account at the
detector, these advanced detection techniques quickly become
unmanageable and low-complexity solutions are required. The
conceptually simplest solution is to let the detector work with a
truncated version of the channel response. However, as expected,
such a strategy  often yields poor performance unless the truncated part
of the channel response has negligible power.
Channel shortening, already described in Chapter~\ref{ch:CS}, 
can be an alternative.
This chapter generalizes the  analysis in \cite{RuPr12}
to  maximum-a-posteriori (MAP) detection for nonlinear satellite channels. 
In \S\ref{sec:system_model}, we briefly 
review the system model for the satellite channel and the underlying
detection algorithm assumed in this thesis. In \S\ref{sec:CS}, we extend the channel shortening technique and 
in \S\ref{sec:num_results} we assess its performance by numerical simulations.

\section{System model and considered detector}\label{sec:system_model}
\begin{figure}
 \begin{center}
	\includegraphics[width=0.99\columnwidth]{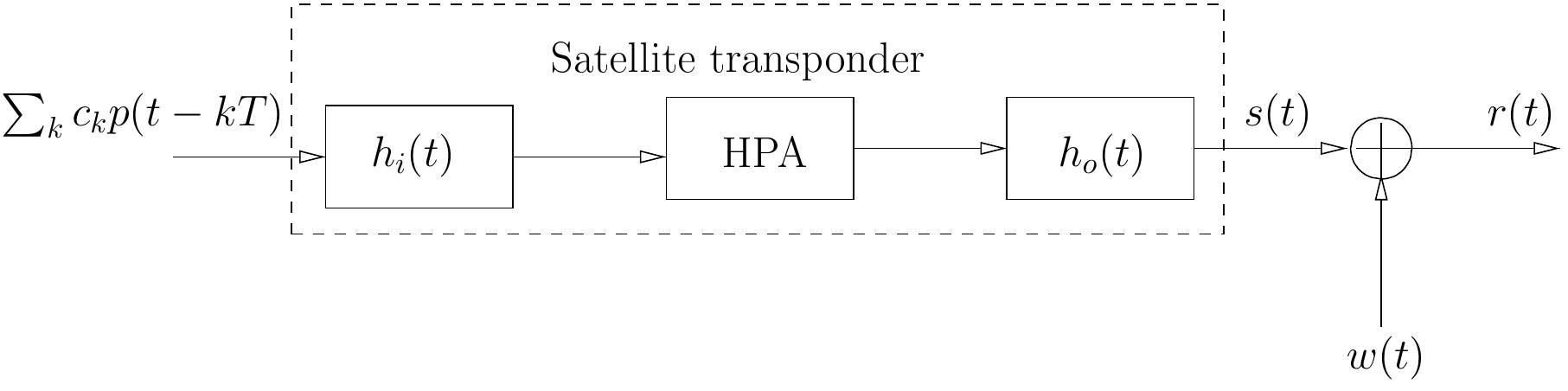}
	\caption{Block diagram of the satellite channel.} \label{fig:block_diagram}
 \end{center}
\end{figure}

We consider a linear modulation with shaping pulse $p(t)$, symbol time~$T$, and uniformly and identically distributed input symbols $\{c_k\}$ 
belonging to an $M$-ary 
constellation, properly normalized such that ${\rm E}\{|c_k|^2\}=1$. 
The nonlinear satellite channel, considering a single-channel-per-transponder scenario, is depicted in Figure~\ref{fig:block_diagram}. 
It includes an IMUX filter $h_i(t)$ which removes the adjacent channels,
a HPA, and an OMUX filter $h_o(t)$ aimed at reducing the spectral broadening
caused by the nonlinear amplifier.
Although the HPA is a nonlinear memoryless device, the
overall  system has memory due to the presence of IMUX and OMUX filters.
The received signal is further corrupted by additive white Gaussian noise whose low-pass equivalent  $w(t)$ 
has power spectral density $N_{0}$. The 
complex baseband representation of the received signal has thus the
following expression
\begin{equation}
 r(t)=s(t)+w(t) \, ,
\end{equation}
where $s(t)$ is the signal at the output of the 
OMUX filter.


In \cite{CoPi12}, it is shown that a suitable approximate model for
the signal $s(t)$ is based on the following $v$th-order (with $v$
being any odd integer) \textit{simplified} Volterra-series expansion
\begin{equation}
 s(t)\simeq \!\!\sum_{k} \sum_{i=0}^{N_V-1} \!c_k \left[ |c_k|^{2i} h^{(2i+1)}\!(t-kT) \right], \label{eq:volterra_apprx} 
\end{equation}
where $N_V=(v+1)/2$, and $h^{(2i+1)}\!(t)$ are complex waveforms given by  linear combinations of the the original $N_V$ Volterra kernels. This 
simplified Volterra-series expansion is obtained from the classical one by neglecting some 
suitable terms. For further details, the reader can refer to~\cite{CoPi12}.
We point out that the approximation (\ref{eq:volterra_apprx}) is used
only for the receiver design and not for generating the
received signal $r(t)$. 

It is easy to show that MAP symbol detection based on this simplified model can be performed through a bank of filters followed by a conventional BCJR detector~\cite{BaCoJeRa74} with proper branch metrics and working on a trellis whose number of states exponentially depends on the channel memory.
When the actual channel memory is large, we have to resort to
complexity reduction techniques. A possible approach is the use of reduced-state techniques (e.g., see~\cite{FeBaCo07}) or 
the use of the
graph-based technique described in~\cite{CoPi12} whose complexity
linearly depends on the channel memory. However, to obtain a further complexity reduction, all these techniques can be combined with the CS technique described in~\cite{RuPr12} properly extended to the channel at hand. 

We will separately consider the cases of phase-shift keying (PSK) modulations and amplitude/phase shift keying (APSK) modulations typically employed in satellite transmissions.

\subsection*{PSK modulations}
It can be seen that the condition $|c_k|^2=1$ implies that the signal (\ref{eq:volterra_apprx}) simplifies to a linear modulation 
\begin{equation}
	s(t) \simeq \sum_k c_k \bar{h}(t-kT)
\end{equation}
where
$\bar{h}(t)=\sum_{i=0}^{N_V-1} h^{(2i+1)}(t)$~\cite{CoPi12}. In this case, detection can be perfomed using the samples $\{r_k\}$ at
the output of a filter matched to $\bar{h}(t)$ as described in \S\ref{sec:opt_det}, and the application
of CS can be carried out as described in \S\ref{sec:cont_time_cs} for linear channels.

\subsection*{APSK modulations}
The samples at the output of a bank of filters matched
to the pulses $h^{(2i+1)}(t)$, for \mbox{$i=0,...,N_V-1$} form a set of sufficient statistics for detection.
Namely, considering an $v$-th order expansion, we have $N_V$ matched filters whose output, sampled at discrete time $k$ 
can be collected in a $N_V\times 1$ vector that can be expressed as
\begin{equation}
  \bm{r}_k = \sum_{i} \bm{G}_{i}\bm{c}_{k-i} + \bm{n}_k \,, 
\end{equation}
where $\bm{c}_k=\left[c_k,~ c_k|c_k|^2,...,c_k|c_k|^{v-1}\right]^T$, 
\begin{equation}
  \bm{G}_{i} = \left(    
        \begin{array}{cccc}
         g_i^{(1,1)} & g_i^{(1,3)} & \cdots & g_i^{(1,v)}\\
         g_{-i}^{(1,3)*} & g_i^{(3,3)} & \cdots & g_i^{(3,v)} \\
         \vdots & & \ddots & \vdots \\
         g_{-i}^{(1,v)*} & g_{-i}^{(3,v)*} & \cdots & g_i^{(v,v)}
        \end{array}
  \right) \, ,
\end{equation}
having defined $g^{(m,l)}_i=\int_{-\infty}^{\infty}
h^{(n)}(t)h^{(m)*}(t-lT)\mathrm{d}t $, and $\bm{n}_k$ is a Gaussian vector with  
\begin{equation}
{\rm E}\{\bm{n}_{k+i}\bm{n}_k^\dagger\}=N_0\bm{G}_{i}\,.
\end{equation}
Vectors $\{\bm{r}_k\}$ can be collected into a single vector
\begin{equation}
  \mathbf{r} =  \mathbf{G} \mathbf{c}  + \mathbf{n}\,,  \label{eq:block_channel}
\end{equation}
where $\mathbf{G}$ is a block Toeplitz matrix constructed from the matrices~\{$\bm{G}_i$\},
whereas $\mathbf{c}$ and $\mathbf{n}$ are block vectors from \{$\bm{c}_k$\} and \{$\mathbf{n}_k$\}.
The channel is fully
described through its conditional probability density function of the
output given the input symbols, which reads
\begin{equation}
p(\mathbf{r}|\mathbf{c}) \propto \exp\left( \frac{ 2\Re(\mathbf{c}^\dagger \mathbf{r})-\mathbf{c}^\dagger\mathbf{G}\mathbf{c}}{ N_0} \right)  \,. \label{eq:channel_law}
\end{equation}

\begin{figure}
  \includegraphics[width=0.99\columnwidth]{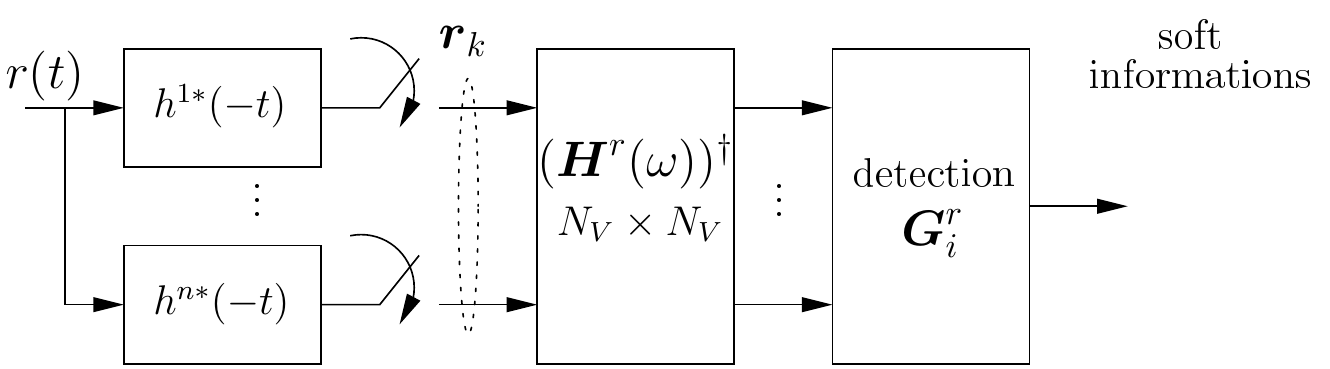}
  \caption{Block diagram of the suboptimal receiver for the nonlinear satellite channel.}\label{fig:receiver_cs}
\end{figure}

According to the CS approach, a \mbox{low-complexity} detector works on a mismatched channel law~\cite{RuPr12}
\begin{equation}
  q(\mathbf{r}|\mathbf{c}) \propto \exp\left( 2\Re(\mathbf{c}^\dagger (\mathbf{H}^r)^\dagger \mathbf{r}) - \mathbf{c}^\dagger\mathbf{G}^r\mathbf{c} \right) \, , \label{eq:mis_channel_law}
\end{equation}
where $\mathbf{H}^r$,$\mathbf{G}^r$ are block Toeplitz matrices constructed from the sequences $\{\bm{H}_i^r\}$ and $\{\bm{G}_i^r\}$, 
respectively, being $\{\bm{H}_i^r\}$  
the channel shortener operating on $\mathbf{r}$, and $\{\bm{G}_i^r\}$ the target response, to be properly designed.
Without loss of generality we absorb the noise variance $N_0$ into the two matrices in (\ref{eq:mis_channel_law}). 
In order to reduce the detection complexity, we constrain $\{\bm{G}_i^r\}$ to
\begin{equation}
 \bm{G}^r_i={\bf 0} ~~~ |i|>L \label{eq:G_constraint_apsk}
\end{equation}
which implies that the memory after CS is $L$ instead
of the true memory of the channel. 
The resulting receiver is suboptimal since it assumes (\ref{eq:mis_channel_law}) rather than the actual law (\ref{eq:channel_law}), and is depicted in Figure \ref{fig:receiver_cs}.

\section{Channel shortening}\label{sec:CS}
The achievable information rate (AIR) $I_{R}$ of a mismatched detector that works with (\ref{eq:mis_channel_law}) is given by
\begin{eqnarray}
    I_{\mathrm{R}} & = & \mathfrak{h}(\mathbf{r}) - \mathfrak{h}(\mathbf{r}|\mathbf{c}) \\
     & = & \lim_{N\rightarrow \infty} {1 \over N}  \mathrm{E}\left\{  \log_2{ q(\mathbf{r}|\mathbf{c}) \over q(\mathbf{r})} \right\}  \, \mathrm{[bit/ch. use]} \label{eq:AIR_sat}
\end{eqnarray}
where $N$ is the number of transmitted symbols and
the average is carried out w.r.t. $\mathbf{r}$ and $\mathbf{c}$, according to the actual channel \mbox{(see \cite{ArLoVoKaZe06}}).

The CS technique finds the optimal $\mathbf{H}^r$,$\mathbf{G}^r$ solving the following optimization problem
\begin{equation}
  \arg \max_{\mathbf{H}^r,\mathbf{G}^r} I_{\mathrm{R}} \label{eq:opt_problem_sat}
\end{equation}
under the constraints specified in (\ref{eq:G_constraint_apsk}).

Problem (\ref{eq:opt_problem_sat}) for a discrete alphabet is a complicated task. However
it can be solved in closed form under the assumption that $\mathbf{c}$ is composed of Gaussian random variables.
Although this assumption is not even approximately true, since
the actual symbols are functions of each other, we will show 
in the simulation results that a very good performance is still achieved.

Defining the correlation matrix $\bm{V}=\mathrm{E}\{ \bm{c}_k\bm{c}_k ^\dagger \}$, the optimal matrix-valued front-end filter $\{\bm{H}_i^r\}$
and target response
$\{\bm{G}_i^r\}$ are obtained in closed form through the following steps:
\begin{itemize}
  \item Compute the DTFT matrix $\bm{G}(\omega)$ of {$\bm{G}_i$} and use the spectral decomposition to find $\bm{L}(\omega)$, i.e., decompose $\bm{G}(\omega)=\bm{L}^\dagger(\omega)\bm{L}(\omega)$. Compute
        \begin{align}
         \bm{B}(\omega) & = N_0\bm{V}\bm{L}^\dagger(\omega)  \nonumber \\
           & \cdot \left[\bm{L}(\omega)\bm{V}\bm{L}^\dagger(\omega)+ N_0\bm{I} \right]^{-1}(\bm{L}^{\dagger}(\omega))^{-1} \,
        \end{align}
        where $\bm{I}$ is the identity matrix. The anti trasform yields the matrix sequence \{$\bm{B}_i$\}
        having size $N_V \times N_V$.
  \item Find 
   \begin{equation}
      \mathbfcal{C}= \bm{B}_0 - \mathbf{\underline{B}} \mathbf{B}^{-1}\mathbf{\underline{B}}^\dagger 
   \end{equation}
  where we defined the block matrix $\mathbf{\underline{B}}=[\bm{B}_1,...,\bm{B}_L]$ with size\footnote{ Here,
  the size for block matrix means the number of scalar entries, as well as for standard matrix.
  This notation is different from the one adopted in \cite{CoMoRu12}.}
  $N_V\times N_VL$ and 
  the block Toeplitz $\mathbf{B}$ with size $N_VL \times N_VL$ constructed on \{$\bm{B}_i$\}.
  \item Define the sequence \{$\bm{U}_k$\} where $\bm{U}_0$ is the Cholesky decomposition of $\mathbfcal{C}$, namely 
        $\mathbfcal{C}=\bm{U}_0^\dagger\bm{U}_0$, and $\bm{U}_i$ for $1 \leq i \leq L$ is
        the $(1,i)$ matrix entry of \mbox{$\mathbf{\underline{U}}=-\bm{U}_0\mathbf{\underline{B}}\mathbf{B}^{-1}$}.
  \item Set 
    \begin{equation}
      \bm{G}^r_i= \sum_{k=\max(0,i)}^{\min(L,L+i)} \bm{U}_{k-i}^\dagger \bm{U}_{k} - \bm{V} \delta_i
    \end{equation}
    where 
    $\delta_i$ is the Kronecker delta.
  \item The optimal front-end filter is given by
    \begin{align}
        &(\bm{H}^r(\omega))^\dagger= \left( \bm{G}^r(\omega)+\bm{V}^{-1}\right) \nonumber \\
         &   \cdot  \bm{V} \bm{L}^{\dagger}(\omega)\left[\bm{L}(\omega)\bm{V}\bm{L}^\dagger(\omega)+ N_0\bm{I} \right]^{-1} (\bm{L}^{\dagger}(\omega))^{-1}  \,.  
    \end{align}
\end{itemize}
The algorithm is carried out by observing that
the channel equation (\ref{eq:block_channel}), under the Gaussian assumption,
is the Ungerboeck observation
model of the MIMO-ISI channel (\ref{eq:forney_block_matrix}).
Thus the proof is the one showed in Appendix \ref{ch:app_cs} for MIMO-ISI channels.

We point out that by analogy to \cite{RuPr12} for linear channels, when $L=0$ the optimal channel shortener 
is a special case of the MMSE filter of \cite{BeBi83} applied to (\ref{eq:volterra_apprx}). 

\section{Numerical results}\label{sec:num_results}
We consider 8PSK and 16APSK modulations. The shaping pulse $p(t)$
has a root-raised-cosine (RRC) spectrum with roll-off 0.05. The IMUX and OMUX filters
have frequency characteristics specified in \cite{DVB-S2-TR} with a
3dB bandwidth of $0.94/T$ and  $0.85/T$ respectively. 
The nonlinear transfer characteristic is the Saleh model \cite{Sa81} with parameters $\alpha_a=2.1322$, $\alpha_\phi=1.7054$, $\beta_a=1.0746$,
and $\beta_\phi=1.5072$.
A 5th-order Volterra expansion is considered at the receiver.
We report all results as functions of the ratio between the normalized power at the
saturation $P_{\mathrm{sat}}$ and the noise power spectral density $N_0$.

\begin{figure}
	\begin{center}
		\includegraphics[width=0.75\columnwidth]{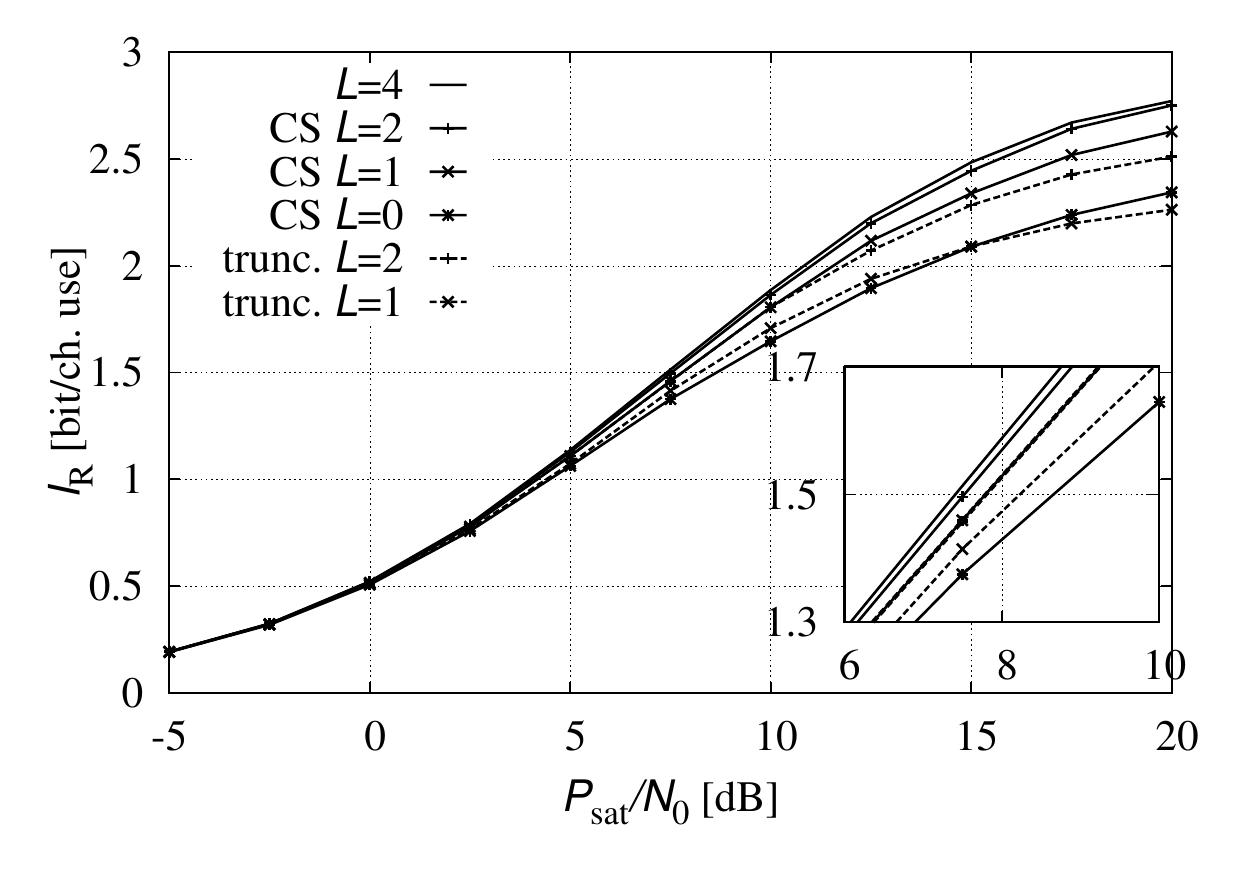}
		\caption{AIR for 8PSK modulation on the nonlinear satellite channel with IBO=0 dB.}\label{fig:ase_8psk} 
	\end{center}
\end{figure}
\begin{figure}
	\begin{center}
		\includegraphics[width=0.75\columnwidth]{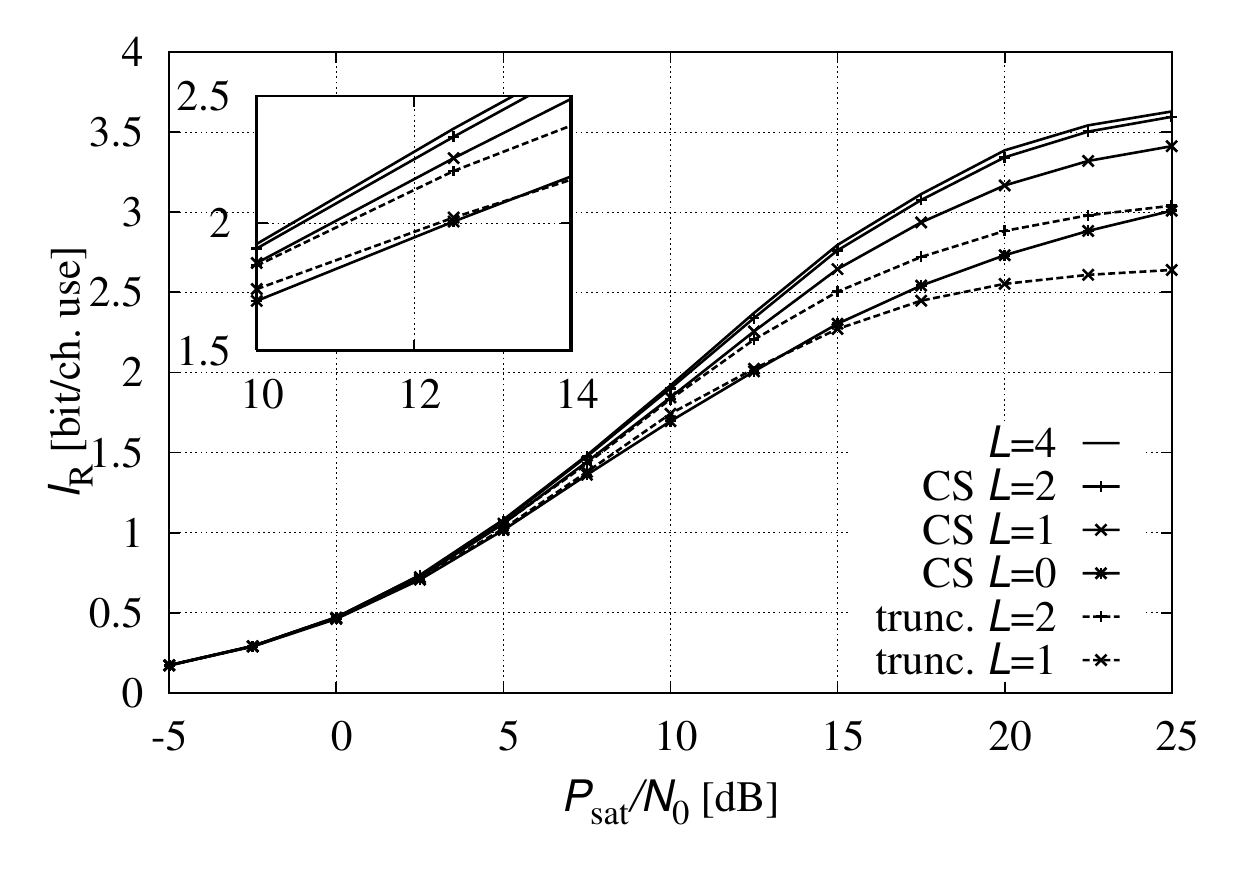}
		\caption{AIR for 16APSK modulation on the nonlinear satellite channel with IBO=3 dB.}\label{fig:ase_16apsk} 
	\end{center}
\end{figure}

The AIR in eq. (\ref{eq:AIR_sat}) can be computed using the Monte Carlo method described in \cite{ArLoVoKaZe06}. 
Figure \ref{fig:ase_8psk} shows the AIR values when CS is employed in
combination with a 8PSK modulation, and an input back-off (IBO) equal to zero.
Results are shown for different values of the detector memory $L$ and an optimization
of the noise variance at the receiver has been carried out to further improve the approximate model.
For comparison, we show also the AIR values when a simple truncation of the ISI
at the detector is adopted. 
The detector with $L=4$ can be considered as effective as a full complexity one,
since most of the ISI is taken into account\footnote{The pulse with RRC spectrum gives an infinite memory
of the channel. However, based on investigations beyond those presented in this thesis, we may assume that $L=4$ is almost optimal.}.
It can be seen that CS has higher AIR
than a simple truncation of the  ISI response, even though
it is designed for a vector ${\bf c}$ with Gaussian components.
CS with memory $L=2$ gives only a minimal performance degradation for all $P_{\mathrm{sat}}/N_0$ values.
Similar conclusions hold for the 16APSK, depicted in Figure \ref{fig:ase_16apsk}. We found similar CS gains also with other modulations (QPSK and 32APSK) and other transponder characteristics (e.g., the HPA
in \cite{DVB-S2-TR}).

\begin{figure}
	\begin{center}
		\includegraphics[width=0.75\columnwidth]{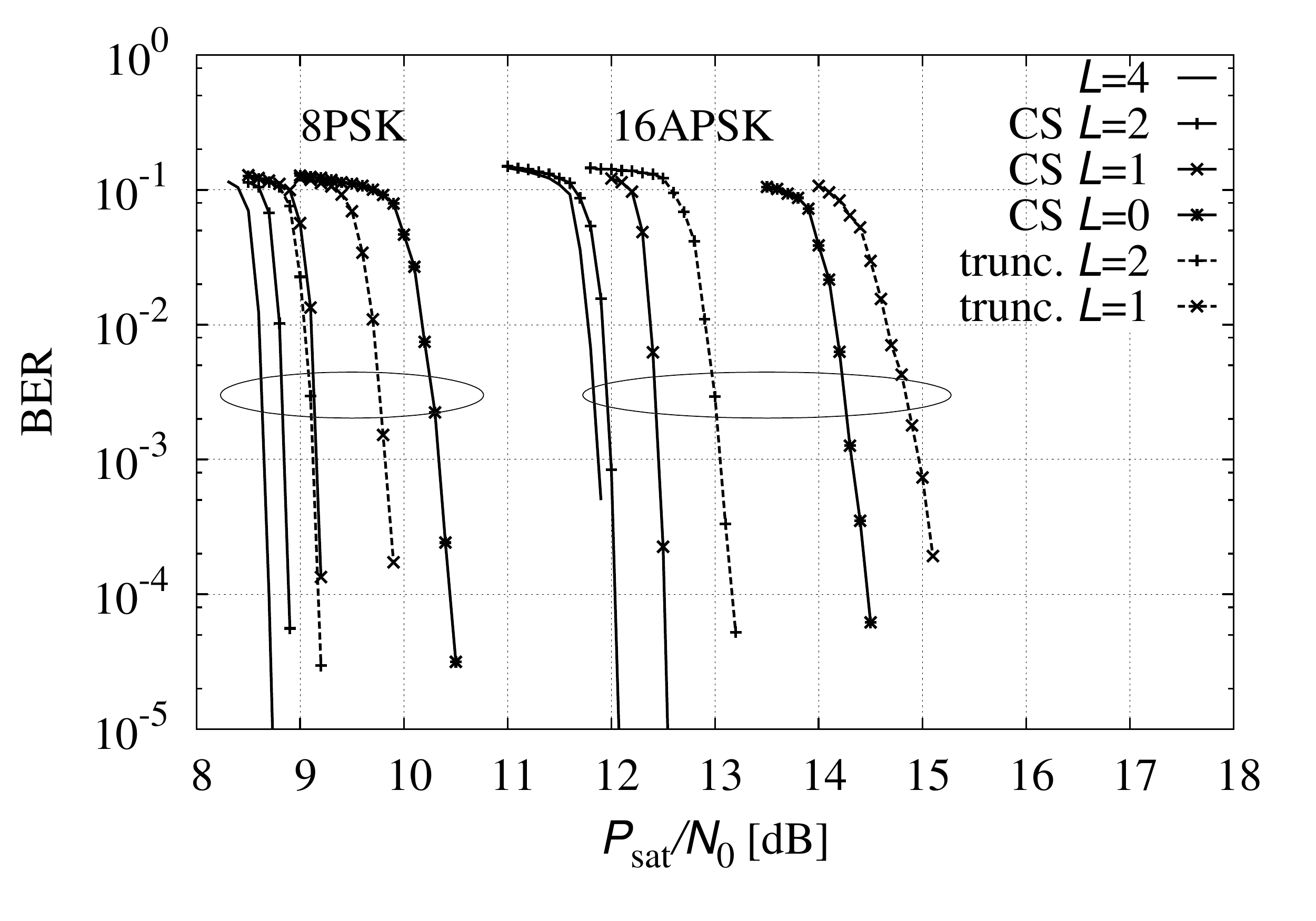}
		\caption{BER for 8PSK and 16APSK modulation, with the DVB-S2 LDPC code having rate 1/2.}\label{fig:ber_8psk_16apsk} 
	\end{center}
\end{figure}

The AIRs  can be approached in practice with proper modulation and coding (MODCODs) formats. 
In Figure~\ref{fig:ber_8psk_16apsk} we report   the bit error rate (BER) of some MODCODs based on the DVB-S2 low-density parity-check code (LDPC) with rate 1/2.
We performed iterative detection and decoding with a maximum of 50 global iterations. 
We note that the MODCODs performance
reflects the AIRs well.

\clearpage
\section{Conclusions}
We generalized the CS technique to the case of nonlinear satellite channels. We showed that when the memory $L$
is lower than the channel memory, an optimization of the mismatched channel law at the detector 
yields significantly better performance than a truncation of the channel impulse response. 
      \chapter{Spectrally efficient communications over the satellite channel}\label{ch:spec_tpack}

\lettrine{I}{n} this chapter, we apply the time-frequency packing (TF packing) technique to nonlinear satellite channels. In particular, we design highly efficient schemes by choosing 
the time and frequency spacings which give the maximum value of SE. 
We assume a realistic satellite channel where nonlinear distortions originate from the presence of a high-power amplifier (HPA).
The considered system is also affected by intersymbol interference (ISI), due to the presence of input and output multiplexing (IMUX and OMUX) filters placed before and after the HPA and intentionally introduced by the adoption of the time packing technique 
as well. We limit our investigation to systems in single-carrier-per-transponder operation (i.e., each transponder is devoted to the amplification of the signal coming from only one user\footnote{However, the actual 
transponders can be hosted on two or more co-located satellites.}). Although interchannel interference (ICI) is also present due to frequency packing of signals coming from different transponders, a single-user detector is considered at the receiver. 
We consider two different approaches 
to detection for nonlinear channels,
namely the use of a detector taking into account the nonlinear effects and a more 
traditional scheme based on predistortion and memoryless detection.
In the case of predistortion, we consider the dynamic data predistortion technique described in~\cite{KaSa91,CaDeGi04}, whereas in the case of advanced detection we 
consider the receiver described in the previous chapter employing the channel shortening technique
(CS) for nonlinear satellite channels. It should be noted that we apply the TF
packing technique to nonlinear satellite channels for which, usually,  even by using RRC pulses there is still ISI at the receiver.

As mentioned, with respect to Chapter~\ref{chap:time_pack}, we here consider nonlinear satellite channels instead of linear ones. 
Our aim here is to show the benefits that can be obtained by employing time-frequency packing jointly with a channel shortening 
receiver.


The proposed TF packing technique promises to provide increased SEs at least for low-order modulation formats. In fact, when dense constellations with shaping are employed, 
we fall in a scenario similar to that of the Gaussian channel with Gaussian inputs for which orthogonal signaling with no excess bandwidth (rectangular shaping pulses) is optimal (although this is mainly true for the linear channel 
and not in the presence of a nonlinear HPA, 
since shaping increases the peak-to-average power ratio). Improving the achievable SE without increasing the constellation order can be considerably convenient since it is well known that low-order constellations are more robust 
to channel impairments such as time-varying phase noise and non-linearities. It is expected that the use of low-order modulations in conjunction with TF packing provides similar advantages in terms of robustness against channel 
impairments.

The proposed approach to improve the SE is very general and the  case of satellite systems for broadband and broadcasting applications must be thus considered just as an example to illustrate the benefits that can be obtained 
through the application of the TF packing paradigm coupled with advanced receiver processing.

The remainder of this chapter is organized as follows. In \S\ref{s:system_model}, we introduce the system model. 
The framework that we use to evaluate the SE of satellite systems is detailed in \S\ref{s:spectral efficiency}, whereas different approaches to the detection 
for the considered channel are described in \S\ref{s:auxiliary channels}. Numerical results are reported in \S\ref{s:results}, where we show how the 
proposed technique can improve the SE of DVB-S2 systems. Finally, conclusions are drawn in \S\ref{s:conclusions}.

\section{System Model}~\label{s:system_model}
We consider the forward link of a transparent satellite system, where synchronous users employ the same linear modulation format, shaping pulse $p(t)$, and symbol interval 
(or time spacing) $T$, and access the channel according to a frequency division multiplexing scheme. 
The transmitted signal in the uplink can be expressed as 
\begin{equation}\label{e:transmitted signal}
x(t)=\sum_\ell \sum_k c^{(\ell)}_{k} p(t-kT)e^{j2\pi \ell F_u t}\, ,
\end{equation}
where $c^{(\ell)}_{k}$ is the symbol transmitted by user $\ell$ during the $k$-th symbol interval, and $F_u$ is the frequency spacing between adjacent channels.\footnote{In this scenario, we will use the terms,
``channels'', ``users'', and ``carriers'' interchangeably.}
The transmitted symbols belong to a given zero-mean $M$-ary complex constellation. Notice that, in order to leave out border effects, the summations 
in~\eqref{e:transmitted signal} extend from $-\infty$
to $+\infty$, namely an infinite number of time epochs and carriers
are considered. 
In \mbox{DVB-S2} standard the base pulse $p(t)$ is an RRC-shaped pulse with roll-off factor $\alpha$ (equal to 0.2, 0.3, or 0.35 depending on the service requirements). We denote by $W$ the bandwidth of pulse $p(t)$. In case 
of pulses employed in DVB-S2, it is $W=(1+\alpha)/T$ since orthogonal signaling is considered. When time packing is adopted, $W$ becomes a further degree of freedom, as described later.

As commonly assumed for broadband and broadcasting systems, on the feeder uplink (between the gateway and the satellite) the impact of thermal noise can be neglected due to a high transmit signal strength. Hence, in our analysis, 
we have considered a noiseless feeder uplink.
Although the TF packing can be applied to other and more general scenarios,  we consider here a single-carrier-per-transponder scenario, where different carriers undergo independent amplification by different transponders on board 
of satellite, each of which works with 
a single carrier occupying its entire bandwidth. This case is
particularly relevant for digital broadcasting services since it allows a more efficient use of on-board resources (in particular the HPA can work closer to saturation).\footnote{The multiple-carriers-per-transponder 
scenario is conceptually similar to that considered in this chapter, the only difference being the fact that more adjacent carriers are amplified by the same HPA. Thus, the effects of nonlinear ICI (intermodulation distortion) 
become more relevant and proper multicarrier detection or predistortion algorithms could be also considered~\cite{BeSe10,Be11,PiShZeRoGrZiGrHeCi12}.}  
The transponder model for user $\ell$ is composed of an IMUX filter which selects the $\ell$-th carrier, an HPA, 
and an OMUX filter which reduces the out-of-band power due to the spectral regrowth after nonlinear amplification~\cite{DVB-S2-TR}.
The HPA is a nonlinear memoryless device defined through its AM/AM and AM/PM characteristics, describing the amplitude and phase distortions caused on the signal at 
its input. 

\begin{figure}
	\begin{center}
		\includegraphics[width=88mm]{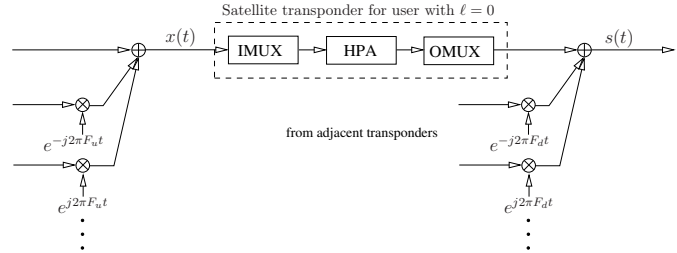} 
		\caption{System model.{\label{f:system_model}}} 
	\end{center}
\end{figure}

The outputs of different transponders are multiplexed again in the downlink to form the signal $s(t)$, and we assume that the adjacent users have a frequency 
separation of $F_d$, usually equal to that in the uplink. Fig.~\ref{f:system_model} shows the system model highlighting the satellite transponder for user with $\ell=0$.
The useful signal at the user terminal is still the sum of independent contributions, one for each transponder (although these contributions are no more, rigorously, 
linearly modulated due to the nonlinear transformation of the on-board HPA).
The received signal is also corrupted by the downlink AWGN, whose low-pass equivalent  $w(t)$ 
has power spectral density (PSD) $N_{0}$. The low-pass equivalent of the received signal has thus expression
\begin{equation}
 r(t)=s(t)+w(t) \, .
\end{equation}
We remark that, in the simulation results, this system has been simulated with realistic assumptions. In other words, the satellite receives the entire signal (\ref{e:transmitted signal}). 
Each carrier is then selected by the IMUX filter of its own transponder,\footnote{Being the IMUX a non ideal filter and due to a possible overlap among different carriers, 
this filtering will not be perfect and ICI will occur.} amplified by its own HPA, filtered again by the OMUX filter, and then the signals at the output of all transponders are multiplexed again on air.

We evaluate the ultimate performance limits of this communication system when single-user detection is employed at the receiver side. 
The proposed technique consists of allowing interference in time and/or frequency by reducing the values of $T$, $F_u$, and $F_d$, (partially) coping with it at the 
receiver, in order to increase the SE. 
In other words, $T$, $F_u$, and $F_d$ are chosen as the values that give the maximum value of the SE. 
These values depend on the employed detector---the larger the interference that the receiver can cope with, the larger the SE and the lower the values of time and 
frequency spacings.
Notice that, since we are considering single-user detection, the receiver is not able to deal with the interference due to the overlap of different channels.
In this case, the optimization of the frequency spacings is actually an optimization of the frequency guard bands generally introduced in satellite systems to avoid 
the nonlinear cross-talk, since a single-user receiver can tolerate only a very small amount of ICI.

The considered nonlinear satellite channel reduces, as a particular case, to the linear channel, provided that the HPA is driven far from saturation. Hence, all the considerations in this chapter can be straightforwardly 
extended to the linear channel case. A few results can be found on Chapter~\ref{chap:time_pack}.

\section{Optimization of the spectral efficiency}\label{s:spectral efficiency}
We describe the framework used to evaluate the ultimate performance limits of the considered satellite system and to perform the optimization of the time and 
frequency spacings. 
To simplify the analysis, we will assume $F_u=F_d=F$. 
We perform this investigation by constraining the complexity of the employed receiver. In particular, as mentioned, we assume that a single-user detector is used. 
For this reason, without loss of generality we only consider the detection of symbols $\bm{c}^{(0)}=\{c^{(0)}_k\}$ of user with $\ell=0$.\footnote{Assuming a system with 
an infinite number of users, the results do not depend on a specific user.} In addition, we also consider low-complexity receivers taking into account only a portion 
of the actual channel memory. Under these constraints, we compute the IR, i.e., the average mutual information when the channel inputs are 
i.u.d. random variables belonging to a given constellation. Provided that a proper \textit{auxiliary channel} can be defined 
for which the adopted low-complexity receiver is optimal, the computed IR represents an achievable lower bound of the IR of the actual channel, according to 
mismatched detection~\cite{MeKaLaSh94}.

Denoting by $\bm{r}^{(0)}$  a set of sufficient statistics for the detection of $\bm{c}^{(0)}$, the achievable IR, measured in bit per channel use, can be obtained as
\begin{equation}\label{e:IR}
I_{\mathrm{R}} =\lim_{N\rightarrow \infty}\frac{1}{N} E
\Bigg\{\log \frac{q(\bm{r}^{(0)} \arrowvert \bm{c}^{(0)} )}{q(\bm{r}^{(0)})}  \Bigg\}\, ,
\end{equation}
where $N$ is the number of transmitted symbols. 
The probability density functions $q(\bm{r}^{(0)} \arrowvert \bm{c}^{(0)})$
and $q(\bm{r}^{(0)})$ are computed by using the optimal maximum-a-posteriori (MAP) symbol detector for the auxiliary channel, while the expectation in~(\ref{e:IR}) 
is with respect to the input and output sequences generated according to the actual channel model~\cite{ArLoVoKaZe06}. In the next section, we will discuss two 
different low-complexity detectors for nonlinear satellite channels and we will define the corresponding auxiliary channels.

We can define the user's bandwidth as the frequency separation $F$ between two adjacent carriers. The achievable SE is thus
\begin{equation}
\eta=\frac{I_{\mathrm{R}}}{FT}  \quad[
\textrm{b/s}/\textrm{Hz} ] . \label{e:eta}
\end{equation}
The aim of the proposed technique is to find the values of $F$ and $T$ providing, for each value of the signal-to-noise ratio (SNR), the maximum value of SE achievable by that particular receiver, optimal for the considered specific 
auxiliary channel. Namely, we compute
\begin{equation}
\eta_\text{M}=\max_{F,T>0} \eta(F,T) \, . \label{e:eta_M}
\end{equation}
Typically, the dependency on the SNR value is not critical, in the sense that we can identify, for each shaping pulse and modulation format, two or at most three SNR regions for which the  optimal spacings practically have the same 
value. 

Having removed the constraint of orthogonal signaling,
one more degree of freedom in the SE optimization is represented by the bandwidth $W$ of the shaping pulse $p(t)$. Hence, guided by the same idea behind the TF packing technique, we can also optimize $W$, further increasing both ICI 
and ISI due to the adjacent users and to the IMUX and OMUX filters, respectively.
Whereas on the AWGN channel this optimization is implicit in TF packing, in the sense that we can obtain the same ICI by fixing $F$ and increasing $W$ or by fixing $W$ and decreasing $F$, 
this is no more true for our nonlinear channel since IMUX and OMUX bandwidths are kept fixed. Hence, an increased value of $W$ also increases the ISI.
The benefit of the bandwidth optimization is twofold: it can be used  as an alternative to frequency packing (e.g., in cases where the transponder frequency plan cannot be modified and hence frequency packing is not an option),
or it can be used to improve the results of TF packing. In this case, we thus compute 
\begin{equation}
\eta_\text{M}=\max_{F,T,W>0} \eta(F,T,W) \, . \label{e:eta_M2}
\end{equation} 

For fair comparisons in terms of SE, we need a proper definition of the SNR. We define the SNR as the ratio  $P_{\mathrm{sat}}/ N_0F$ between the peak power $P_{\mathrm{sat}}$ at the output of an  
HPA in response to a continuous wave input, denoted as the amplifier saturation power, and the noise power in the bandwidth assigned to each carrier, which coincides with the frequency spacing $F$ between two adjacent carriers. 
This is because  in a satellite forward link, the two main resource constraints are the available frequency spectrum and the radiated power on-board of the satellite.
The adopted SNR definition is independent of the transmit waveform and its parameters. This provides a common measure to compare the performance of different solutions in a fair manner. 
In the following, we also define the output back-off (OBO) for each waveform (or modulation scheme) as the power ratio (in dBs) between the unmodulated carrier at saturation and the modulated carrier after the OMUX. 

Without loss of generality, $T$ and $F$ in (\ref{e:eta})-(\ref{e:eta_M2}) can be normalized to some reference values $T_B$ and $F_B$.
We will denote $\nu=F/F_B$ and $\tau=T/T_B$. In the numerical results, we will choose $T_B$ and $F_B$ as the symbol time and the frequency spacing adopted in the DVB-S2 standard, which is considered as a benchmark scenario. 

\section{Considered detectors and corresponding auxiliary channel models}\label{s:auxiliary channels}

The system model described in~\S\ref{s:system_model} is representative of the considered scenario and has been employed in the information-theoretic analysis and in the simulations results. In this section, we describe 
the employed auxiliary channel models and the corresponding optimal MAP symbol detectors. As explained in~\S\ref{s:spectral efficiency}, they are used to compute two lower bounds on the SE for the considered 
channel~\cite{ArLoVoKaZe06}. Since these lower bounds are achievable by those receivers, we will say that the computed lower bounds are the SE values of the considered channel when those receivers are employed.

\subsection*{Memoryless model and predistortion at the transmitter}\label{ss:predistortion}
When the HPA AM/AM and AM/PM characteristics are properly estimated and fed back to the transmitter, the sequence of symbols $\{c^{(\ell)}_k\}$ can be properly 
predistorted to form the sequence  $\{c'^{(\ell)}_k\}$ that is transmitted instead, in order to compensate for the effect of the non-linearity and possibly to reduce the ISI. 
Here we consider the dynamic data predistortion technique described in~\cite{KaSa91,CaDeGi04} and also suggested for the application in DVB-S2 systems~\cite{DVB-S2-TR}, where the symbol $c'^{(\ell)}_k$ transmitted by user $\ell$ 
at time $k$ is a function of a sequence of $2L_p+1$ input symbols, i.e.,  $c'_k=f(c_{k-L_p},\ldots,c_k,\ldots,c_{k+L_p})$. 
The mapping  $f$ at the transmitter is implemented through a look-up table (LUT), which is computed through an iterative
procedure performed off-line and described in~\cite{KaSa91,CaDeGi04} for each modulation format, setting of the system parameters, and SNR value.  This procedure searches the best trade-off
between the interference reduction and the increase of the OBO.
The complexity at the transmitter depends on the number of symbols accounted for through the parameter $L_p$. 
The transmitted signal is thus
\begin{equation}
 x(t)=\sum_\ell\sum_k c'^{(\ell)}_kp(t-kT)e^{j2\pi\ell Ft}
\end{equation}
whereas, at the receiver, a simple single-user memoryless channel is assumed corresponding to the auxiliary channel (for user with $\ell=0$)
\begin{equation}\label{e:auxiliary_channel_pred}
  r^{(0)}_k=c^{(0)}_k+n_k
\end{equation}
where $n_k$ is a zero-mean circularly symmetric white Gaussian
process with PSD $(N_0+N_I)$, $N_I$ being a design parameter which
can be optimized through computer simulations---an increase of the assumed noise variance can improve 
the computed achievable lower bound on the SE~\cite{BaFeCo09b}.

\subsection*{Model with memory and advanced detection}\label{ss:advanced_detection}
A valid alternative to nonlinear compensation techniques at the transmitter relies upon the adoption of advanced detectors which can manage the nonlinear 
distortions and the ISI. 
In this work, we consider detection based on an approximate signal model described in~\cite{CoPi12}, which comes from a simplified 
Volterra series expansion of the nonlinear channel. 
To limit the receiver complexity with a limited performance degradation, we also apply a CS technique~\cite{FaMa73}.
In fact, when the memory of the channel is too large to be taken into account by a full complexity detector, an excellent performance can be achieved 
by properly filtering the received signal before adopting a reduced-state detector~\cite{FaMa73}.

A very effective CS technique for general linear channels is described in~\cite{RuPr12}, while its extension to nonlinear satellite systems is reported 
in Chapter~\ref{ch:cs_sat}. 
We will denote the memory taken into account by the advanced detection scheme as $L_r$.

\section{Numerical results}\label{s:results}
\subsection*{Spectral efficiency}
Considering the DVB-S2 system as a benchmark scenario, we now
show the improvement, in terms of SE, that can be obtained by adopting the TF packing technique joint with an advanced processing at the receiver. In the following, we assume as reference the time and frequency spacings 
of DVB-S2, i.e., $1/T_B=27.5$~Mbaud and $F_B=41.5$ MHz, and use these values to normalize all time and frequency spacings. Let us consider a typical DVB-S2 scenario where, at the transmitter, the shaping 
pulse $p(t)$ has a RRC spectrum, whereas the IMUX and OMUX filters and the nonlinear characteristics of the HPA are those reported in~\cite[Figs.~H.12 and H.13]{DVB-S2-TR}. 
The standard considers the following modulation formats: QPSK, 8PSK, 16APSK, and 32APSK. 
To combat ISI and nonlinear distortions, a data predistorter is employed at the transmitter whereas at 
the receiver a symbol-by-symbol detector is assumed.
Here, we consider the predistorter described in~\S\ref{ss:predistortion}, with $L_p=2$ for QPSK, 8PSK 16APSK, and $L_p=1$ for 32APSK.
The SE results have been obtained by computing the IR in~(\ref{e:IR}) by means of the Monte Carlo method described in~\cite{ArLoVoKaZe06}. For each case, i.e., for each modulation format, employed detection algorithm, 
and choice of the system parameter, we also performed a coarse optimization of the noise variance to be set at 
receiver~\cite{BaFeCo09b}, 
and of the amplifier operation point through the OBO. Unless otherwise specified, the roll-off factor of the RRC pulses is $\alpha=0.2$, which is the lowest value considered in the standard. 

\begin{figure}
	\begin{center}
		\includegraphics[width=88mm]{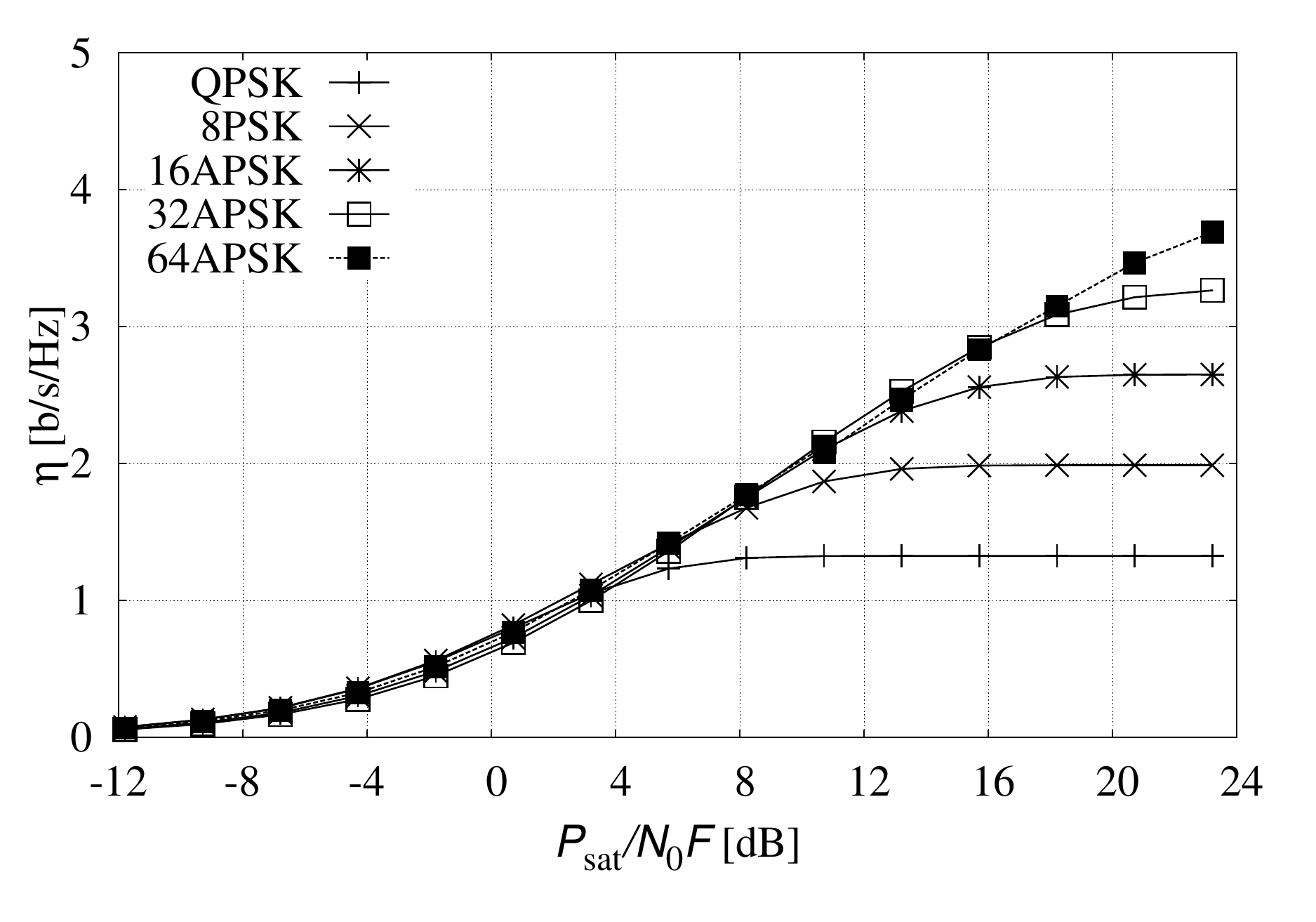}
		\caption{Spectral efficiency of DVB-S2 modulations with roll-off 0.2, data predistortion, and memoryless detection. Comparison with a constellation of increased cardinality (64APSK).}\label{f:ASE_DVB} 
	\end{center}
\end{figure}

We first consider the achievable SE of our benchmark scenario, and in Fig.~\ref{f:ASE_DVB} we report $\eta$ as a function of $P_{\mathrm{sat}} /N_0 F$ for the four modulation formats of the standard (QPSK, 8PSK, 16APSK, and 32APSK). 
In this case, it is $W=(1+\alpha)/T_B$.
We verified that comparable SE values can be also obtained by using, instead of the predistorter, the advanced detection scheme of~\S\ref{ss:advanced_detection} with $L_r=0$ (MMSE detection). We also consider two alternative 
ways that, at least in the case of a linear channel, can be used to improve the SE without resorting to TF packing.\footnote{On a nonlinear satellite channel, due to the increased peak-to-average power ratio, their application must 
be carefully considered since not necessarily produces the expected benefits.} The simplest approach relies on the increase of the modulation cardinality, and in Fig.~\ref{f:ASE_DVB} we also show the SE for the 64APSK 
modulation~\cite{LiAl08}.
It can be seen that the 64APSK modulation, due to the higher impact of the non-linearities, allows to increase the SE only at high SNR values and it seems there is no hope to improve the SE in the low and medium SNR regions.

\begin{figure}
	\begin{center}
		\includegraphics[width=88mm]{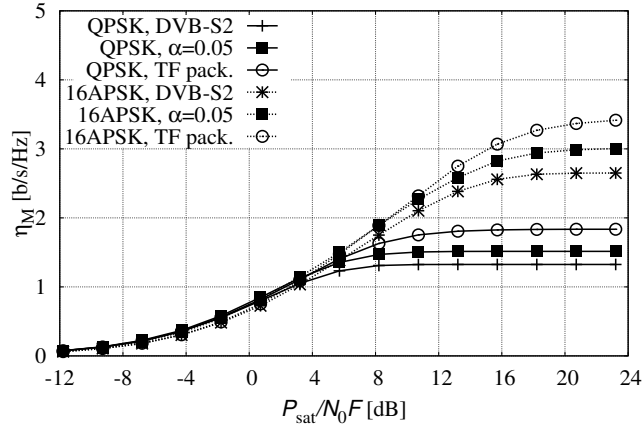}
		\caption{Improvements, in terms of spectral efficiency, that can be obtained for QPSK and 16APSK modulations by reducing of the roll-off ($\alpha=0.05$) or by adopting the TF packing technique. 
		In all cases, predistortion at the transmitter and memoryless detection at the receiver are employed.}\label{f:ASE_TF_pre} 
	\end{center}
\end{figure}

An alternative way of improving the SE is based on a reduction of the roll-off factor. In Fig.~\ref{f:ASE_TF_pre}, we consider QPSK and 16APSK modulations in a scenario where predistortion at the transmitter and symbol-by-symbol 
detection at the receiver are still employed. We show the SE improvement that can be obtained by reducing the roll-off to $\alpha=0.05$.\footnote{We properly modify the transmitted signal such that it occupies the same bandwidth 
as that of the signal with roll-off 0.2. We verified that, in this particular case, no improvement can be obtained by resorting to a more sophisticated receiver based on linear or nonlinear equalization in addition to or in 
substitution of the predistorter.}
We can observe that the roll-off reduction improves the SE with respect to DVB-S2 for all SNR values. On the other hand, as shown in Fig.~\ref{f:ASE_TF_pre}, better results can be obtained by allowing 
TF packing. The values of $T$ and $F$ are chosen as those providing the largest SE. This search is carried out by evaluating~(\ref{e:eta_M}) on grid of values of $T$ and $F$ (coarse search), followed by interpolation 
of the obtained values (fine search). We point out that these curves have been obtained without 
reducing the roll-off factor, which is still $\alpha=0.2$, and employing the same predistorter and symbol-by-symbol receiver adopted in the DVB-S2 system~\cite{KaSa91,CaDeGi04}. 

\begin{figure}
	\begin{center}
		\includegraphics[width=88mm]{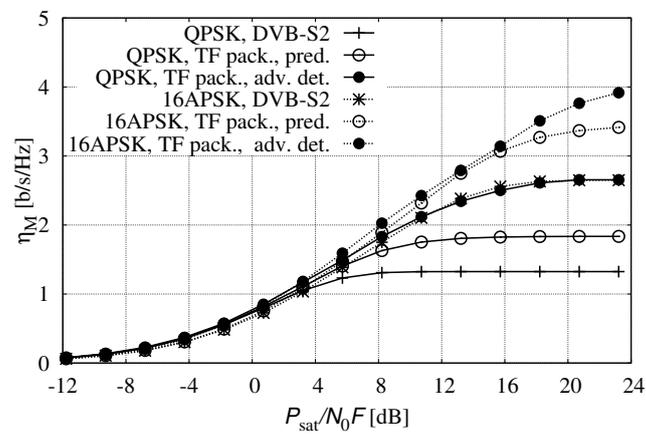}
		\caption{Spectral efficiency for QPSK and 16APSK modulations with TF packing and advanced detector (TF pack., adv. det.). Comparison with the DVB-S2 scenario and the case of TF packing when a predistorter and a symbol-by-symbol detector are adopted (TF pack., pred.).}\label{f:ASE_TF_CS} 
	\end{center}
\end{figure}

With the aim of further improving the performance, we now consider TF packing and a system without predistortion at the transmitter but using the advanced detection algorithm described in~\S\ref{ss:advanced_detection}, joint 
with CS  ($L_r=1$, to reduce the receiver complexity, since in this case the BCJR algorithm has only $M$ states). 
The assumed order of the Volterra model~(\ref{eq:volterra_apprx}) is $v=5$ since, in this case, when considering a single-carrier transmission (no adjacent users) the minimum mean square error between the model and the actual 
signal is very low. The results for QPSK and 16APSK modulations are reported in Fig.~\ref{f:ASE_TF_CS}, where we also show the DVB-S2 benchmark curves  discussed above and the curves related to TF packing 
when predistortion at the transmitter and memoryless detection at the receiver are used. These results show the impressive improvement achievable by TF packing combined with the considered advanced receiver, which, with a memory of 
only one symbol, can cope with much more interference than the schemes employing the predistorter and a memoryless detector.

\begin{figure}
	\begin{center}
		\includegraphics[width=88mm]{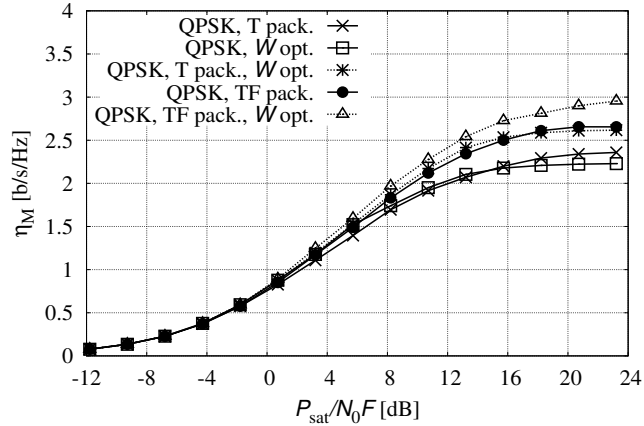}
		\caption{Spectral efficiency of QPSK modulation with TF packing and bandwidth optimization by adopting the advanced receiver with CS ($L_r=1$).}\label{f:ASE_B_PACK}
	\end{center}
\end{figure}

In the previous figures we considered, as mentioned, a bandwidth $W=1.2/T_B$. We also considered the possibility of optimizing the bandwidth $W$, as described in~\S\ref{s:spectral efficiency}.
In Fig.~\ref{f:ASE_B_PACK}, we consider QPSK modulation and the advanced receiver with $L_r=1$. As expected, the combination of TF packing with the bandwidth optimization gives the best results. We also show the results in case only 
time packing or only the bandwidth optimization are adopted. Interestingly, the SE of time packing with bandwidth optimization is quite similar to that achievable by TF packing.

\begin{figure}
	\begin{center}
		\includegraphics[width=88mm]{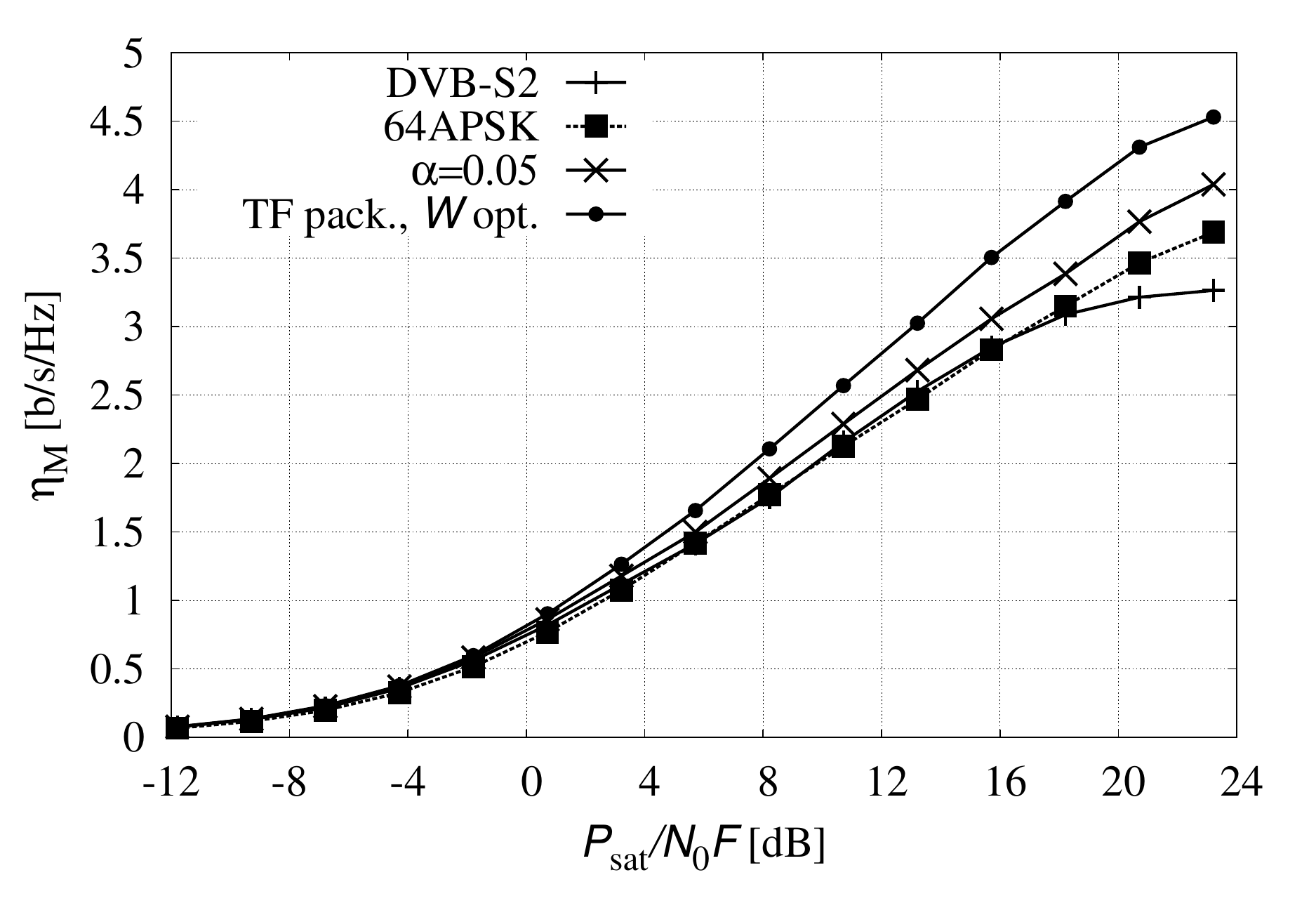}
		\caption{Spectral efficiency of TF packing with bandwidth optimization (TF pack., $W$ opt.). Comparison with DVB-S2, 64APSK and roll-off reduction.}\label{f:ASE_TFB}
	\end{center}
\end{figure}

Finally, to summarize the results, Figure~\ref{f:ASE_TFB} shows the SE for all DVB-S2 modulations (QPSK, 8PSK, 16APSK and 32APSK) with TF packing, bandwidth optimization, and the advanced receiver. For clarity, 
we show only one curve which, for each abscissa, reports only the largest value of the four curves (the ``envelope''). In the same figure, we also plot three other SE curves obtained by using predistortion and a memoryless receiver. 
The lowest one is that corresponding to the DVB-S2 scenario (one curve which is the ``envelope'' of all four curves in Fig.~\ref{f:ASE_DVB}), the SE curve for the 64APSK modulation, and the SE curve in  case of roll-off $\alpha=0.05$ 
reduction. In this latter case, we considered all modulations with cardinality up to 64, and hence this curve represents the effect 
of both roll-off reduction and cardinality increase with respect to DVB-S2. The figure shows that TF packing and advanced receiver processing allows a SE improvement of around 40$\%$ w.r.t. DVB-S2 at high SNR.
This gain is partly due to the fact that current DVB-S2 standard does not support higher order modulations or lower roll-off values. However, there is still considerable SE improvements at lower SNR values.

\subsection*{Modulation and coding formats}
What information theory promises can be approached by using proper coding schemes. All the considered modulation and coding formats (MODCODs) use the low-density parity-check (LDPC) codes with length 64800 bits of the DVB-S2 
standard.  We adopt the optimized values for $T$, $F$, and $W$ and the advanced detector described in~\S\ref{ss:advanced_detection}.
Due to the soft-input soft-output nature of the considered detection algorithm, we can adopt iterative detection and decoding. We distinguish between local iterations, within the LDPC decoder, and global iterations,
between the detector and the decoder. Here, we allow a maximum of 5 global iterations and 20 local iterations.

\begin{figure}
	\begin{center}
		\includegraphics[width=88mm]{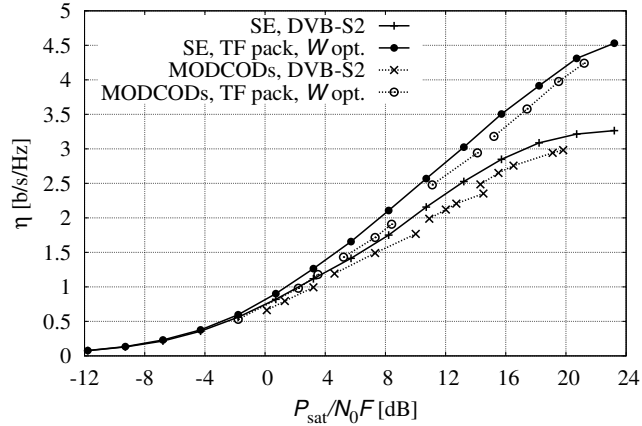}
		\caption{Modulation and coding formats of the DVB-S2 standard and comparison with those designed for the proposed TF packing technique with optimized bandwidth.}\label{f:modcods}
	\end{center}
\end{figure}

BER results have been computed by means of Monte Carlo simulations and are reported in the
SE plane in Figure ~\ref{f:modcods} using, as reference, a BER of $10^{-6}$. In the same figure, the performance of the DVB-S2 MODCODs is also shown for comparison. 
We recall that for them predistortion at the transmitter and symbol-by-symbol detection at the receiver are adopted. Moreover, for them we have $\tau=1$ and $\nu=1$.
The details of the considered MODCODs are reported in Tables~\ref{t:modcods_det} and~\ref{t:modcods_dvbs2}.
These results are in perfect agreement with the theoretical analysis and confirm that the TF packing technique can provide an impressive performance improvement w.r.t. the DVB-S2 standard. 

\begin{table}
\caption{Details of the MODCODs based on TF packing.}\label{t:modcods_det}
    \centering
    \begin{tabular}{c|c|c|c|c|c|c|}
	\cline{2-7}
	& rate & $\tau$ & $\nu$ & $W_{\rm opt}$  &  $\!P/N_0 F\!$  & $\eta$ \\
	&   &   &   &    &    [dB] &   [b/s/Hz]\\
	\hline
	\hline
	\multirow{3}{*}{QPSK} & 1/3 & 0.833  & 1.00 & +20\% & -1.7 & 0.53 \\
	& 1/2 & 0.750 & 0.90 & +20\% & 2.2 & 0.98  \\
	& 3/5 & 0.750 & 0.90 & +20\% & 3.6 & 1.18 \\ 
	\hline
	\multirow{3}{*}{8PSK} & 1/2 & 0.731 & 0.95 & +30\%  & 5.3 & 1.43  \\
	& 3/5 & 0.731 & 0.95 & +30\%  & 7.4 & 1.72  \\
	& 2/3 & 0.731 & 0.95 & +30\% & 8.5 & 1.91  \\
	\hline
	\multirow{2}{*}{16APSK} & 2/3 & 0.792 & 0.90 & +20\% & 11.1 & 2.48  \\
	& 3/4 & 0.750 & 0.90& +20\%  & 14.1 & 2.94  \\
	\hline
	\multirow{4}{*}{32APSK} & 2/3 & 0.731 & 0.95 & +30\% & 15.3 & 3.18  \\
	& 3/4 & 0.731 & 0.95 & +30\%  & 17.5 & 3.58 \\ 
	& 5/6 & 0.731 & 0.95 & +30\%  & 19.5 & 3.98 \\ 
	& 8/9 & 0.731 & 0.95 & +30\%  & 21.2 & 4.24 \\ 
	\hline
    \end{tabular}
\end{table}

\begin{table}
   
     \caption{Details of the DVB-S2 MODCODs. In this case, $\tau=1$ and $\nu=1$.}\label{t:modcods_dvbs2}
     \centering
    \begin{tabular}{c|c|c|c|}
	\cline{2-4}
	& rate &  $P/N_0 F$ [dB] & $\eta$ [b/s/Hz]\\
	\hline
	\hline
	\multirow{3}{*}{QPSK} & 1/2 & 0.1 & 0.66 \\
	& 3/5  & 1.4 & 0.79  \\
	& 3/4 & 3.2 & 0.99 \\ 
	\hline
	\multirow{3}{*}{8PSK} & 3/5 & 4.6 & 1.19 \\
	& 3/4 & 7.3 & 1.49  \\
	& 8/9 & 10.0 & 1.77  \\
	\hline
	\multirow{2}{*}{16APSK} & 3/4 & 10.9 & 1.99  \\
	& 4/5 & 12.0 & 2.12  \\
	& 5/6 & 12.7 & 2.21  \\
	& 8/9 & 14.6 & 2.35  \\
	\hline
	\multirow{4}{*}{32APSK} & 3/4  & 14.3 & 2.48  \\
	& 4/5 & 15.6 & 2.65 \\ 
	& 5/6 & 16.5 & 2.76 \\ 
	& 8/9 & 19.2 & 2.94 \\ 
	& 9/10 & 19.9 & 2.98 \\ 
	\hline
    \end{tabular}
\end{table}

\section{Conclusions}\label{s:conclusions}
We have investigated the TF packing technique, jointly with an advanced processing at the receiver, to improve the spectral efficiency of a nonlinear satellite system employing linear modulations with finite constellations. 
As a first step, through an information-theoretic analysis, we computed the spectral efficiency achievable through this technique showing, with reference to the DVB-S2 specifications, that without an advanced processing at 
the receiver, the potential gains are very limited. On the other hand, a detector which takes into account a memory of only one symbol, and thus with a very limited complexity increase, it is possible to obtain a gain up to 
40\% in terms of spectral efficiency with respect to the conventional use of the current standard. This impressive gain is partly due to optimized carrier spacing of adjacent transponders. Although this assumption 
may not be applicable to all satellite communication systems, the results of this chapter indicates possible new system design directions to further improve the spectral efficiency. 
All these considerations can be extended to other channels and scenarios. 

\cleardoublepage
\appendix
\renewcommand{\chaptermark}[1]{\markboth{{\appendixname}\ \thechapter.\hspace{1em}#1}{}}
	\chapter{Toeplitz matrix, circulant matrix and Szeg\"{o} theorem}\label{ch:toeplitz_matrix}
\thispagestyle{empty}

Toeplitz matrix and circulant matrix are both useful structure
to represent channels with memory.
In this section we denote a $N\times N$ Toeplitz matrix by $\bm{T}_N$,
where the subscript denotes explicitly its dimension. 
A Toeplitz matrix has elements
$(\bm{T}_N)_{ij}=t_{i-j}$, being $\{t_i\}$ a sequence, and reads
\begin{equation}
	\bm{T}_N=\left( 
		\begin{array}{cccc}
			t_0 & t_{-1} & \dots & t_{-(N-1)} \\
			t_1 & t_{0} & \dots & t_{-(N-2)} \\
			\vdots &  & \ddots & \vdots \\
			t_{N-1} & t_{N-2} & \dots & t_{0}
		\end{array}
	\right) \,. \label{eq:T_toeplitz}
\end{equation}
The most common application of Toeplitz matrix in this thesis,
is to represent a filtering.
Namely, if we consider a sequence $\{x_i\}$
filtered by  $\{t_i\}$, the sequence at the output of the filter reads
\begin{equation}
	y_k= \sum_i t_i x_{k-i} \,. \label{eq:yk_conv}
\end{equation}
The convolution (\ref{eq:yk_conv}) can be also written by means of the following matrix notation
\begin{equation}
	\bm{y}= \bm{T}_N\bm{x} \label{eq:T_matrix}
\end{equation}
where $\bm{x}=[x_0,\dots,x_{N-1}]^T$.

A circulant matrix $\bm{C}_N$ is a special case of Toeplitz matrix constructed
on a sequence such that $t_{-k}=t_{N-k}$.
With such sequence, the matrix reads
\begin{equation}
	\bm{C}_N=\left( 
		\begin{array}{ccccc}
			t_0 & t_{N-1} & t_{N-2} & \dots & t_{1} \\
			t_1 & t_{0} & t_{N-1} & \dots & t_{2} \\
			t_2 & t_1 & t_{0} & & t_3 \\
			\vdots &  &  & \ddots & \vdots \\
			t_{N-1} & t_{N-2} & \dots & \dots &  t_{0}
		\end{array}
	\right) \,. \label{eq:C_circ}
\end{equation}
In other words, in a circulant matrix all rows are cyclic shift of the first row.
One important property of a circulant matrix is related to its eigenvalue decomposition~\cite{HoJo85}.
Said $\bm{F}$ the $N\times N$ Fourier matrix, with elements $(\bm{F})_{ik}=e^{-j2\pi ik/N}$,
it can be shown that any circulant matrix can be decomposed as
\begin{equation}
	\bm{C}_N=\bm{F}^\dagger \bm{\Lambda} \bm{F}
\end{equation}
where $\bm{\Lambda}$ is the diagonal matrix containing the eigenvalues $(\bm{\Lambda})_{ii}=C_i$,
being $C_i$ the eigenvalues of $\bm{C}_N$.
It can be shown that these eigenvalues are given by the discrete fourier trasform (DFT)\footnote{The DFT must not be confused with the discrete time Fourier transform (DTFT).}
\begin{equation}
	C_i = \sum_{k=0}^{N-1} t_k e^{-j2\pi k i/N} \,.
\end{equation}

From the two definitions of Toeplitz matrix (\ref{eq:T_toeplitz}) and circulant matrix (\ref{eq:C_circ}) it can be expected
that Toeplitz matrix and circulant matrix can have similar properties.
In particular, by using na\"{\i}ve words, we can expect that a Toeplitz matrix $\bm{T}_N$, for $N\rightarrow \infty$
{\it behaves} like a circulant matrix, and thus also its eigenvalues are related to the Fourier transform of \{$t_i$\}.
More formally this relation is given by the Szeg\"{o} theorem.

\begin{thm}[Szeg\"{o} theorem]
 Let \{$\bm{T}_N$\} be a sequence of $N\times N$ Toeplitz matrix such that $\{t_i\}$
 is absolutely summable. Let $\{\tau_{N,i}\}_{i=0}^{N-1}$ be the eigenvalues
 of $\bm{T}_N$ and s any positive definite integer. Then
\begin{equation}
	\lim_{N\rightarrow\infty} \frac{1}{N} \sum_{i=0}^{N-1} \tau_{N,i}^s = \frac{1}{2\pi} \int_{0}^{2\pi} T^s(\omega) \mathrm{d}\omega
\end{equation}
where $T(\omega)$ is the discrete time Fourier transform (DTFT)
\begin{equation}
	T(\omega)= \sum_{i=0}^{N-1}t_i e^{-j\omega i}\,.
\end{equation}

Furthermore, if $T(\omega)$ is real, with essential infimum $m$ and essential supremum $M$, 
then for any continuous function $f:[m,M] \rightarrow [0,\infty)$  
\begin{equation}
 \lim_{N\rightarrow\infty} \frac{1}{N}\sum_{i=0}^{N-1} f(\tau_{N,i}) = \frac{1}{2\pi} \int_{0}^{2\pi} f(T(\omega)) \mathrm{d}\omega .
\end{equation}
\end{thm}
A simple application of the theorem, is given by setting $f$ as the logarithm. This gives
the identity
\begin{equation}
	\lim_{N\rightarrow\infty}  \frac{1}{N} \log \det \bm{T}_N = \frac{1}{2\pi} \int_{0}^{2\pi} \log(T(\omega)) \mathrm{d}\omega .
\end{equation}
If we now consider as example the matrix channel (\ref{eq:matrix_ungerboeck}),
its capacity is given by $\log_2 \det(\bm{I}+\bm{G}/N_0)$,
which for $N\rightarrow\infty$ tends to
\begin{equation}
	\lim_{N\rightarrow\infty} \frac{1}{N}\log_2 \det\left(\bm{I}+\frac{\bm{G}}{N_0}\right) = \frac{1}{2\pi}\int_{0}^{2\pi} \log\left(1+ \frac{G(\omega)}{N_0}\right) \mathrm{d}\omega 
\end{equation}
as shown in \cite{Hi88}.

For further details with a very clean explation, further consequences and applications, the reader can see \cite{Gr06}.

\chapter{CS for channels represented by a block Toeplitz matrix}\label{ch:app_cs}
\thispagestyle{empty}

In this appendix we derive the CS solution when the channel model reads
\begin{equation}
  \mathbf{r} = \mathbf{H}\mathbf{c}+ \mathbf{w} \,, 
\end{equation}
where $\mathbf{H}$ is block lower triangular and Toeplitz matrix with size
$KN\times KN$ built from a sequence of matrices $\{\bm{H}_i\}_{i\geq0}$
with size $K\times K$.
Namely it reads
\begin{equation}
	\mathbf{H}=\left( 
		\begin{array}{cccc}
		 \bm{H}_0 & \bm{0} & \dots & \bm{0} \\ 
		 \bm{H}_1 & \bm{H}_0 & \dots & \bm{0} \\ 
		 \vdots &  & \ddots & \vdots \\
		 \bm{H}_{N-1} & \bm{H}_{N-2} & \hdots  & \bm{H}_0 
		\end{array}
	\right) \,.
\end{equation}
$\mathbf{c}$ is assumed to be a block vector of complex Gaussian random variables, 
with mean zero and  autocorrelation matrix \mbox{$\mathbf{V}=\mathrm{E}\{\mathbf{c}\mathbf{c}^\dagger\}$}.
We constrain the autocorrelation matrix $\mathbf{V}$ to be block diagonal as
\begin{equation}
	\mathbf{V}=\left( 
		\begin{array}{cccc}
		 \bm{V} & \bm{0} & \dots & \bm{0} \\ 
		 \bm{0} & \bm{V} & \dots & \bm{0} \\ 
		 \vdots &  & \ddots & \vdots \\
		 \bm{0} & \bm{0} & \hdots  & \bm{V} 
		\end{array}
	\right) \,,
\end{equation}
where $\bm{V}$ is a positive definite matrix.
The CS detector considers a mismatched channel law
\begin{equation}
	q(\mathbf{r}|\mathbf{c}) \propto \mathrm{exp}\left\{ \Re \left( \mathbf{c}^\dagger (\mathbf{H}^r)^\dagger \mathbf{r}\right) - \mathbf{c}^\dagger\mathbf{G}^r\mathbf{c}    \right\} \,. 
\end{equation}
where the channel shortener $\mathbf{H}^r$ and the target response $\mathbf{G}^r$
are block Toeplitz matrices such that the AIR is maximized for a given memory $L$ taken into account at the detector.
The target response has constraint
\begin{equation}
(\mathbf{G}^r)_{ij}=\bm{0}  \quad \forall |i-j|>L
\end{equation}
where with $(\mathbf{G}^r)_{ij}$ we mean the $(i,j)$ block, and $\bm{0}$ is $K\times K$ matrix of all zeros.

To derive the optimal channel shortener and target response, as first
step we need a closed formula for the AIR
\begin{eqnarray}
	I_\mathrm{R} & = & \mathfrak{h}(\bm{r})-\mathfrak{h}(\bm{r}|\bm{c}) \, \\
	& = & \mathrm{E}\left\{ \log_2  \frac{q(\bm{r}|\bm{c})}{q(\bm{r})} \right\} \,.
\end{eqnarray}
$q(\bm{r})$ is found to be
\begin{eqnarray}
  q(\mathbf{r}) & = & \frac{1}{\pi^{KN} \det(\mathbf{V} )} \int q\left(\mathbf{r} |\mathbf{c}\right)\exp \left\{ -\mathbf{c}^\dagger \mathbf{V}^{-1} \mathbf{c} \right\}\mathrm{d}\mathbf{c} \\
  & = & \frac{1}{\det(\mathbf{G}^r\mathbf{V}+\mathbf{I})} \exp\{ \mathbf{d}^\dagger \left( \mathbf{G}^r+\mathbf{V}^{-1} \right)^{-1} \mathbf{d} \} 
\end{eqnarray}
where $\mathbf{d}=\mathbf{H}^r \mathbf{r}$.
Therefore,
\begin{equation}
   \mathfrak{h}(\bm{r}) =  \log \det(\mathbf{G}^r\mathbf{V}+\mathbf{I})  
      - \mathrm{Tr}\left( (\mathbf{H}^r)^\dagger \left[ \mathbf{H}\mathbf{V}\mathbf{H}^\dagger+ N_0\mathbf{I}  \right] \mathbf{H}^r(\mathbf{G}^r+\mathbf{V}^{-1})^{-1} \right)  
\end{equation}
and
\begin{equation}    
\mathfrak{h}(\bm{r}|\bm{c})  =  \mathrm{Tr}(\mathbf{G}^r\mathbf{V}) - 2\Re\left( \mathrm{Tr} \left( (\mathbf{H}^r)^\dagger\mathbf{H}\mathbf{V}   \right) \right)   \,.
\end{equation}
The derivative of $I_{\mathrm{R}}$ w.r.t. $(\mathbf{H}^r)^\dagger$ is
\begin{equation}
  \frac{\partial I_{\mathrm R} }{ \partial (\mathbf{H}^r)^\dagger}  =   \left( \mathbf{H}\mathbf{V} \right)^T  -\left(\left[ \mathbf{H}\mathbf{V}\mathbf{H}^\dagger+ N_0\mathbf{I}  \right] \mathbf{H}^r(\mathbf{G}^r+\mathbf{V}^{-1})^{-1} \right)^T \,. 
\end{equation}
By setting the derivative to zero, we obtain that 
the optimal filter $\mathbf{H}^r$ is
\begin{equation}
 \mathbf{H}^r = \left[\mathbf{H}\mathbf{V}\mathbf{H}^\dagger+ N_0\mathbf{I} \right]^{-1} \mathbf{H} \mathbf{V}   \left( \mathbf{G}^r+\mathbf{V}^{-1}\right)  \,.  \label{eq:opt_frontend}
\end{equation}
Using (\ref{eq:opt_frontend}), the $I_{\mathrm{R}}$ is
\begin{equation}
  I_{\mathrm{R}} = \frac{1 }{ N} \left(  \log(\det(\mathbf{U}^\dagger\mathbf{U}\mathbf{V})) - \mathrm{Tr}\left( \mathbf{U}\mathbf{B}\mathbf{U}^\dagger\right) +KN \right)  \label{eq:Ir_proof}
\end{equation}
where $\mathbf{U}$ is obtained from the Cholesky decomposition \mbox{$\mathbf{G}^r+\mathbf{V}^{-1}= \mathbf{U}^\dagger \mathbf{U}$} and 
\begin{equation}
    \mathbf{B} = \mathbf{V}- \mathbf{V}\mathbf{H}^\dagger \left[\mathbf{H}\mathbf{V}\mathbf{H}^\dagger+ N_0\mathbf{I} \right]^{-1}\mathbf{H}\mathbf{V} \,. \label{eq:B_proof}
\end{equation}
The $(m,n)$ entries of $\mathbf{U}$ and $\mathbf{B}$ will be denoted by $\bm{U}_{mn}$ and $\bm{B}_{mn}$, respectively.
Since  $\det(\mathbf{U}^\dagger\mathbf{U}\mathbf{V})$ depends only on the diagonal elements of $\mathbf{U}$, 
we can optimize $I_{\mathrm{R}}$ over the diagonal of $\mathbf{U}$ and the
off-diagonal elements separately.
We define $\mathbf{\underline{U}}_n=[\bm{U}_{n\,n+1},..,\bm{U}_{n\,\min(n+L,N)}]$, $\mathbf{\underline{B}}_n=[\bm{B}_{n\,n+1},..,\bm{B}_{n\,\min(n+L,N)}]$,
\begin{equation} 
  \mathbf{B}_n = \left[ 
    \begin{array}{ccc}
      \bm{B}_{(n+1)\,(n+1)} & \cdots & \bm{B}_{\min(n+1,L)\,(n+L)} \\
       \vdots & \ddots & \vdots \\ 
      \bm{B}_{\min(n+L,N)\,(n+1)} & \cdots & \bm{B}_{\min(n+1,L)\,\min(n+1,L)}
    \end{array}
  \right]
\end{equation}
and finally 
\begin{equation} 
  \mathbfcal{C}_n = \bm{B}_{nn} - \mathbf{\underline{B}}_n\mathbf{B}_n^{-1}(\mathbf{\underline{B}}_n)^\dagger \,. 
\end{equation}
Now the trace $\mathrm{Tr}\left(\mathbf{U}\mathbf{B}\mathbf{U}^\dagger \right)$ can be rewritten as
\begin{equation} 
\sum_n\mathrm{Tr}
 \left( 
      [\bm{U}_{nn}\, \mathbf{\underline{U}}_n] 
      \left[  
        \begin{array}{cc}
          \bm{B}_{nn} & \mathbf{\underline{B}}_n \\
          \mathbf{\underline{B}}_n^{\dagger} & \mathbf{B}_n
        \end{array}
      \right]  
      \left[ \begin{array}{c}                       
        \bm{U}_{nn}^ \dagger \\                                                                                                   
        \mathbf{\underline{U}}_n^\dagger
      \end{array}
      \right]      
 \right) \,.
 \vspace{-1mm}
\end{equation}
Setting its derivative w.r.t. $\mathbf{\underline{U}}_n^\dagger$ to zero gives
\begin{equation}
  \frac{\partial  }{ \partial \mathbf{\underline{U}}_n^\dagger} \mathrm{Tr}\left(  \mathbf{U}\mathbf{B}\mathbf{U}^\dagger \right)= (\bm{U}_{nn}\mathbf{\underline{B}}_n )^{T} + (\mathbf{\underline{U}}_n \mathbf{B}_n )^{T} = \bm{0}
\end{equation}
which gives
\begin{equation}
  \mathbf{\underline{U}}_{n}= -\bm{U}_{nn}  \mathbf{\underline{B}}_n\mathbf{B}_{n}^{-1} \,. \label{eq:u_opt}
\end{equation}
Replacing (\ref{eq:u_opt}) in (\ref{eq:Ir_proof}) we find
\begin{equation}
  I_{\mathrm{R}}=\frac{1}{ N} \log\det(\mathbf{V}) + K 
   + \frac{1 }{N} \sum_n \log(\det(\bm{U}_{nn}^\dagger\bm{U}_{nn})) - \mathrm{Tr}\left( \bm{U}_{nn}\mathbfcal{C}_n\bm{U}_{nn}^\dagger  \right) \label{eq:IR_partial_opt}
\end{equation}
that can be maximized by setting its derivative w.r.t. $\bm{U}_{nn}^\dagger$ equal to zero.
This gives that 
\begin{equation}
  \frac{ \partial I_\mathrm{R}  }{ \partial \bm{U}_{nn}^\dagger} = (\bm{U}_{nn}^*)^{-1} -(\bm{U}_{nn}\mathbfcal{C}_n)^T =\bm{0}
\end{equation}
and the optimal $\bm{U}_{nn}$ is given by the Cholesky decomposition
\begin{equation}
 \mathbfcal{C}_n^{-1} = \bm{U}_{nn}^\dagger \bm{U}_{nn} \label{eq:Copt}
\end{equation}
Inserting (\ref{eq:Copt}) into (\ref{eq:IR_partial_opt}), the AIR for
Gaussian symbols is
\begin{equation}
  I_{\mathrm{R}}= \frac{1}{ N} \log\det(\mathbf{V}) + \frac{1}{N} \sum_{n} \log(\det(\mathbfcal{C}_n^{-1})) \,.
\end{equation}
When $N \rightarrow \infty$, all $\mathbfcal{C}^{-1}_n$ are the same, and we obtain the stationary solution
\begin{equation}
	I_{\mathrm{R}} = \log\det(\bm{V})+ \log\det(\mathbfcal{C}^{-1})
\end{equation}
and (\ref{eq:opt_frontend}), (\ref{eq:B_proof}) become stationary as
\begin{eqnarray}
	 \bm{H}^r(\omega) & = & \left[\bm{H}(\omega)\bm{V}\bm{H}^\dagger(\omega)+ N_0\bm{I} \right]^{-1} \bm{H}(\omega) \bm{V}   \left( \bm{G}^r(\omega)+\bm{V}^{-1}\right)  \, \\
	\bm{B}(\omega) & = & \bm{V}- \bm{V}\bm{H}^\dagger(\omega) \left[\bm{H}(\omega)\bm{V}\bm{H}^\dagger(\omega)+ N_0\bm{I} \right]^{-1}\bm{H}(\omega)\bm{V} \,.
\end{eqnarray}

\chapter{Proof of Theorem \ref{thm:opt_pulse}}\label{ch:app_th_opt_p}

We first note that $P(\omega)$ only enters the
optimization through its square magnitude, and we therefore make the
variable substitution $S_p(\omega)=|P(\omega)|^2$ and optimize over
$S_p(\omega)$ instead. 

The proof will consist of three steps
\begin{itemize}
\item A formula for stationary points.
\item The observation that some of these do not have strictly positive spectrum.
\item Fixing the problem identified in the previous bullets.
\end{itemize}
Let us now start with the first bullet.
From Cramer's rule~\cite{HoJo85}, we get that 
$${\bm B}^{-1} = \frac{1}{\det({\bm B})}[C_{ij}],$$
where $C_{ij}$ is the cofactor of entry $(i,j)$ in ${\bm B}$.
This implies that in (\ref{cc}) we can express ${\bm b}{\bm B}^{-1}{\bm b}^{\dagger}$ as
\begin{equation} \nonumber
\frac{\sum_{m=1}^M \alpha_m
  b_0^{\phi_{m,0}}b_1^{\phi_{m,1}}(b_1^*)^{\phi_{m,2}} \cdots b_L^{\phi_{m,2L-1}} (b_L^*)^{\phi_{m,2L}} }{\sum_{n=1}^N \beta_n
  b_0^{\psi_{n,0}}b_1^{\psi_{n,1}}(b_1^*)^{\phi_{m,2}}\cdots b_{L-1}^{\psi_{n,2L-3}} (b^*_{L-1})^{\psi_{n,2L-2}}  },
\end{equation}
where $M$ and $N$ are finite constants that depend on $L$, $\alpha_m,\beta_m\in\{\pm 1\}$, and both
$\phi_{m,\ell}$ and $\psi_{n,\ell}$ are non-negative integers which
satisfy
$$\sum_{\ell=0}^{2L} \phi_{m,\ell}=L+1\quad \textrm{and}\quad \sum_{\ell=0}^{2L-2} \psi_{n,\ell}=L\,.$$

We next introduce the variable substitution 
$$y(\omega)=\frac{N_0}{|H(\omega)|^2S_p(\omega)+N_0},\,\, S_p(\omega) =\frac{N_0}{|H(\omega)|^2}\left[\frac{1}{y(\omega)}-1\right].$$
The constraint $\int S_p(\omega)\mathrm{d}\omega=2\pi$ translates into
$$e[y(\omega)] \triangleq \int_{-\pi}^{\pi} \frac{1}{y(\omega)|H(\omega)|^2}\mathrm{d}\omega =
\int_{-\pi}^{\pi}\frac{1}{|H(\omega)|^2}\mathrm{d}\omega +\frac{2\pi}{N_0}.$$
Furthermore, we have
$$b_i = \frac{1}{2\pi}\int_{-\pi}^{\pi} y(\omega) e^{j\omega i}\mathrm{d}\omega.$$
The constrained Euler-Lagrange equation~\cite{Ch10} becomes
$$\frac{\delta \mathcal{C}}{\delta y} = \lambda \frac{\delta e}{\delta y}=-\frac{\lambda}{|H(\omega)|^2y^2(\omega)}.$$
The functional derivative $\delta b_k^s/\delta y$ equals
\begin{eqnarray}
 \frac{\delta b_i^s}{\delta y}& =& \frac{\delta \left[\int_{-\pi}^{\pi} y(\omega)e^{j\omega i}\mathrm{d}\omega\right]^s}{\delta y}\nonumber\\
				& =& s\left[\int_{-\pi}^{\pi} y(\omega)e^{j\omega i}\mathrm{d}\omega\right]^{s-1} e^{j\omega i}\nonumber \\
				& =& sb_i^{s-1}e^{j\omega i}. \nonumber
\end{eqnarray}
We now note that $b_i$, raised to any power, is a \emph{constant} that
    depends explicitly on $y$. Therefore,
    by an application on the quotient rule for the derivative and the chain rule to (\ref{cc}),
  we obtain an expression of the form
$$\frac{\delta \mathcal{C}}{\delta y} = 1 -\frac{\sum_{\ell=-L}^L
  A_{\ell}[y(\omega)]e^{j\ell\omega}}{C[y(\omega)]},$$
where the constants $A_{\ell}[y(\omega)]$ and $C[y(\omega)]$ explicitly depend on $y(\omega)$, e.g., $$C[y(\omega)]=\left[\sum_{n=1}^N \beta_n
  b_0^{\psi_{n,0}}b_1^{\psi_{n,1}}\cdots b_{L-1}^{\psi_{n,2L-3}} (b^*_{L-1})^{\psi_{n,2L-2}}\right]^2\,.$$
By manipulation of the Euler-Lagrange equation and by introducing a new set of
  constants $\{B_{\ell}[y(\omega)]\}$, we obtain
$$y(\omega) =\frac{1}{\sqrt{|H(\omega)|^2[\sum_{\ell=-L}^L
    B_{\ell}[y(\omega)]e^{j\ell \omega}]}}.$$
This translates into a general form of the optimal $S_p(\omega)$ which
    reads
\begin{equation}
 \label{DegradedSerieA} S_p^{\mathrm{opt}}(\omega) =
\frac{N_0}{\sqrt{|H(\omega)|^2}}\sqrt{\sum_{\ell=-L}^L A_{\ell}e^{j\ell \omega}}-\frac{N_0}{|H(\omega)|^2} 
\end{equation}
where coefficients $A_{\ell}$ must have a Hermitian symmetry.

We have now found a general form for any stationary
point. Unfortunately, for a given $H(\omega)$, this stationary point
may lie outside the domain of the optimization. The
optimal spectrum $S_p(\omega)$ must therefore lie on the boundary of the
optimization domain, which in this case implies that \mbox{$S_p(\omega)=0$} for
$\omega\in\mathcal{I}_0\subset [-\pi,\pi].$ Let us define $\mathcal{I}_+$ as the subset $[-\pi,\pi]$ where $S_p(\omega)>0$ except for the endpoints of $\mathcal{I}_+$ where $S_p(\omega)=0$ due to the assumption of a continuous spectrum. Note that $\mathcal{I}_+$  may be the union of several disjoint sub-intervals of $[-\pi,\pi]$. We can now rewrite the constraint and the expressions of $b_k$ as

$$e[y(\omega)]=
\int_{\mathcal{I}_+}\frac{1}{|H(\omega)|^2}\mathrm{d}\omega +\frac{2\pi}{N_0}$$
and
$$b_i = \frac{1}{2\pi}\int_{\mathcal{I}_+} y(\omega) e^{j\omega i}\mathrm{d}\omega.$$
From the first part of the proof, i.e., identifying a necessary condition for stationary points, we have that 
(\ref{DegradedSerieA}) must hold within the interval $\mathcal{I}_+$, and the constants $\{A_{\ell}\}$ must be such that $ S_p^{\mathrm{opt}}(\omega)=0$ at the end-points of each sub-interval within $\mathcal{I}_+$. Hence, no matter what $\mathcal{I}_+$ is,  we can express the optimal $S_p^{\mathrm{opt}}(\omega)$ as in (\ref{eq:Pw_opt}).

\chapter{Proof of theorem \ref{th:memory}}\label{ch:app_cs_memory}

The waterfilling algorithm provides a
transmit filter that satisfies \cite{Hi88}
\begin{equation}
 |P(\omega)|^2 = \max\left(0,\theta-\frac{N_0}{|H(\omega)|^2}\right) \label{eq_Pw}
\end{equation}
for some power constant $\theta$. In view of Theorem 1,
$|P(\omega)|^2$ in (\ref{eq_Pw}) must also satisfy (\ref{eq:Pw_opt}).
Equating (\ref{eq_Pw}) and (\ref{eq:Pw_opt}) yields
\begin{equation}
\label{eq:equality_water_opt} \theta-\frac{N_0}{|H(\omega)|^2}=\frac{N_0}{\sqrt{|H(\omega)|^2}}\sqrt{\sum_{\ell=-\nu_C}^{\nu_C} A_{\ell} e^{j\ell \omega}}-\frac{N_0}{|H(\omega)|^2}. 
\end{equation}
From (\ref{eq:equality_water_opt}), it can be seen that we must have 
\begin{equation}
\sum_{\ell=-\nu_C}^{\nu_C} A_{\ell} e^{j\ell \omega} = \gamma |H(\omega)|^2\,, 
\end{equation}
for some constant $\gamma$. However, 
\begin{equation}
|H(\omega)|^2 = \left|\sum_{\ell=0}^{\nu} h_\ell e^{-j\ell \omega} \right|^2= \sum_{\ell=-\nu}^{\nu} g_{\ell}e^{-j\ell \omega}\,,  
\end{equation}
where 
\begin{equation}
g_\ell=\sum_k {h_k h^*_{k-\ell}}\,. 
\end{equation}
Clearly,  to satisfy
$$\sum_{\ell=-\nu_C}^{\nu_C} A_{\ell} e^{j\ell \omega} =\gamma\left[\sum_{\ell=-\nu}^{\nu} g_{\ell}e^{-j\ell \omega}\right],$$
$\nu_C$ must at least equal  $\nu$. 

\cleardoublepage
\addcontentsline{toc}{chapter}{\numberline{}References}

\bibliographystyle{ieeetr}    
\bibliography{Refs,Refs_andre}     
\cleardoublepage
\addcontentsline{toc}{chapter}{\numberline{}Acknowledgement}
      \chapter*{Acknowledgement}
\markboth{Acknowledgement}{Acknowledgement}
\thispagestyle{empty}

Probably the {\it acknowledgement} is my favorite part.
The reason is two fold: first, it means that I finished my thesis (yes!).
Second, it is the only part of my thesis written with the heart,
and not the reason.

So, let me start from Lena. Let be $f : \mathbb{R}\rightarrow \mathbb{N}$, such that
for a given time $t$, it is the number of thanks that she deserves, it holds
\begin{equation}\nonumber
	\forall M>0, M \in \mathbb{N}, \exists t_0>0\,\, \mathrm{s.t.}\,\, f(t)>M, \forall t>t_0   \, .
\end{equation}
(I hope that finally you will understand the limit of a function).
I cannot avoid to cite my family and co-family: my parents, my sister, my brothers-in-law with their {\it lumachina}
eating-pizza-all-time, and their crazy frog (impossible write their actual names without a typo, so I will not at all).

Moving to the University: Amina (also well known as $\mathrm{RNH_2}$) does not want to get any acknowledgement, thus I will not. 
I just let know that she was for me
like a co-advisor during my whole PhD.
Then, without a particular order, my thanks to Ale, Tommy, Nicolo (this thesis is written in English, thus
without {\it accento}) and the other guys of {\it Pal 2}.

Cannot forget all the Lund-University guys: Fredrik, Dzevdan, Rohit, Martina, Egle, Taimoor,
and everyone else made my stay in Lund wonderful.

Last, but not least my closest friends: Giubbos, {\it Ingegner duca conte} Pier,
Ruben, {\it Ragionier} Villazzi, and all the other guys of {\it the company} as well. Also need to cite Maria Paola, lost somewhere in the huge Milan, but
without forgetting us in Parma.

And that's all... just kidding! I did not forget you Giulio! But you know, I'm not leaving now, and this is not a farewell,
so let time pass and {\it put off till tomorrow what you can do today}. Mmm, maybe it was different, but it does not matter.

\begin{flushright}
\vspace{1.0cm}
 {\it\Large Andrea}
 \vspace{1.0cm}
\end{flushright}

\begin{center}
	\includegraphics[width=0.55\paperwidth]{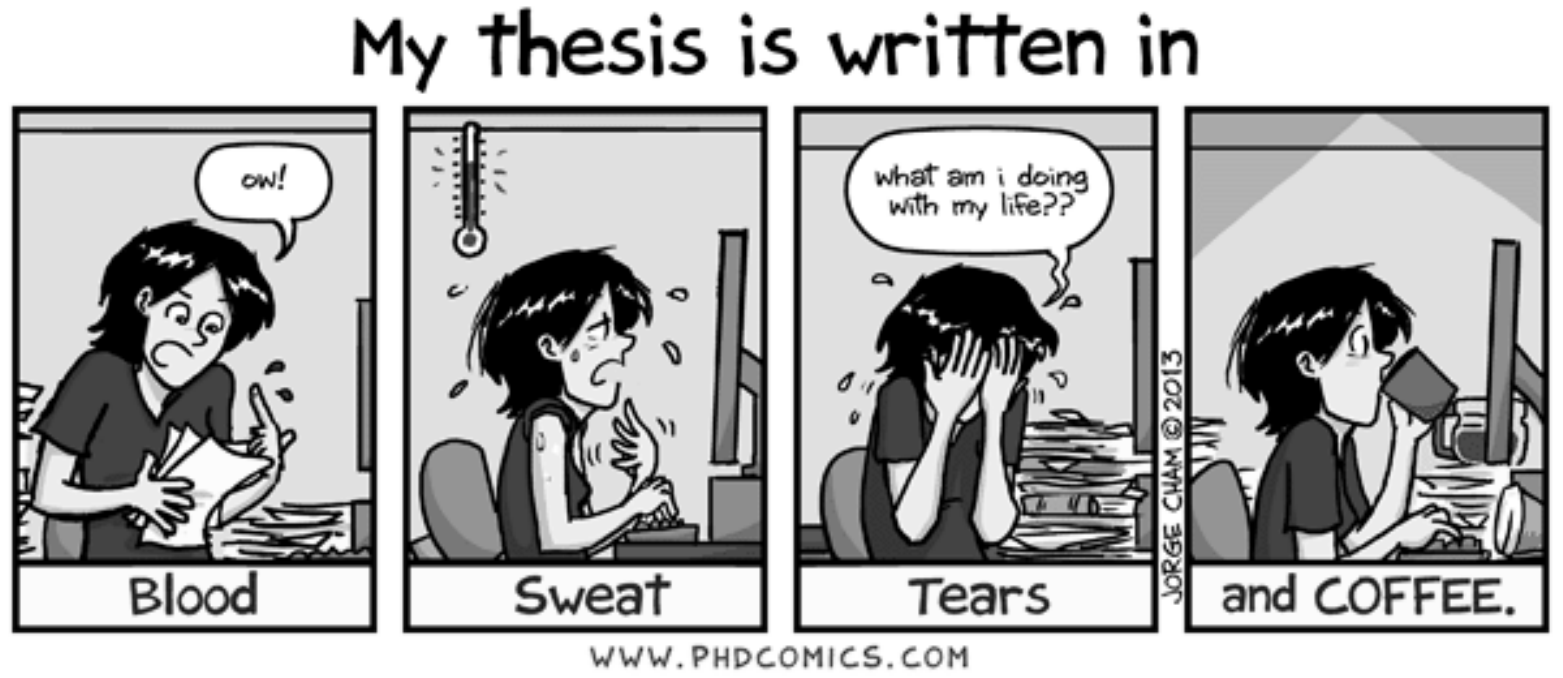}\\
	{\tiny \textcopyright Piled Higher and Deeper by Jorge Cham, www.phdcomics.com. Permission for publication granted
	to the author of this thesis, in december 2013.}
	\vspace{1.0cm}
\end{center}

\end{document}